\documentclass[fleqn,usenatbib]{mnras}

\usepackage{newtxtext,newtxmath}

\usepackage[T1]{fontenc}

\DeclareRobustCommand{\VAN}[3]{#2}
\let\VANthebibliography\thebibliography
\def\thebibliography{\DeclareRobustCommand{\VAN}[3]{##3}\VANthebibliography}

\usepackage{graphicx}	
\usepackage{amsmath}	
\usepackage{xcolor}

\usepackage{CJKutf8}    
\newcommand{\chinese}[1]{\begin{CJK}{UTF8}{bsmi}#1\end{CJK}}

\newcommand{\feh}{\ensuremath{[\mathrm{Fe/H}]}}

\newcommand{\alphafe}{\ensuremath{[\mathrm{\alpha/Fe}]}}
\newcommand{\hbetao}{\ensuremath{\mathrm{H}\beta_{\mathrm{o}}}}

\newcommand{\mh}{\ensuremath{[\mathrm{M/H}]}}
\newcommand{\ebv}{\ensuremath{E(B-V)}}

\newcommand{\angstrom}{\mbox{\normalfont\AA}}
\newcommand{\vlos}{$V_{\rm LOS}$}
\newcommand{\dispersion}{$\sigma$}
\newcommand{\instdispersion}{$\sigma_{\rm inst}$}
\newcommand{\truedispersion}{$\sigma_{\rm{true}}$}
\newcommand{\logage}{$\log \rm{age}$}

\newcommand{\galcraft}{\textsc{GalCraft}}
\newcommand{\glx}{\textsc{Galaxia}}
\newcommand{\eglx}{\textsc{E-Galaxia}}
\newcommand{\ppxf}{\textsc{pPXF}}
\newcommand{\ulyss}{\textsc{ULYSS}}

\newcommand{\kms}{{\rm km s}$^{-1}$}
\newcommand{\perpixel}{$\text{pix}^{-1}$}
\newcommand{\pergyrdex}{$\text{Gyr}^{-1} \text{dex}^{-1}$}
\newcommand{\pergyr}{$\text{Gyr}^{-1}$}
\newcommand{\perdex}{$\text{dex}^{-1}$}
\newcommand{\solarmass}{\ensuremath{\mathrm{M}_{{\odot}}}}
\newcommand{\solarlum}{\ensuremath{\mathrm{L}_{{\odot}}}}

\title[Validating \ppxf{} using mock Milky Way IFS data]{Validating full-spectrum fitting with a synthetic integral-field spectroscopic observation of the Milky Way}

\author[Zixian Wang et al.]{
Zixian Wang (\chinese{王梓先}),$^{1,2,3}$\thanks{E-mail: wang.zixian.astro@gmail.com}
Sanjib Sharma,$^{1,2,4}$\thanks{These authors contributed equally to this work.}
Michael R. Hayden,$^{1,2,5,6}$\footnotemark[2]
Jesse van de Sande,$^{1,2,6}$
\newauthor
Joss Bland-Hawthorn,$^{1,2,7}$
Sam Vaughan,$^{1,2,8,9}$
Marie Martig,$^{10}$
Francesca Pinna$^{11,12,13}$
\
\\
$^{1}$Sydney Institute for Astronomy, School of Physics, A28, The University of Sydney, NSW, 2006, Australia\\
$^{2}$ARC Centre of Excellence for All Sky Astrophysics in Three Dimensions (ASTRO-3D)\\
$^{3}$Department of Physics and Astronomy, University of Utah, Salt Lake City, UT 84112, USA\\
$^{4}$Space Telescope Science Institute, 3700 San Martin Drive, Baltimore, MD 21218, USA\\
$^{5}$Homer L. Dodge Department of Physics \& Astronomy, University of Oklahoma, 440 W. Brooks St., Norman, OK 73019, USA\\
$^{6}$School of Physics, University of New South Wales, NSW, 2052, Australia\\
$^{7}$Miller Professor, Miller Institute, UC Berkeley, Berkeley CA 94720\\
$^{8}$Astronomy, Astrophysics and Astrophotonics Research Centre, Macquarie University, Sydney, NSW 2109, Australia\\
$^{9}$Centre for Astrophysics and Supercomputing, School of Science, Swinburne University of Technology, Hawthorn, VIC 3122, Australia\\
$^{10}$Astrophysics Research Institute, Liverpool John Moores University, 146 Brownlow Hill, Liverpool L3 5RF, UK\\
$^{11}$Max-Planck-Institut für Astronomie, Königstuhl 17, D-69117 Heidelberg, Germany\\
$^{12}$Instituto de Astrofísica de Canarias, Calle Vía Láctea s/n, E-38205 La Laguna, Tenerife, Spain\\
$^{13}$Departamento de Astrofísica, Universidad de La Laguna, Av. del Astrofísico Francisco Sánchez s/n, E-38206, La Laguna, Tenerife, Spain
}

\date{Accepted 2024 September 12. Received 2024 August 26; in original form 2023 October 27}

\pubyear{2022}

\begin{document}
\label{firstpage}
\pagerange{\pageref{firstpage}--\pageref{lastpage}}
\maketitle

\begin{abstract}
Ongoing deep IFS observations of disk galaxies provide opportunities for comparison with the Milky Way (MW) to understand galaxy evolution.
However, such comparisons are marred by many challenges such as selection effects, differences in observations and methodology, and proper validation of full-spectrum fitting methods. 
In this study, we present a novel code \galcraft{} to address these challenges by generating mock IFS data cubes of the MW using simple stellar population models and a mock MW stellar catalog derived from \eglx{}. 
We use the widely adopted full-spectrum fitting code \ppxf{} to investigate the ability to recover kinematics and stellar populations for an edge-on mock MW IFS observation. 
We confirm that differences in kinematics, mean age, \mh{}, and \alphafe{} between thin and thick disks can be distinguished.
However, the age distribution is overestimated in the ranges between $2-4$ and $12-14$~Gyr compared to the expected values. 
This is likely due to the age spacing and degeneracy of SSP templates. We find systematic offsets in the recovered kinematics due to insufficient spectral resolution and the variation of line-of-sight velocity distribution with age and \mh{}. 
With future higher resolution and multi-\alphafe{} SSP templates, \galcraft{} will be useful to validate key signatures such as \alphafe{}-\mh{} distribution at different $R$ and $|z|$ and potentially infer radial migration and kinematic heating efficiency to study detailed chemodynamical evolution of MW-like galaxies.
\end{abstract}




\section{Introduction}
\label{sec:intro}

How galaxies form and evolve with time remains one of the outstanding questions facing astrophysics today. 
Over the last decade, integral-field spectroscopy (IFS) instruments have enabled us to obtain integrated light observations of galaxies across different regions of the same object, which have much better spatial coverage than single- or multi-slit spectroscopy.
Several IFS surveys have already been carried out such as SAMI \citep{2012MNRAS.421..872C}, CALIFA \citep{2012A&A...538A...8S}, and MaNGA \citep{2015ApJ...798....7B} to measure properties of thousands of galaxies to understand their evolution history.
In extragalactic analysis, an integrated spectrum is assumed to be a weighted superposition of many simple stellar population (SSP) spectra and the spectral features are reshaped by line-of-sight velocity distributions, dust attenuation and gas emission \citep{2013ARA&A..51..393C}.
To decompose these complex components and measure the kinematics and stellar population properties, many stellar population synthesis software have been developed such as \ppxf{} \citep{2004PASP..116..138C, 2017MNRAS.466..798C, 2023MNRAS.526.3273C}, \textsc{STARLIGHT} \citep{2005MNRAS.358..363C}, \textsc{STECMAP} \citep{2006MNRAS.365...46O}, \textsc{Pipe3D} \citep{2016RMxAA..52...21S}, and \textsc{Firefly} \citep{2017MNRAS.472.4297W} to perform full-spectral fitting to the observational spectra.
They have been broadly used to provide spatially resolved galaxy properties in data releases for a wide variety of scientific topics.

Due to our unique vantage point, the Milky Way (MW) is by far the best-studied galaxy in the Universe.
It provides an ideal test bed for our understanding of galaxy formation and evolution, and is often used to compare with extragalactic studies. 
Precise astrometric data from Gaia \citep{2023A&A...674A...1G} and accurate chemical abundances of individual stars from large spectroscopic surveys such as LAMOST \citep{2012RAA....12..723Z}, GALAH \citep{2021MNRAS.506..150B} and APOGEE \citep{2017AJ....154...94M} have been analysed in the last decade for Galactic archaeologists to reveal the detailed chemodynamical picture of the Milky Way \citep{2002ARA&A..40..487F, 2016ARA&A..54..529B}.
This includes the \alphafe{}-\feh{} bimodal distribution (e.g., \citealt{2014ApJ...796...38N, 2015ApJ...808..132H, 2021MNRAS.507.5882S}), radial migration (e.g., \citealt{2009MNRAS.396..203S}), accretion history and the interplay of chemical and dynamical processes (e.g., \citealt{2022Natur.603..599X}).
In the past decades, many Galactic chemical evolution (GCE) models have been developed to reproduce the \alphafe{} bimodality and uncover its origin (e.g., \citealt{2018A&A...618A..78H, 2019A&A...623A..60S, 2015MNRAS.449.3479S, 2021MNRAS.507.5882S, 2023MNRAS.523.3791C}).
However, because all of these models can reproduce consistent chemodynamical distributions with observations, the origin of \alphafe{}-bimodality is still under debate. 
Moreover, the MW is only one galaxy, whether the formation theories apply to other disk galaxies or whether the MW is unique in the Universe is still an open question.
Therefore, it is essential to combine the MW with other Milky Way analogous (MWAs) to address all these questions \citep{2021NatAs...5..879V}.

Several studies compared face-on MWAs from extragalactic IFS surveys with the MW:
\cite{2020MNRAS.491.3672B} selected 62 MWAs from MaNGA with the criteria of stellar masses and bulge-to-total ratios. They found most of these galaxies have flatter stellar and gas-phase metallicity gradients due to different disc scale lengths. A greater consistency can be found when scaling gradients by these lengths;
\cite{2023MNRAS.521.5810Z} revisited 138 MaNGA galaxies by fitting the spectra with a semi-analytic chemical evolution model \citep{2022MNRAS.513.5446Z} and measured their star formation and chemical enrichment histories. They detected similar \alphafe{} bimodality as the Galactic APOGEE observations \citep{2022ApJS..259...35A} in many of the galaxies.  
Compared to face-on galaxies, edge-on MWAs are more useful in providing elemental abundance distributions at different $R$ and $|z|$.
Several studies of nearby edge-on MWAs and lenticular galaxies using MUSE found similar kinematics and \alphafe{} differences between the thin and thick disks (e.g., \citealt{2019A&A...623A..19P, 2019A&A...625A..95P, 2021ApJ...913L..11S, 2021MNRAS.508.2458M}). 
In particular, \cite{2021ApJ...913L..11S} demonstrated that UGC 10738 has similar \alphafe{}-\feh{} distributions at different projected $R$ and $|z|$ with the MW observation in \cite{2015ApJ...808..132H} and model predictions in \cite{2021MNRAS.507.5882S}, which supports the commonness of MW's chemical distributions.

Despite the efforts above, a direct comparison of the MW with its analogs in kinematics and chemistry is still challenging because the observables and methods used for studying the MW are different from those for external galaxies, i.e., utilizing properties of individual stars as opposed to integrated quantities from stellar populations with projection effects. 
Therefore, the comparisons may be impacted by systematics or biases \citep{2019MNRAS.489.5030F, 2020MNRAS.491.3672B}.
In addition, some key processes such as radial migration and kinematic heating have not been extensively explored like the MW on external galaxies, which are also essential to identify whether the MW's formation and evolution is distinct from the general population.
Therefore, to take the MW as an ideal laboratory and test its evolution theories on other MWAs, one needs to remove these observational and methodological biases, transfer the observables of MW and external systems into the same definition, and apply models that consider both chemical enrichment and kinematic processes.

In addition, even though the full-spectrum fitting methods have been useful in providing kinematics and stellar population properties of different components of external galaxies, due to the high degeneracy of all the parameters, it is still difficult to verify how close the measured values are to the ``actual'' galaxy properties and whether any potential bias is introduced by the measuring methods.
This could cause different fitting methods or strategies to obtain different results and lead to different conclusions.
Therefore, every individual study has to validate the reliability of their measurements (e.g., \citealt{2018MNRAS.480.1973K}) and the conclusions are often only made qualitatively.
Several studies examined the robustness of different full-spectrum fitting software and investigated effects due to different spectral models (e.g., \citealt{2018MNRAS.478.2633G, 2019MNRAS.485.1675G, 2024MNRAS.530.4260W}).
However, they only tested mock spectra generated using one or two stellar populations, or only tested on global stellar population properties without spatial distribution.
Until now, there has not been a comprehensive exploration to perform full-spectrum fitting on a realistic mock galaxy with chemical enrichment and kinematic processes and investigate the recovery ability of spatially resolved properties in detail.
All the above considerations lead to the development of tools and studies presented in this work.

In this study, to address the challenges of comparing the MW to other galaxies, we present a novel code \galcraft{} to generate mock integral-field spectrograph (IFS) data cubes of the MW with integrated spectra using simple stellar population (SSP) models and mock stellar catalog from \eglx{} (Sharma et al. in perp.), which is based on the chemodynamical model of \cite{2021MNRAS.507.5882S} (hereafter S21) that was verified to be consistent to the current Galactic observations in both kinematics and chemistry across a range of $R$ and $|z|$. 
This mock data cube is in the same format as extragalactic IFS observations. Therefore, extragalactic data analysis methods (e.g., the GIST pipeline \citealt{2019A&A...628A.117B} with \ppxf{} \citealt{2004PASP..116..138C,  2017MNRAS.466..798C, 2023MNRAS.526.3273C}) can be applied to the mock data cube to measure directly comparable parameters in (age, \mh{}, \alphafe{}, \vlos{}, \dispersion{}, $h_3$, $h_4$, light/mass fraction distributions). 
To address the reliability of full-spectrum fitting methods in measuring galaxy properties, we apply \ppxf{} on \galcraft{}-generated mock edge-on IFS cubes and compare the measured kinematics and stellar population properties to the input values according to the GCE model S21.
This paper provides a comprehensive test of full-spectrum fitting methods on a mock galaxy that considers physical chemodynamical processes at different galaxy components.
The results will be useful as a reference for future extragalactic data analysis using similar spectrum fitting algorithms.

We describe the ingredients used in \galcraft{} and detailed procedures of this code in Section~\ref{sec:mwcubeegeneration}. 
In Section~\ref{sec:galaxymap}, we test the ability of broadly used spectral fitting tool \ppxf{} to recover kinematics, stellar population parameters, and light/mass fraction distributions. 
In Section~\ref{sec:discussion}, we discuss the causes of deviations between measured and true (input) values and address potential reasons and future improvements on current algorithms for obtaining better results. 
We explore the effect of different fitting strategies and provide some suggestions when using such spectral fitting tools for future studies. We also give some caveats about using mock cubes generated from \galcraft{} code.
In Section~\ref{sec:future}, we talk about future plans on the improvements of \galcraft{} along with spectral fitting methods and prospect some scientific topics that can be done by using \galcraft{}.
A summary is presented in Section~\ref{sec:summary}.

\section{Data Cube Generation}
\label{sec:mwcubeegeneration} 

The purpose of \galcraft{} is to take the mock stellar catalog of the MW obtained from \eglx{} (Sharma et al. in prep.), and transform it into a mock data cube in three dimensions (two in spatial and one in spectral) as observed by integral-field spectrographs (IFS) such as PMAS/PPak \citep{2005PASP..117..620R, 2006PASP..118..129K}, AAT/SAMI \citep{2012MNRAS.421..872C}, Sloan/MaNGA \citep{2015ApJ...798....7B}, and VLT/MUSE \citep{2010SPIE.7735E..08B}. 
The \galcraft{} code takes as input the following set of user-specified input parameters: galaxy distance ($d$), inclination ($i$), extinction, SSP templates, instrumental properties, as well as a few additional parameters (see the full list in Table~\ref{tab:mwdatacube_setup}).
The produced mock data cube can be processed in the same way as real IFS observations data by many methods, particularly codes like Voronoi binning \citep{2003MNRAS.342..345C}, Penalized Pixel-Fitting (\ppxf{}, \citealt{2004PASP..116..138C, 2017MNRAS.466..798C, 2023MNRAS.526.3273C}), line-strength indices (e.g., \citealt{1994ApJS...95..107W, 2007ApJS..171..146S, 2011MNRAS.412.2183T, 2018MNRAS.475.3700M}), or a combination of them (e.g., the GIST pipeline, \citealt{2019A&A...628A.117B}). 
The results can be compared directly with those from IFS observations of MWAs in terms of mass- or light-weighted parameter maps, line-of-sight velocity distribution (LOSVD), and mass fraction distributions. The ingredients, flexibility, procedures, and computational expenses of this code are described in detail in the following sub-sections.

\subsection{The ingredients}
\label{sec:mwcubeingreds}

\subsubsection{Galactic chemical evolution (GCE) model}
\label{sec:mwcubeingreds:chemicalevo}

We apply the analytical chemodynamical model of the MW developed by \citealt{2021MNRAS.507.5882S} (hereafter S21) which can predict the joint distribution of position~$(\mathbf{x})$, velocity~$(\mathbf{v})$, age~$(\tau)$, extinction~\ebv{}, photometric magnitude in several bands and chemical abundances~(\feh{},  \alphafe{}) of stars in the MW.
Compared with previous models (e.g., \citealt{1997ApJ...477..765C, 2019A&A...625A.105H}), this model included a new prescription for the evolution of \alphafe{} with age and \feh{} and a new set of relations describing the velocity dispersion of stars \citep{2021MNRAS.506.1761S}.
This model shows for the first time that it can reproduce the \alphafe{}-\feh{} distribution of APOGEE observed stars \citep{2015ApJ...808..132H} at different radius~$R$ and height~$|z|$ across the Galaxy.
In this model, the origin of two \alphafe{} sequences is due to two key processes: 
the delay between the first star formation and sequential occurrence of SN Ia causes the sharp transition from high-\alphafe{} to low-\alphafe{} at around 10.5~Gyr ago, which is likely to create a valley between the two sequences; the radial migration or so-called churning is responsible for the large spread of the low-\alphafe{} sequence along the \feh{} axis. This model also showed that without churning it is not sufficient to reproduce the two sequences (see their Fig.~6).
Because this chemical evolution model is purely analytical, it is easy to be inserted into the forward-modeling tool \eglx{} to generate mock stellar catalogs for further analysis. In addition, several free parameters such as radial migration and kinematic heating efficiency are adjustable, which will be useful to implement similar analysis on external galaxies.

\subsubsection{\eglx{}}
\label{sec:mwcubeingreds:galaxia}

To mock the MW IFS data cube, we need a comprehensive stellar catalog from observations with well-measured parameters such as position $(\mathbf{x})$, velocity $(\mathbf{v})$, age $(\tau)$ and chemical abundances. 
However, this is impractical as the MW has hundreds of billions of stars being unexplored and the dust in the disk obscures distant light.
An alternative way is to employ N-body/hydrodynamical simulations (e.g., EAGLE \citealt{2015MNRAS.446..521S}, FIRE-2 \citealt{2018MNRAS.480..800H}), but most of the simulations only contain $\sim10^6$ stellar particles, which is not enough to generate integrated spectra because each spatial bin would only contain less than a hundred particles. This in turn would increase the sampling noise of the integrated spectrum and make observables derived from spectra noisy. 
Even though oversampling can solve this problem, it is still challenging to find a simulation that is observably identical to the MW in all aspects quantitatively, especially the \alphafe{} bimodal trends.

Therefore, here we generate a mock catalog to represent the MW. We use \eglx{} (Sharma et al. in prep.), which is a tool in accordance with the chemical evolution model of S21 and can create a catalog with the user-defined number of stellar particles ($N_p$) containing parameters including position $(\mathbf{x, y, z})$, velocity $(\mathbf{v_x, v_y, v_z})$, age $(\tau)$, metallicity (\mh{}) and \alphafe{}, and the parameter distributions are chemodynamically consistent to the MW observations.
In \eglx{}, each stellar particle is equivalent to a single-stellar population (SSP) with an initial birth mass of 1 \solarmass{}, so one can directly assign spectra using SSP templates under the default unit without transferring from the mass of initial cloud to remaining stars.
Other codes like \textsc{TRILEGAL} \citep{2005A&A...436..895G}, \textsc{BESANCON} \citep{2005A&A...436..895G}, and \glx{} \citep{2011ApJ...730....3S} can also create mock catalogs. 
However, compared to \eglx{}, the underlying models of these codes lack the information of \feh{} and \alphafe{}, and do not have the processes of radial migration and kinematic heating.
Furthermore, the ability of \eglx{} to control the observed properties by regulating the underlying physical process is important for future analysis of external galaxies, whose formation history and dynamical processes are expected to be different from the MW. 
One caveat is that SSP particles in the current version of \eglx{} are distributed only in the disk and there is no bulge/bar, halo, or nuclear disk structure. 
This is because the chemical and kinematic distribution functions predicted by S21 are extrapolated from observations in the solar neighborhood.
Nevertheless, particle parameter distributions in \eglx{} are still consistent with APOGEE observations in the range of $3<R_{\rm{gc}}<15$~kpc \citep{2021MNRAS.507.5882S}.

\subsubsection{Spectral libraries}
\label{sec:mwcubeingreds:ssp}

To build a mock data cube, one important procedure is to turn particles in \eglx{} into stellar spectra based on their properties, so a spectra library is needed.
Given each particle is an SSP in \eglx{} and the integrated spectrum is a sum of many stellar populations
We will employ SSP spectral libraries in \galcraft{}.
The SSP spectrum describes the spectral energy distribution (SED) of a stellar population with a single age, metallicity, and chemical abundance patterns. An initial mass function (IMF) is assumed, and the stellar population is evolved using a given isochrone \citep{2013ARA&A..51..393C}.
\galcraft{} currently supports MILES \citep{2010MNRAS.404.1639V, 2015MNRAS.449.1177V} and PEGASE-HR \citep{2004A&A...425..881L}, both of which will be used in this study. 

The MILES SSP library ($\rm{FWHM}=2.51\Angstrom$, $3500\Angstrom<\lambda<7500\Angstrom$) is based on the model of \cite{1999ApJ...513..224V}. It uses Padova 2000 \citep{2000A&AS..141..371G} or BaSTI \citep{2004ApJ...612..168P, 2006ApJ...642..797P} isochrones and IMF in Unimodal/Bimodal \citep{1996ApJS..106..307V}, Kroupa Universal/Revised \citep{2001MNRAS.322..231K} and Chabrier \citep{2003PASP..115..763C} shapes. For Padova 2000 isochrones, the template grids cover 7 metallicity bins between (-2.32, 0.22)~dex, 50 age bins between (0.063, 17.78)~Gyr and one scaled-solar \alphafe{} bin \citep{2010MNRAS.404.1639V}. 
As for BaSTI isochrones, the template grids cover 12 metallicity bins between (-2.27, 0.40)~dex, 53 age bins between (0.03, 14.00)~Gyr and two \alphafe{} bins in 0.0 and 0.4 dex \citep{2015MNRAS.449.1177V}. Since \alphafe{} enhancement is essential in this study, for most of the cases we will use the $\alpha$-variable templates.

The PEGASE-HR library ($\rm{FWHM}=0.55\Angstrom$, $3900\Angstrom<\lambda<6800\Angstrom$) is based on the Elodie 3.1 stellar library \citep{2001A&A...369.1048P,2007astro.ph..3658P}. 
The SSP models are computed using Padova 1994 \citep{1994A&AS..106..275B} isochrones with a Salpeter \citep{1955ApJ...121..161S}, Kroupa, or top-heavy \citep{2003MNRAS.338..817E} IMF. The templates grid consists of 7 metallicity bins between (-2.30, 0.70)~dex, 68 age bins between (0.001, 20)~Gyr, and one scaled-solar \alphafe{} bin. 
In this work, the PEGASE-HR templates are mainly used to explore the effect of spectral resolution on \ppxf{} fitting. Therefore, it is still useful even if it lacks variable \alphafe{}.
Revised grids interpolated by \ulyss{} \citep{2009A&A...501.1269K} in the same way as \cite{2018MNRAS.480.1973K} for PEGASE-HR are also available. The new grids contain 15 steps in metallicity between -2.3 and 0.7~dex and 50 steps in age between 10~Myr and 14~Gyr.

\subsection{Configurations of \galcraft{}}
\label{sec:mwcubeflex}

\begin{table*}
\caption{Configuration parameters in the \galcraft{} code.}
\label{tab:mwdatacube_setup}
    \begin{tabular}{lll}
    \hline
    Parameters & Description & Unit\\
    \hline
    \textbf{GALPARAMS:} & \textbf{Observation parameters of the Galaxy} & \\
    \qquad distance & Distance of the Galaxy center to the observer  & kpc \\
    \qquad $\theta_{zx}$ & Angles that the Galaxy rotates along the Y axis in the direction from $z$ to $x$ & deg \\
    \qquad $\theta_{yz}$ & Angles that the Galaxy rotates along the X axis in the direction from $y$ to $z$ & deg \\
    \qquad $l$ & Galactic longitude of the center of the Galaxy & deg \\
    \qquad $b$ & Galactic latitude of the center of the Galaxy & deg \\
    \textbf{SSPPARAMS:} & \textbf{Parameters of the SSP templates} & \\
    \qquad model & The SSP model to be used for the spectrum (MILES or PEGASE-HR) & \\
    \qquad isochrone & Isochrones used to generate the SSP templates & \\
    \qquad IMF & Initial-mass function used to generate the SSP templates & \\
    \qquad single\_alpha & If use single \alphafe{} or $\alpha$-variable templates for the spectrum & bool \\
    \qquad factor & Oversampling factor of the templates & \\
    \qquad FWHM\_gal & Spectral resolution in FWHM of the output data cube & $\Angstrom$ \\ 
    \qquad dlam & Bin width of the wavelength sampling of the output data cube & $\Angstrom$ \\ 
    \qquad age\_range & Optional age range (inclusive) in for the SSP models & Gyr \\
    \qquad metal\_range & Optional metallicity range (inclusive) in for the SSP models & dex \\
    \qquad wave\_range & Optional wavelength range (inclusive) in for the SSP models & $\Angstrom$ \\
    \qquad spec\_interpolator & Interpolation method to assign the spectrum to the particle given its (age, metallicity, \alphafe{}) & \\
    \textbf{INSPARAMS:} & \textbf{Instrument properties} & \\
    \qquad instrument & Instrument name, when using this parameter, the other parameters will be automatically set & \\
    \qquad spatial\_resolution & Spatial resolution of the data cube & arcsec \\ 
    \qquad spatial\_nbin & Number of spatial bins (spaxels) in two coordinates, in the format of [$n_x$, $n_y$] &  \\ 
    \qquad FWHM\_spatial & Full-Width Half-Maximum of the PSF function to model the atmosphere & arcsec \\ 
    \hline
    \end{tabular}
\end{table*}

To meet different research purposes, we incorporated a wide range of adjustable parameters in \galcraft{}. Specifically, the adjustable parameters are divided into three groups: \texttt{GALPARAMS}, \texttt{SSPPARAMS} and \texttt{INSPARAMS}, as listed with descriptions in Table~\ref{tab:mwdatacube_setup}. 
The user can set up their preferred parameters to obtain the expected data cubes. 
\texttt{GALPARAMS} is for setting the distance, inclination, and position of the mock MW using coordinates transformation; 
\texttt{SSPPARAMS} governs the spectral properties, such as the selection of SSP model, single or variable \alphafe{}, spectral resolution, age, \mh{} and wavelength ranges and interpolation method to be used to assign the spectrum according to particles parameters (see details in Section~\ref{sec:mwcubesteps}). 
Many interpolation options could be used such as ``nearest'', ``linear'', etc.
The \texttt{INSPARAMS} controls the instrument spatial sampling, atmospheric effects (PSF), and the number of spatial bins in each coordinate. 
We also provide an option to derive the data cube in the format of a specified instrument. 
Alternatively, the user can also design a hypothetical instrument that does not exist by manually setting up these parameters, which will be useful for future instrument designs.

\subsection{Procedures of making a mock data-cube}
\label{sec:mwcubesteps}

\begin{figure*}
\includegraphics[width=2\columnwidth]{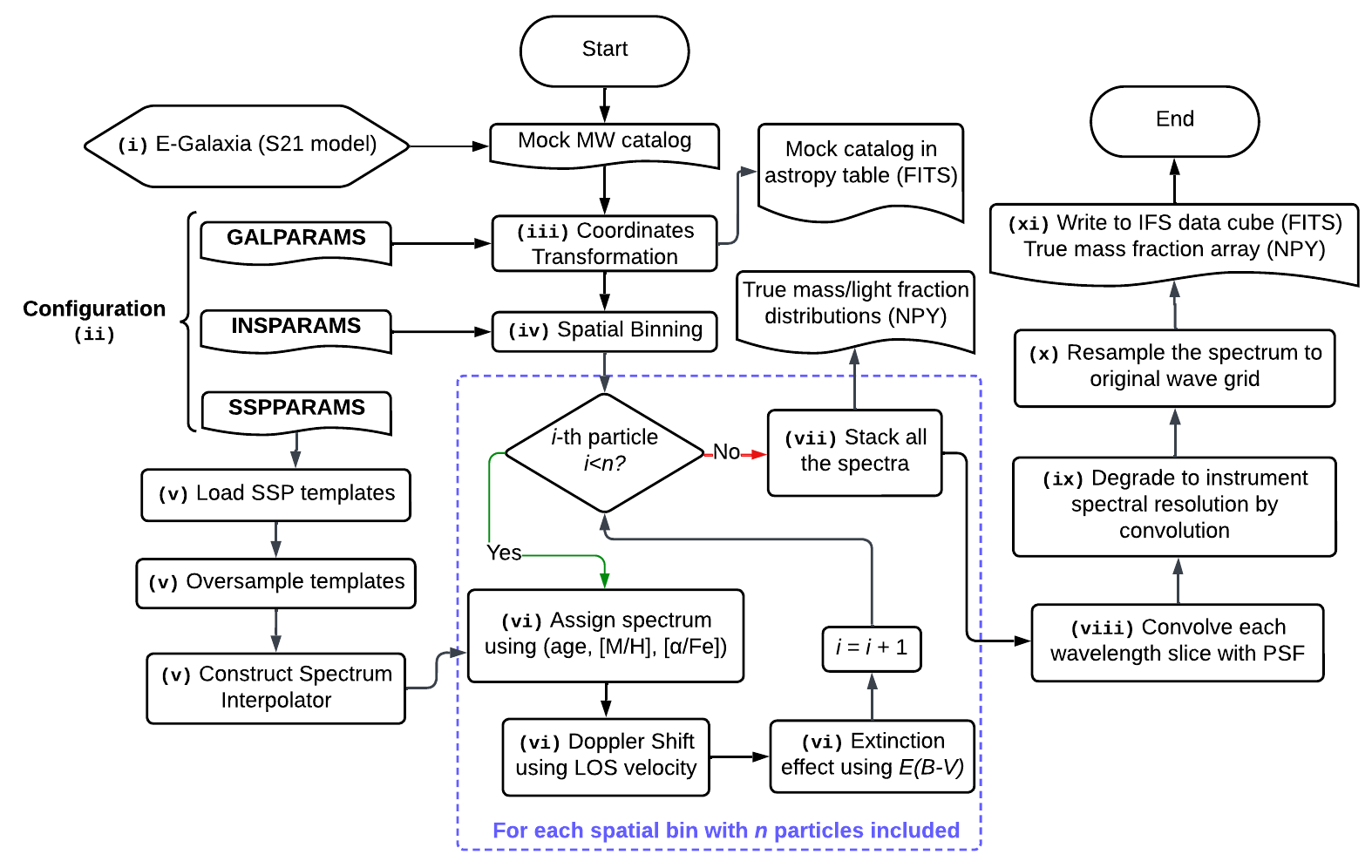}
\caption{
A flowchart demonstrating the workflow of \galcraft{} code.
The numbers on some processes correspond to the steps listed in Section~\ref{sec:mwcubesteps}.}
\label{fig:galcraftflow}
\end{figure*}

The procedures of generating mock MW data cubes are described in detail as the following steps, along with a flowchart in Fig.~\ref{fig:galcraftflow}:
\begin{enumerate}
 \item Use \eglx{} to generate a mock stellar catalog of the MW, with particles containing position $(\mathbf{x})$, velocity $(\mathbf{v})$, age $(\tau)$, extinction~\ebv{}, metallicity (\feh{}) and \alphafe{}, transfer \feh{} to \mh{}.
 \item Load the configurations of Table~\ref{tab:mwdatacube_setup} provided by the user, as well as a list of data cubes to be generated with their center coordinates in $(l, b)$ or $(ra, dec)$.
 \item Apply coordinates transformation based on \texttt{GALPARAMS} parameters to project this Galaxy on the Celestial sphere with a certain distance and inclination.
 \item Apply the spatial binning based on IFS instrument properties given by \texttt{INSPARAMS}, and find the particles included in each bin, then note the bin index for each particle for later use.
 \item Load the SSP templates with defined age and \mh{} ranges by \texttt{SSPPARAMS}, then oversample the spectra by a factor of \texttt{SSPPARAMS:factor} using \texttt{spline} interpolation with the order of three. Next, construct the interpolator to be used to generate integrated spectra in the next step.
 \item Select particles in a spatial bin, assign each particle an SSP spectrum based on its age, \mh{}, and \alphafe{} by interpolating the templates with the method defined by  \texttt{SSPPARAMS:spec\_interpolator}. Multiply the spectrum by the particle's stellar mass because SSP spectral templates are normalized to 1~\solarmass{}. Shift the spectrum due to its line-of-sight velocity (\vlos{}) using the Doppler equation. Then apply a flux calibration due to the particle's distance and extinction. 
 \item Loop the procedure (vi) over all the spatial bins to stack integrated spectra. In the meantime, generate the mass/light-fraction distribution of each spatial bin. The light weights are obtained within the wavelength range given by \texttt{SSPPARAMS:wave\_range}.
 \item After generating integrated spectra for all the spatial bins, apply the atmosphere effects by convolving each wavelength slice with a point spread function (PSF), this can be either a Gaussian or Moffat kernel function with the given \texttt{INSPARAMS:FWHM\_spatial}.
 \item Degrade the stacked spectrum to the instrument resolution given by \texttt{SSPPARAMS:FWHM\_gal} using convolution with a Gaussian line-spread-function.
 \item Re-bin the oversampled flux array into the original wave grid or the user-defined wavelength interval and wavelength range, according to \texttt{SSPPARAMS:dlam} and \texttt{SSPPARAMS:wave\_range}.
 \item Write the data cube flux array in the format of FITS file with \galcraft{} configuration parameters in the header.
\end{enumerate}

\begin{figure}
\centering
\includegraphics[width=1\columnwidth]{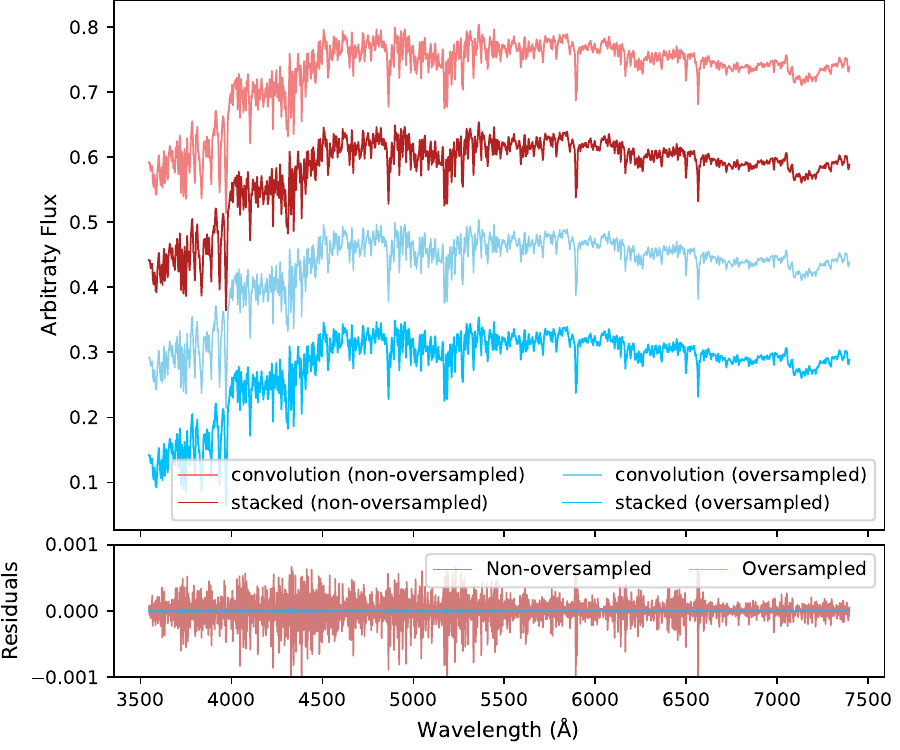}
\caption{
Comparison of the mock integrated spectrum generated by using the non-oversampled (in red and light-red) and the oversampled (in blue and light-blue) MILES templates. The light-color lines are spectra convolved by a Gaussian kernel with given $(\mu, \sigma)$. The dark-color lines are spectra stacked by 2000 shifted spectrum with different line-of-sight velocities which follow a Gaussian distribution with the same $(\mu, \sigma)$. The bottom panel shows residuals of convolved and stacked spectra. We find that oversampling can reduce the deviation between the convolved and stacked spectrum significantly by a factor of $\sim25$.
}
\label{fig:original_vs_oversampling}
\end{figure}

Other than the above procedures, there are a few points that need to be clarified to the users as follows:
\begin{itemize}
    \item The original mock stellar catalog in step (i) should have two coordinate systems, Cartesian coordinates $(x, y, z)$ and Galactic coordinates $(l, b)$, where $(l, b)$ are overlapped with $(y, z)$, respectively. Therefore, by adjusting $(\theta_{zx}, \theta_{yz})$, users can rotate the mock MW into a defined inclination. This is convenient when compared with real observations.
    \item The oversampling in step (v) has two purposes: one is to ensure the validity of degrading - when degrading the SSP templates from the original spectral resolution to instrumental resolution (e.g., from MILES to MUSE), the $\sigma$ of the Gaussian kernel to be convolved with the templates will be smaller than wavelength interval $\Delta \lambda$ without oversampling.
    In this case, the kernel function array becomes invalid with only one element having a value of 1, and the rest having values of 0.
    Then the degraded spectrum will be still in its original resolution. Another reason is to minimize the particle sampling error when stacking the spectrum with different \vlos{}. Fig.~\ref{fig:original_vs_oversampling} shows an example mock integrated spectrum generated by using the non-oversampled (in red and light-red) and oversampled (in blue and light-blue) MILES SSP templates. The light-color lines are the spectra convolved with a Gaussian kernel by the given mean velocity and dispersion $(\mu, \sigma)$, which is the manner of \ppxf{} during the kinematics measurements. The dark-color lines are stacked from 2000 spectra shifted by Gaussian distributed \vlos{} using the same $(\mu, \sigma)$, which is the manner of \galcraft{}. By calculating the residuals of these two lines (shown in the bottom panel), we find that oversampling can reduce the deviation between the convolved and stacked spectrum significantly by a factor of $\sim25$, which will be helpful to reduce potential issues when \ppxf{} performs Gauss-Hermite convolution to fit on the particle-stacked spectrum.
    \item This package can select a portion of particles within the field of view (FoV) based on the user-defined instrument to generate the data cube, rather than taking a whole catalog into account. It helps to mimic a more realistic IFS observation and reduce computational expenses. 
    \item \galcraft{} can be executed in the batch mode where users can provide a list of cubes with the central coordinates in $(l, b)$ or $(ra, dec)$, and all the cubes can be automatically generated in one execution;
    \item Other than the data cube FITS file, this package also generates some by-products including mass/light-fraction and number of particle distribution arrays for each spatial bin, mass/light-weighted \mh{}, \alphafe{}, age and kinematic properties. These by-products are obtained from properties of particles in \eglx{} and can be used to calculate the expected true answers that \ppxf{} is expected to recover from full-spectrum fitting. This will be described in detail in Section~\ref{sec:galaxymap}.
\end{itemize}

\subsection{Estimate the sampling error}
\label{sec:mwcubesamplingerror}

\begin{figure}
\centering
\includegraphics[width=1\columnwidth]{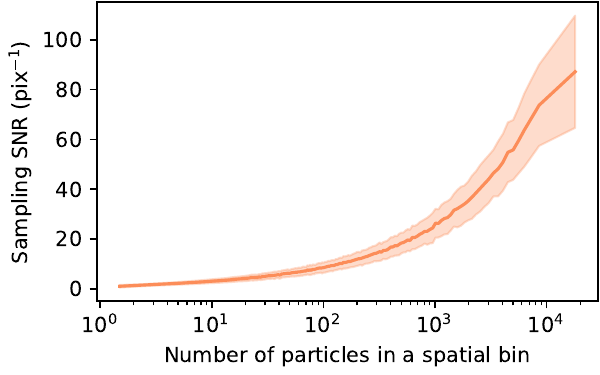}
\caption{
The sampling signal-to-noise ratio (SNR) of an integrated spectrum as a function of the number of particles in a spatial bin. The shaded region is the standard deviation of sampling SNRs for spaxels with the same number of particles. This is calculated by using original and bootstrapped mock data cubes described in Section~\ref{sec:galaxymapmockcube}. This figure provides an estimation of spectral sampling noise due to the limited number of particles in \eglx{}.
}
\label{fig:samplingsnr_nparticles}
\end{figure}

Compared to the real catalogs of the MW (e.g. Gaia), one shortcoming of the \eglx{} mock stellar catalog is the limited number of particles. Although there are $10-100$ times more particles $(10^8)$ in the mock stellar catalog compared to MWAs in N-body/hydrodynamical simulations, some spatial bins inevitably contain too few particles for robust measurements of their properties. 
Even for spaxels or Voronoi bins containing more than $\sim10^4$ particles, 
the spectral noise due to finite star particles is still non-negligible.
We call this type of spectral noise ``sampling error'' $(e_{f, S})$.
One way to reduce sampling error is to generate more particles from \eglx{}. However, this will increase the computational expenses and RAM/memory usage of \galcraft{} significantly, and exceed most of the HPCs' limitation ($\sim22$~GB for a catalog containing $10^8$ particles, hence $\sim220$~GB for $10^9$ particles). 
Therefore, when mocking IFS observations, one has to ensure that for each integrated spectrum, the sampling error $(e_{f, S})$ is smaller than the observational flux error ($e_{f, O}$) at given instrument conditions and exposure time. 
In this way, it is safe to apply the data reduction pipeline on this mock data cube and allow \ppxf{} to derive mathematically reasonable results.
This is particularly important in kinematics because the LOSVD effect in \galcraft{} integrated spectra is implemented by stacking individual spectra of particles with each shifted by its own \vlos{}.
But \ppxf{} employs the Gauss-Hermite function to convolve with SSP templates and determines the kinematics moments.
In Section~\ref{sec:galaxymapdatareduction}, we will provide a detailed example of how to deal with the sampling error for each Voronoi bin.

To estimate the sampling error, the \galcraft{} code provides an option to apply bootstrapping. First, it randomly re-selects particles from the original \eglx{} catalog and generates a bootstrapped catalog with the same particle number ($N_p$). Then, \galcraft{} uses this catalog to generate bootstrapped data cubes by repeating the above procedures a certain number of times, and the sampling error $(e_{f, S})$ for each spaxel can be represented by the standard deviation of these bootstrapped fluxes. 
This sampling flux error will be used as the lower limit of the acceptable Gaussian noise when mimicking the real observations and obtaining the final integrated spectra of Voronoi bins.
Fig.~\ref{fig:samplingsnr_nparticles} illustrates how the sampling $\rm{SNR}$ (flux divided by sampling noise) varies as a function of the number of particles in a spatial bin. The shaded region is the standard deviation of sampling SNRs for spaxels having the same number of particles. 
We obtain this figure by using the original and bootstrapped mock data cubes later in Section~\ref{sec:galaxymapmockcube}.
It can be seen that a spatial bin with $\sim10^3$ particles can generate a spectrum with sampling $\rm{SNR}\sim25$~\perpixel{}, and a spatial bin with $\sim10^4$ particles can generate a spectrum with sampling $\rm{SNR}\sim80$~\perpixel{}.

\subsection{Computational expenses and multiprocessing strategy}
\label{sec:mwcubecpu}

The execution time of \galcraft{} to generate mock data cubes depends mostly on the number of particles included in the instrument FoV and the spectral interpolation method. In general, the execution time is proportional to the number of particles used, and the ``nearest'' interpolation is three times faster than the ``linear'' interpolation.
In this code, we apply \texttt{python-multiprocessing} techniques to speed up the computing. 
For a typical MUSE FoV containing $6\times10^6$ particles, the execution time spent with a 24-core CPU (2.50GHz) is $\sim 1.4$~hour. Based on this, the users can roughly estimate the execution time they will spend.
If taking into account the bootstrapped cubes, the total execution time will be multiplied by the number of bootstrapping times plus 1. Therefore, it is highly recommended to run it on high-performance computers (HPC) or a Cluster where these 21 jobs can be executed simultaneously.

\section{Recovery of the Galaxy Chemodynamical Properties}
\label{sec:galaxymap}

In this section, we take \ppxf{} as a representation software to test the ability of full spectrum fitting methods to recover galaxy properties by applying it to mock cubes generated from \galcraft{}.
We measure kinematics (\vlos{}, \dispersion{}, $h_3$, $h_4$), stellar population parameters (age, \mh{}, \alphafe{}) and light/mass fraction distributions of different structural components. 
The analysis is performed in the same way as extragalactic studies.
Then we compare the results with the input true values that are obtained by properties of SSP particles from \eglx{} catalog. 
This test allows us to access the consistency of parameters measured via broadly applied software in other studies (e.g. \citealt{2019A&A...623A..19P, 2019A&A...625A..95P, 2021ApJ...913L..11S, 2021MNRAS.508.2458M}), which was not possible previously as the true values of external galaxies are unknown. 
Furthermore, it also provides standard references for the future to better understand extragalactic results (e.g., gradient, flaring) by distinguishing real distributions from artificial effects due to the spectral fitting methods, projected view, and integrated light.
We note that our goal is to explore the general performances of the full-spectrum fitting method using template weighting and regularization, i.e., its underlying mathematical framework. We choose to test \ppxf{} because it has been widely used by many previous studies. However, any systematic bias found in this study is not specific to \ppxf{}, but would be equally applicable to other software using the same framework. And we leave testing on other software for future studies.

\subsection{Mock cube generation for MUSE instrument}
\label{sec:galaxymapmockcube}

\begin{figure}
\includegraphics[width=1\columnwidth]{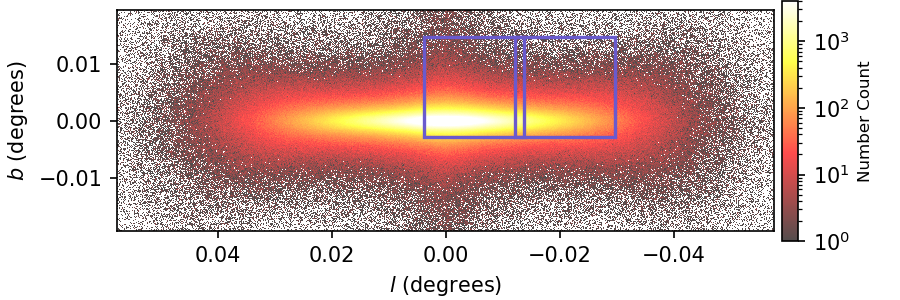}
\caption{
Two mock data cube FoV (in purple) relative to the number distribution of particles in \eglx{} mock stellar catalog. The model was set with a distance of 26.5 Mpc and inclination of $86^{\circ}$. This observation strategy is the same as the MUSE observation on NGC~5746.
}
\label{fig:cube_distrib}
\end{figure}

We generate a mock MUSE observation by \galcraft{}, using the \eglx{} catalog that contains $10^8$ particles.
We remove particles with stellar age less than 0.25~Gyr because their position and kinematics are erroneous in the current version of \eglx{}, and we confirm that removing these particles does not affect our conclusions.
The mock MW catalog is assumed to have a distance of $26.5$~Mpc and inclination of $86^{\circ}$ to the observer, which is the same as the projection of NGC~5746 observed by MUSE with comprehensive analysis in \citealt{2021MNRAS.508.2458M} (hereafter M21). 
We use MILES $\alpha$-variable SSP templates \citep{2015MNRAS.449.1177V} with the BaSTI stellar isochrones \citep{2004ApJ...612..168P} and Kroupa Universal IMF \citep{2001MNRAS.322..231K}. The templates have 53 bins in age, 12 bins in \mh{} and 2 bins in \alphafe{} and we apply a ``linear'' interpolation to assign each particle a spectrum based on its age, \mh{} and \alphafe{}, and then degrade the stacked spectra to MUSE spectral resolution ($\mathrm{FWHM}\sim2.65\Angstrom$). 
Following procedures in Section~\ref{sec:mwcubesteps}, we obtain two mock cubes focusing on the central $(N_{p}=61575676)$ and the disk $(N_{p}=7379847)$ regions, as shown in Fig.~\ref{fig:cube_distrib}. 
This observation strategy is also the same as MUSE observations on NGC~5746 in M21.
The total execution time spent by \galcraft{} on a 24-core CPU for these two cubes is $\sim14.5$ hours. 
We also generate $2\times20$ bootstrapped cubes and use 16th and 84th percentiles to calculate the sampling error of each spaxel. 
We do not apply extinction in these cubes because here we only focus on full-spectrum fitting validation.
Adding extinction would blend all the effects and make it difficult to differentiate their individual impacts.
Therefore, we reserve this topic for future studies.

The next procedure is to add Gaussian flux error to the spectra.
We first derive the observational flux error $(e_{f, O})$ of the mock cubes. The observational flux error depends on many aspects but can be classified into two main categories: the observation conditions (seeing, air-mass, exposure time, etc.) and the instrumental properties (telescope aperture, system efficiency, dark current, read-out noise, etc). 
For simplicity, we ignore the sky conditions, dark current, and read-out noise which only contribute a few percent to total received photons, and assume the spectral SNR and received photons are defined by
\begin{equation}
\begin{split}
& \mathrm{SNR}=\frac{f}{e_{f, O}}=\sqrt{N}, \\
& N=aft, 
\end{split}
\end{equation}
where $f$ is the flux of the target; $e_{f, O}$ is the observational flux error (by MUSE in this case); $N$ is the received number of photons; $t$ is the exposure time; $a$ is an overall reaction of sky transmission, efficiency and telescope aperture. Therefore, $a$ should only depend on wavelength for the same instrument. 
By substituting the above equations, the parameter $a$ can be calculated using $f$, $e_{f, O}$ and $t$ from an observation by 
\begin{equation}
a=\frac{f}{e^2_{f, O}\times t}.
\label{eqn:a_eft}
\end{equation}
In this work, we take all the bulge and disk observations of NGC~5746 from MUSE in M21 and fit equation~\ref{eqn:a_lambda} as a function of wavelength $(\lambda)$ using a 4-degree polynomial, which is described by
\begin{equation}
\begin{split}
a(\lambda) &= 4.34274826^{-18} \lambda^4 - 1.43263443^{-13} \lambda^3 + 1.61141240^{-9} \lambda^2\\ 
& -7.29164505^{-6} \lambda + 1.17213217^{-2},
\end{split}
\label{eqn:a_lambda}
\end{equation}
where $a(\lambda)$ is in the unit of $1/10^{-20}~\rm{erg}~\rm{cm}^{-2}~\angstrom^{-1}$. 
Next, we set the bulge and disk mock cubes to have an exposure time of 1729.39~s and 6221.84~s, respectively, and use the equation~\ref{eqn:a_eft} and \ref{eqn:a_lambda} to estimate the observational flux error $e_{f, O}$ of each spaxel.
Then we use this error to add Gaussian noise to all the spaxels. 
Here the disc exposure time is chosen to satisfy the upper bound of $e_{f, O} \geq e_{f, S}$. The bulge exposure time is then determined by assuming the same bulge-to-disk exposure time ratio in MUSE observations of NGC~5746 by M21, and we confirm that $e_{f, O} \geq e_{f, S}$ also applies.
These two values are slightly smaller than those used by M21.
Finally, the two mock cubes are stitched together.

\subsection{Extracting galaxy properties}
\label{sec:galaxymapdatareduction}

\begin{figure}
\centering
\includegraphics[width=1\columnwidth]{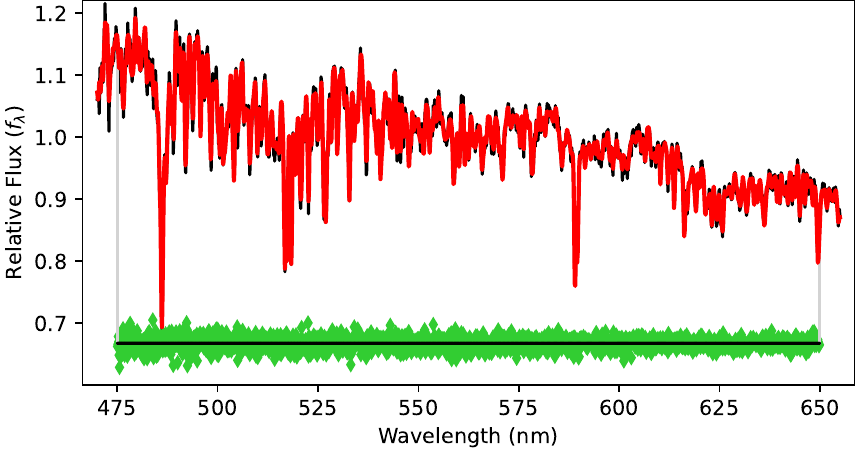}
\caption{
A spectral fit example of a Voronoi bin by \ppxf{} in Section~\ref{sec:galaxymapdatareduction}. The ``observed'' spectrum is in black while the best-fit spectrum is in red. Residuals are in green. The fitting area is within the grey lines on both sides.
}
\label{fig:eg_ppxf_fit}
\end{figure}

We apply the GIST pipeline\footnote{\url{https://gitlab.com/abittner/gist-development}} \citep{2019A&A...628A.117B} on the stitched mock cube to measure the kinematics and stellar population parameters. 
The GIST pipeline combines all the tools needed to process the data and users can obtain final results in a single execution.
Here we use a modified version to implement some functionalities that the current public version (v3.1.0) does not have but are needed in this work. A detailed list of added features is given in Appendix~\ref{appendix:gistrevision}. 

We run the GIST pipeline in the following steps: First, we apply the Voronoi tessellation software \citep{2003MNRAS.342..345C} to spatially re-bin the mock cube and increase the spectral SNR to 80~\perpixel{}, which results in 1477 Voronoi bins.
Most of the Voronoi bins contain $N_p>10^4$. 
For the other Voronoi bins, $N_p$ is also very close to this number.
To ensure the sampling noise $e_{f, S}$ is less than the MUSE flux error $e_{f, O}$, we apply the same Voronoi binning arrangement to all 20 bootstrapped cubes.
Then we calculate the sampling noise $e_{f, S}$ of each Voronoi bin by taking half of the difference between the 16th and 84th percentiles of its 20 bootstrapped integrated spectra.
We confirm that on average $e_{f, O}>e_{f, S}$ applies for all the bins.

Next, we apply the full-spectrum fitting method \ppxf{} (\citealt{2004PASP..116..138C, 2017MNRAS.466..798C, 2023MNRAS.526.3273C}) to measure the stellar kinematics of each Voronoi bin. 
Here we use the same MILES $\alpha$-variable templates \citep{2015MNRAS.449.1177V} as when we generated the mock cubes in the same MUSE spectral resolution. 
In the following analysis, we always use the whole grid of MILES templates in age, \mh{}, and \alphafe{}. We are aware of some unsafe regions as listed in Table.~2 of \cite{2015MNRAS.449.1177V}. However, because no particles from the \eglx{} catalog have ages or \mh{} that fall within these regions, it does not affect our conclusions.
We also confirm that our conclusions remain the same after removing templates in the unsafe regions.
Since there is no emission line or atmosphere effect, we use a wide wavelength range of $(4750, 6550)\Angstrom$ to fit with the Voronoi binned spectra and remove the first and last $75\Angstrom$ to avoid effects caused by spectral oversampling, rebinning and Doppler shift, etc.
During the fitting, the MILES templates are convolved with a line-of-sight velocity distribution (LOSVD) described by a Gauss-Hermite equation to match the Voronoi binned spectrum. We parameterize the LOSVD using four moments, which are mean line-of-sight velocity \vlos{}, line-of-sight velocity dispersion \dispersion{}, and the third- and fourth-order moments $h_3$, $h_4$. 
No regularization is applied in this process, and we include a fourth-order multiplicative Legendre polynomial during the fitting. 

After measuring kinematics, we employ \ppxf{} again to obtain the stellar population parameters for each Voronoi bin. 
We choose the same templates, spectral resolution, and fitting wavelength range as the previous step. 
We use the LOSVDs (\vlos{}, \dispersion{}, $h_3$, $h_4$) measured in the last step as input, and fix them during the fitting to obtain the weight of each template. The best-fit spectrum is the weighted sum of all the templates. 
For the initial fitting, we set no regularization to obtain the initial $\chi^2$.
To avoid the ill-conditioned inverse problem due to severe degeneracies in mathematics (as discussed in Section~3.5 of \citealt{2017MNRAS.466..798C}), we then follow the approach in \cite{2015MNRAS.448.3484M} and iterate the fitting by increasing \texttt{regul} until $\chi^2$ is increased by $\Delta\chi^2=\sqrt{2N}$, where $N$ is the number of wavelength pixels considered for the fit.
This iteration process allows us to obtain the smoothest solution that is still compatible with the data within $1\sigma$ level. 
At this stage, we note \texttt{regul} reaches the maximum regularization \texttt{regul$_{max}$}. 
Next, we choose \texttt{regul}~$=5$ which is between 0 and \texttt{regul$_{max}$} $(30-100)$ to keep smooth solutions while still allowing for a variation on short timescales of the star formation, which will disappear if \texttt{regul} is too large (see similar discussions in \citealt{2019A&A...623A..19P} and M21).
Here we only apply the first order of regularization for all the fittings. 
We employ this fitting strategy using templates normalized to both 1 \solarmass{} and 1 \solarlum{} to obtain mass- and light-weighted results, respectively.
Fig.~\ref{fig:eg_ppxf_fit} is an example of the fitting on one Voronoi binned spectrum.
Finally, using the weights, we can calculate light-weighted and mass-weighted age, \mh{}, \alphafe{}, and light/mass fraction distributions.

To estimate uncertainties of template weights and weighted-mean age, \mh{}, and \alphafe{}, we apply Monte Carlo (MC) realizations similar to the way described in \cite{2018MNRAS.480.1973K}, M21, and \cite{2023MNRAS.526.3273C}.
For each Voronoi-binned spectrum, we first perform a fitting with no regularization and obtain the best-fit spectrum and residuals compared to the initial input spectrum.
Next, we add residuals with signs randomly flipped to the best-fit spectrum 100 times to obtain resampled spectra.
Each resampled spectrum is then fitted by \ppxf{} with \texttt{regul}~$=0.1$ to reduce bias towards smooth solutions.
Finally, we calculate uncertainties of template weights and weighted mean properties by taking the half of difference between the 16th and 84th percentiles of 100 MC realizations.

In addition, we also apply the LS module in the GIST pipeline to compute line-strength indices of each Voronoi bin spectrum in the LIS system \citep{2010MNRAS.404.1639V}  choosing the definitions provided at a spectral resolution of 8.4~\angstrom{}. 
This routine was presented by \cite{2006MNRAS.369..497K} and \cite{2018MNRAS.475.3700M}. 
Next, given the relationship between templates' properties (age, \mh{}, \alphafe{}) and line strengths, the measured line strengths can be matched to SSP-equivalent parameters by using the MCMC implementation from the package \textit{emcee} \citep{2013PASP..125..306F}. 
In this work, we follow M21 and use \hbetao{} as an age indicator and \rm{Fe5015}, \rm{Fe5270} and \rm{Fe5335}, and $\mathrm{Mg}b$ to trace metallicity and \alphafe{}, respectively. 
The Monte-Carlo simulation is run 15 times and each one uses 100 walkers and 1000 iterations to obtain uncertainties.

After measuring all the parameters, in the next sub-sections, we compare the kinematics, light- and mass-weighted stellar population parameters, and light/mass fraction distributions of the mock cube with the true values to verify the recovery ability of full-spectrum fitting method \ppxf{}.

\begin{figure*}
\includegraphics[width=2\columnwidth]{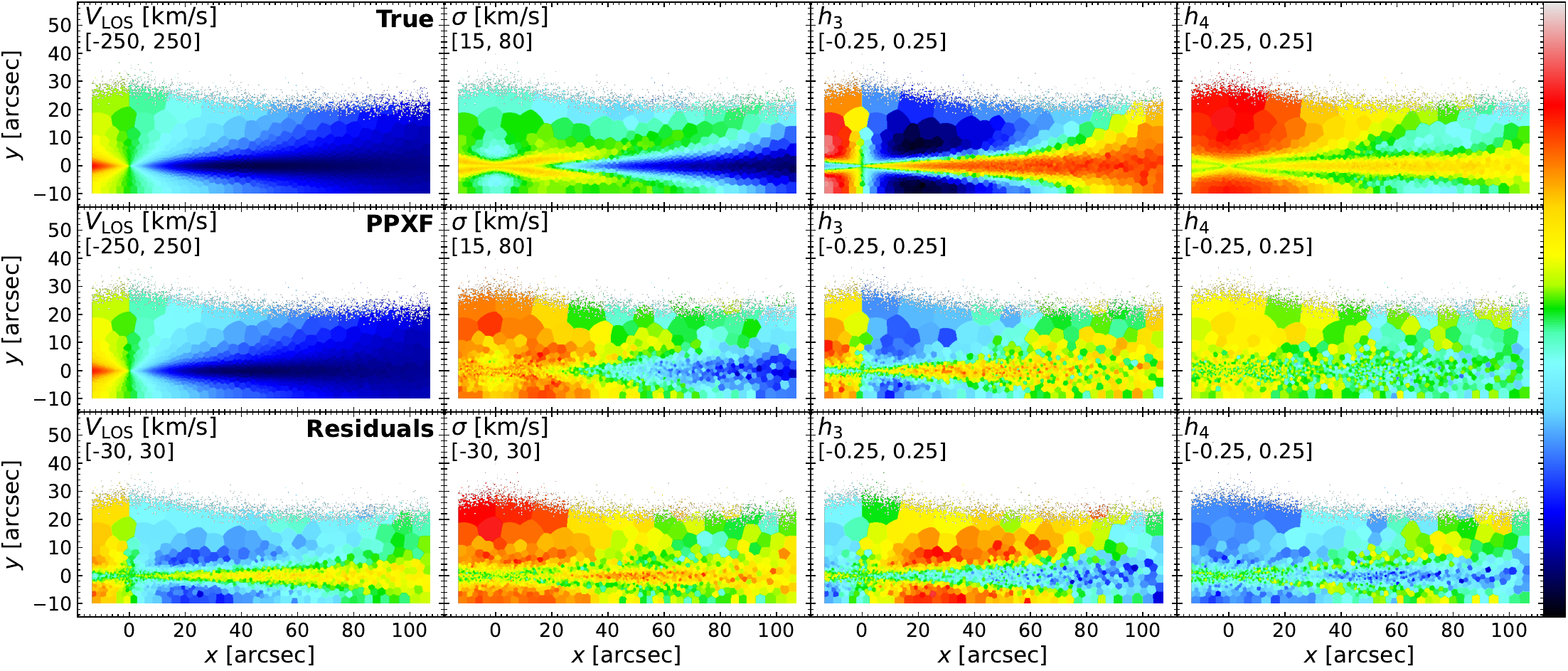}
\caption{
Stellar kinematics maps (\vlos{}, \dispersion{}, $h_3$, $h_4$) of the mock MUSE cube generated in Section~\ref{sec:galaxymapmockcube}. The scale of the color bar is given in the second line of the upper left corner for each panel. \textbf{Top row}: True values calculated by fitting a Gauss-Hermite equation with particles' velocity distribution weighted by their total flux in \ppxf{} fitted wavelength region for each Voronoi bin. \textbf{Middle row}: Results from \ppxf{}. \textbf{Bottom row}: Residuals of the \ppxf{} results and the true values. The residual panels indicate results from \ppxf{} have systematic offsets compared to true values.
}
\label{fig:galaxymapkin}
\end{figure*}

\subsection{Kinematic maps}
\label{sec:galaxymapkin}

Fig.~\ref{fig:galaxymapkin} shows the kinematics maps in four moments (\vlos{}, \dispersion{}, $h_3$, $h_4$) of the mock MUSE cubes. The scale of the color bar is given in the second line of the upper left corner of each panel. 
We calculate the true values shown in the first row with the following procedures:
First, for each Voronoi bin, we calculate the total flux of each particle in the \ppxf{} fitted wavelength region. Then we plot flux-weighted \vlos{} histogram distribution using all the particles included in this bin. Next, we fit this histogram with a Gauss-Hermite equation and obtain four best-fit moments (\vlos{}, \dispersion{}, $h_3$, $h_4$).
This method is consistent with the definition of light-weighted kinematics which \ppxf{} is expected to recover during the fitting process.
The second row is the results from \ppxf{} and the bottom row shows the residuals between \ppxf{} results and the true values. 

In this figure, the kinematic moments obtained by \ppxf{} have the same trend compared to the true values. 
Both show two different structures: one is aligned to $y\sim0$ and the thickness increases with $x$, which has larger absolute \vlos{} and $h_3$, and smaller \dispersion{}; 
the other is in a similar projected radius but vertically higher and thicker, and it has smaller absolute \vlos{} and $h_3$, but larger \dispersion{}.
An anti-correlation of $h_3$ with \vlos{} which are usually associated with disk-like components (e.g. \citealt{2008MNRAS.390...93K, 2016A&A...591A.143G, 2017ApJ...835..104V}) is seen and are similar to MUSE edge-on galaxies studies of \cite{2019A&A...623A..19P, 2019A&A...625A..95P} and M21.
These two components are mostly likely to be associated with thin and thick disks. We will explore this in detail in Section~\ref{sec:sfhmfd}.

However, in the residual panels, all these four moments show systematic offsets. 
Compared to the true values, \vlos{} from \ppxf{} is around $17$~\kms{} lower above and below the very thin mid-plane ($y\sim 0$) and shows a more significant difference around $x\sim [10, 30]$~arcsec. 
Around the galaxy center, the residual of \vlos{} also shows a continuous decrease from negative to positive $x$;
\dispersion{} is generally overestimated everywhere in the galaxy with few light blue residuals.
$h_3$ is overestimated in regions of $x\sim[10, 60]$~arcsec and $y\sim[10, 25]$~arcsec and underestimated in the outer region of $x\sim[60, 110]$~arcsec;
$h_4$ from \ppxf{} has no significant structures like \dispersion{} and $h_3$ maps, which is also seen in real galaxies results (e.g., \citealt{2019A&A...623A..19P, 2019A&A...625A..95P} and M21), but the true $h_4$ map clearly shows kinematic differences.
The clear structures in these residual panels indicate that it is not because of the fitting uncertainties. 
We will investigate this in detail in Section~\ref{sec:discussionkin}.

\subsection{Stellar population property maps}
\label{sec:galaxymapchemppxf}

\begin{figure*}
\centering
\includegraphics[width=1.85\columnwidth]{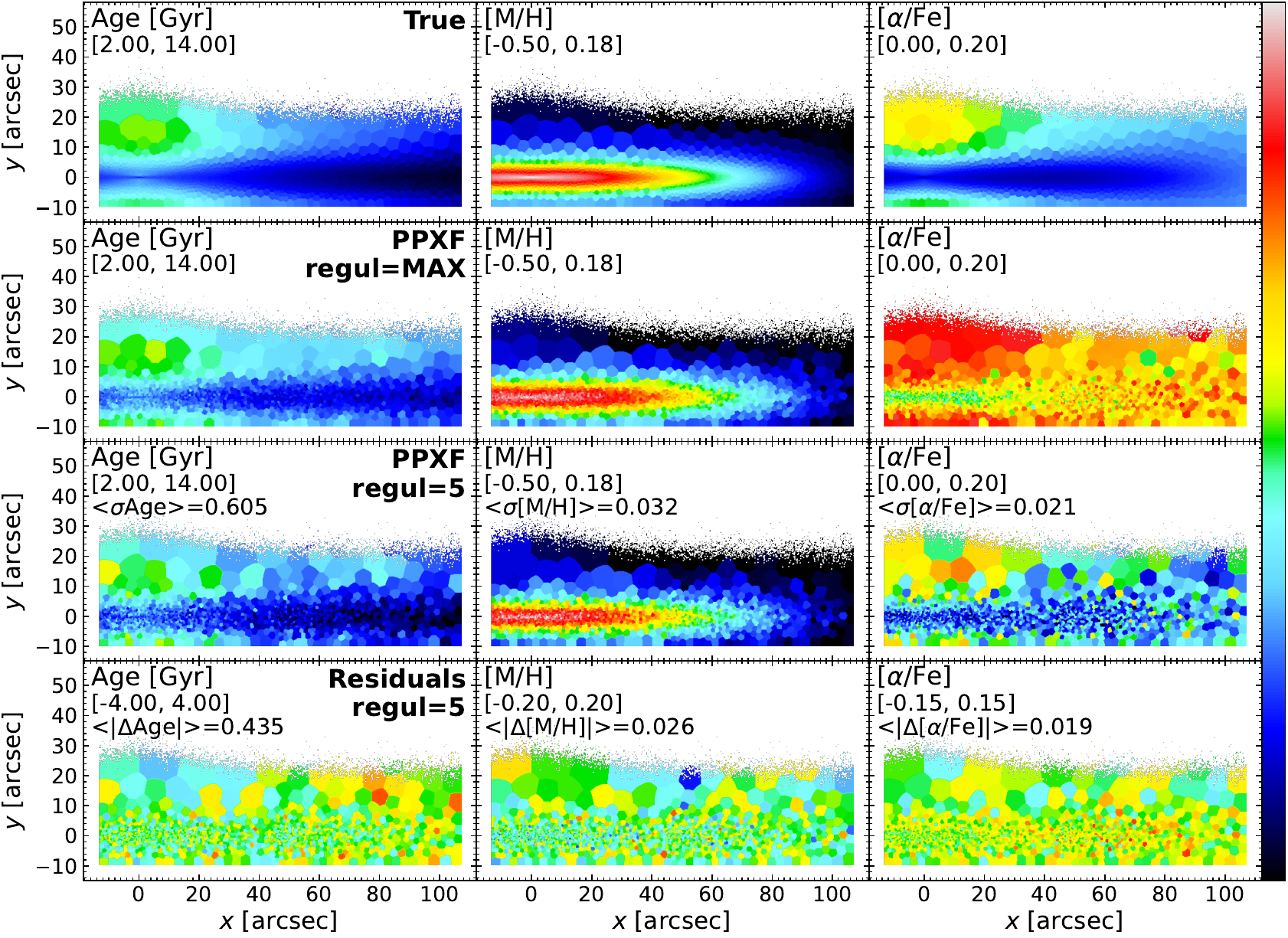}
\caption{
Light-weighted stellar population property maps (age, \mh{} and \alphafe{}) of the mock MUSE cube generated in Section~\ref{sec:galaxymapmockcube}. 
The scale of the color bar is given in the second line of the upper left corner for each panel. 
\textbf{First row}: True values calculated by the light-weighted average of particles' age, \mh{} and \alphafe{} for each Voronoi bin. 
\textbf{Second row}: Results from \ppxf{} with \texttt{regul}~$=$~\texttt{regul$_{max}$}, calculated using strategies of \citealt{2015MNRAS.448.3484M}. 
\textbf{Third row}: Results from \ppxf{} with \texttt{regul}~$=5$ and the average uncertainty of all the Voronoi bins from MC realizations is written in the left top corner.
\textbf{Last row}: Residuals of \ppxf{} results with \texttt{regul}~$=5$ and the true values. The average absolute residual of all the Voronoi bins is written in the left top corner.
This figure indicates that \ppxf{} results with proper regularization can identify different galaxy components by their stellar population parameters, which are consistent with true values and the residuals are within the order of uncertainties on average. 
When applying \texttt{regul}~$=$~\texttt{regul$_{max}$}, the distributions are smoothed and the \alphafe{} panel becomes inconsistent with the true values.
}
\label{fig:galaxymapchem_lw}
\end{figure*}

\begin{figure*}
\centering
\includegraphics[width=1.85\columnwidth]{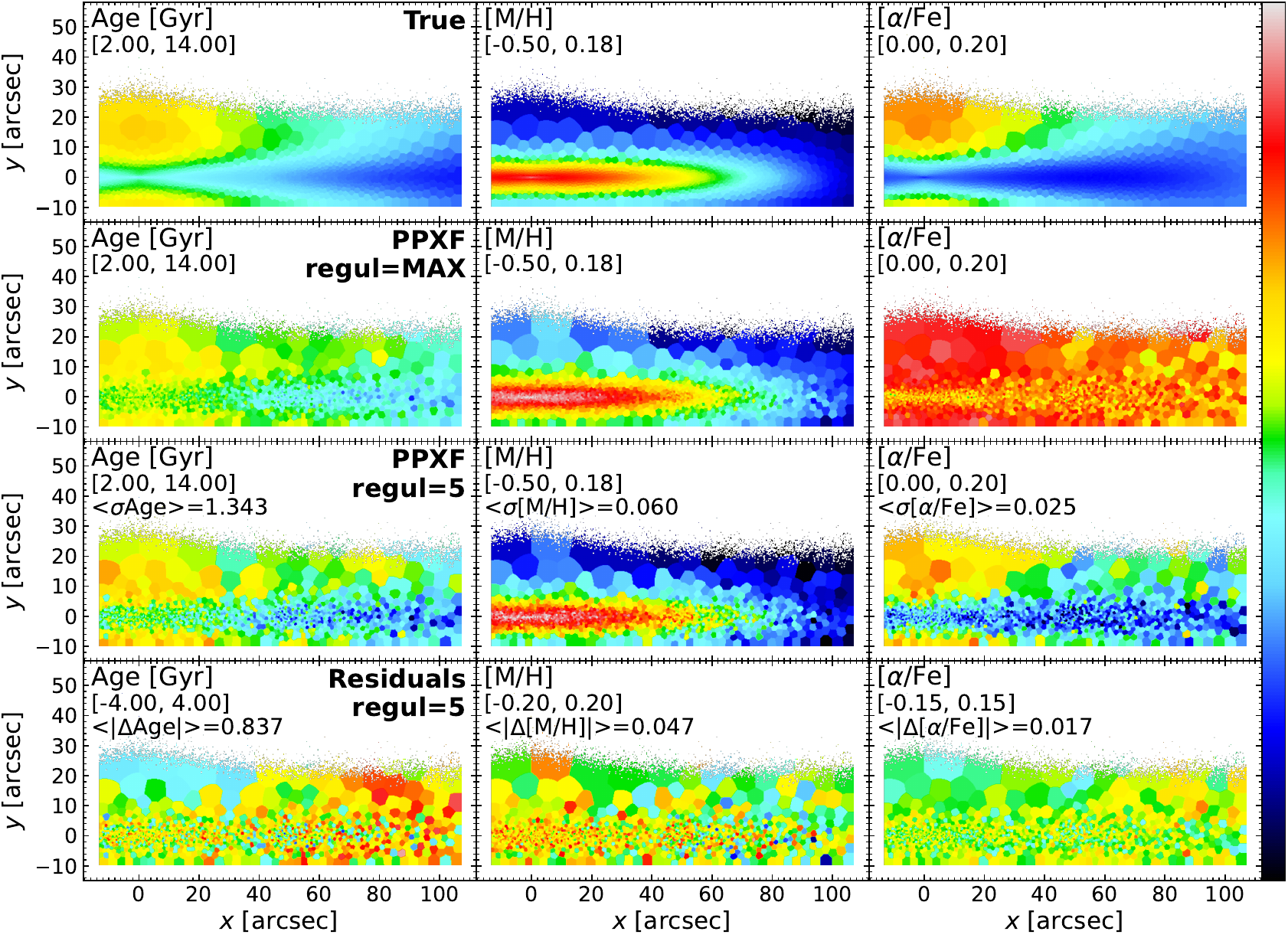}
\caption{
Same as Fig.~\ref{fig:galaxymapchem_lw} but for mass-weighted stellar population properties. 
}
\label{fig:galaxymapchem_mw}
\end{figure*}

Fig.~\ref{fig:galaxymapchem_lw} and Fig.~\ref{fig:galaxymapchem_mw} show the light- and mass-weighted age, \mh{} and \alphafe{} maps of the mock MUSE cubes, respectively.
The first row is the true values by calculating the median age, \mh{}, and \alphafe{} of \eglx{} particles weighted by luminosity or mass, which are equivalent to light- or mass-weighted values. The second and third rows are results from \ppxf{} with \texttt{regul}~$=$~\texttt{regul$_{max}$} and \texttt{regul}~$=5$, respectively. 
We also write the average uncertainty from MC realizations in the top left corner of the third row.
The last row shows the residuals of the \ppxf{} results with \texttt{regul}~$=5$ and the true values, with the average of absolute residual written in the top left corner. 
For both figures, the overall distributions of these three parameters obtained by \ppxf{} are very close to the true values, and the residuals are within the order of uncertainties on average.
This confirms the reliability of spectral fitting methods to measure the weighted age and chemical compositions. 
Especially, the \alphafe{} map from \ppxf{} with \texttt{regul}~$=5$ indicates the capability of \ppxf{} to identify distinct \alphafe{}-rich and \alphafe{}-poor populations in the thick and thin disk, respectively, even though only two \alphafe{} bins are available.
The residuals of \alphafe{} from \ppxf{} with \texttt{regul}~$=5$ and the true values are flat and no systematic pattern is found.

However, mass-weighted age from \ppxf{} is slightly overestimated in the outer regions with more yellow and red Voronoi bins in the residual panel, where the residuals are larger than uncertainties.
This overestimation is much more obvious in mass-weighted results.
In addition, even though residuals of light-weighted \mh{} are mostly close to 0, the mass-weighted \mh{} are overestimated in the central regions.
This means the age gradient from \ppxf{} is underestimated but \mh{} gradient is overestimated, and such effects are more dominant in mass-weighted results.
In addition, the \alphafe{} distribution from \ppxf{} results with \texttt{regul}~$=$~\texttt{regul$_{max}$} is almost uniformly high and much larger than the true values for all the Voronoi bins in both light- and mass-weighted results. 
This is because when \texttt{regul} is very large, the \ppxf{} algorithm forces the result to have very smooth template weights in three parameter dimensions (age, \mh{}, \alphafe{}). 
Since there are only two \alphafe{} grids, regularization will force them to have similar weights to achieve smoothness requirements and does not permit large deviations (e.g., more than 2\%). Therefore, it will be challenging to identify \alphafe{} bimodality.
Results from \ppxf{} with \texttt{regul}~$=$~\texttt{regul$_{max}$} also show much underestimation for age gradients along the x-axis than results with \texttt{regul}~$=5$.
The age and \mh{} gradients are essential properties to help understand the star formation and chemical enrichment processes.
Therefore, a wrong choice of regularization will then easily lead to wrong conclusions.
We will explore the reasons for these offsets in more detail in Section~\ref{sec:sfhmfd} using light and mass fraction distributions and the effect of regularization in Section~\ref{sec:discussionmfd} and Section~\ref{sec:discussionregul}.

\subsection{SSP-equivalent maps from line-strength indices}
\label{sec:galaxymapchemls}

\begin{figure*}
\centering
\includegraphics[width=1.85\columnwidth]{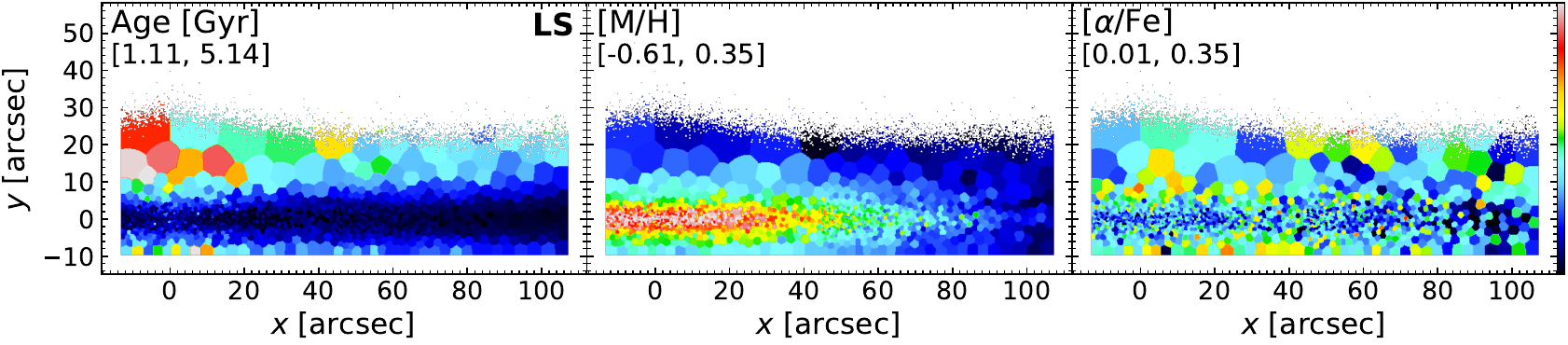}
\caption{
SSP-equivalent age, \mh{} and \alphafe{} maps of the mock MUSE cube generated in Section~\ref{sec:galaxymapmockcube}. The parameters are measured by line-strength indices. The scale of the color bar is given in the second row of the upper left corner for each panel. 
Similar to Fig.~\ref{fig:galaxymapchem_lw}, galaxy components with different ages, \mh{}, and \alphafe{} can be identified. However, the parameter ranges differ from \ppxf{} results and the true values. 
}
\label{fig:galaxymapls}
\end{figure*}

Fig.~\ref{fig:galaxymapls} shows the SSP-equivalent age, \mh{} and \alphafe{} maps of the mock MUSE cubes measured by line-strength indices. 
This figure shows that the main structures we derived from \ppxf{} are also recovered by the line-strength analysis with consistent trends. 
In the age panel, young populations are closer to the mid-plane, and old populations are further to the mid-plane or above/below the central region. 
In the \mh{} panel, we see the metallicity gradient from the inner center to the outer Galaxy. In the \alphafe{} panel, we could see the $\alpha$-rich bins in the center and $\alpha$-poor bins in the outer region, even though the differences are not as obvious as \ppxf{} results in Fig.~\ref{fig:galaxymapchem_lw} and Fig.~\ref{fig:galaxymapchem_mw}.
The main difference compared with \ppxf{} results is that the age panel shows a very low range of $[1-5]$~Gyr. 
This is also seen in M21 (Fig. B1) and because of the Balmer line indices being dominated by young stars. Therefore, the SSP-equivalent ages only reflect the fraction of stars formed within the past Gyr \citep{2007MNRAS.374..769S, 2009MNRAS.395..608T}.
The SSP-equivalent \mh{} and \alphafe{} range are much closer to \ppxf{} results because young populations do not contribute much to the metal lines, which is also indicated in M21. This figure confirms that both line-strength indices and \ppxf{} analysis can identify $\alpha$-rich and $\alpha$-poor populations.

\subsection{Weight fraction distributions of different galaxy components}
\label{sec:sfhmfd}

\begin{figure}
\includegraphics[width=1\columnwidth]{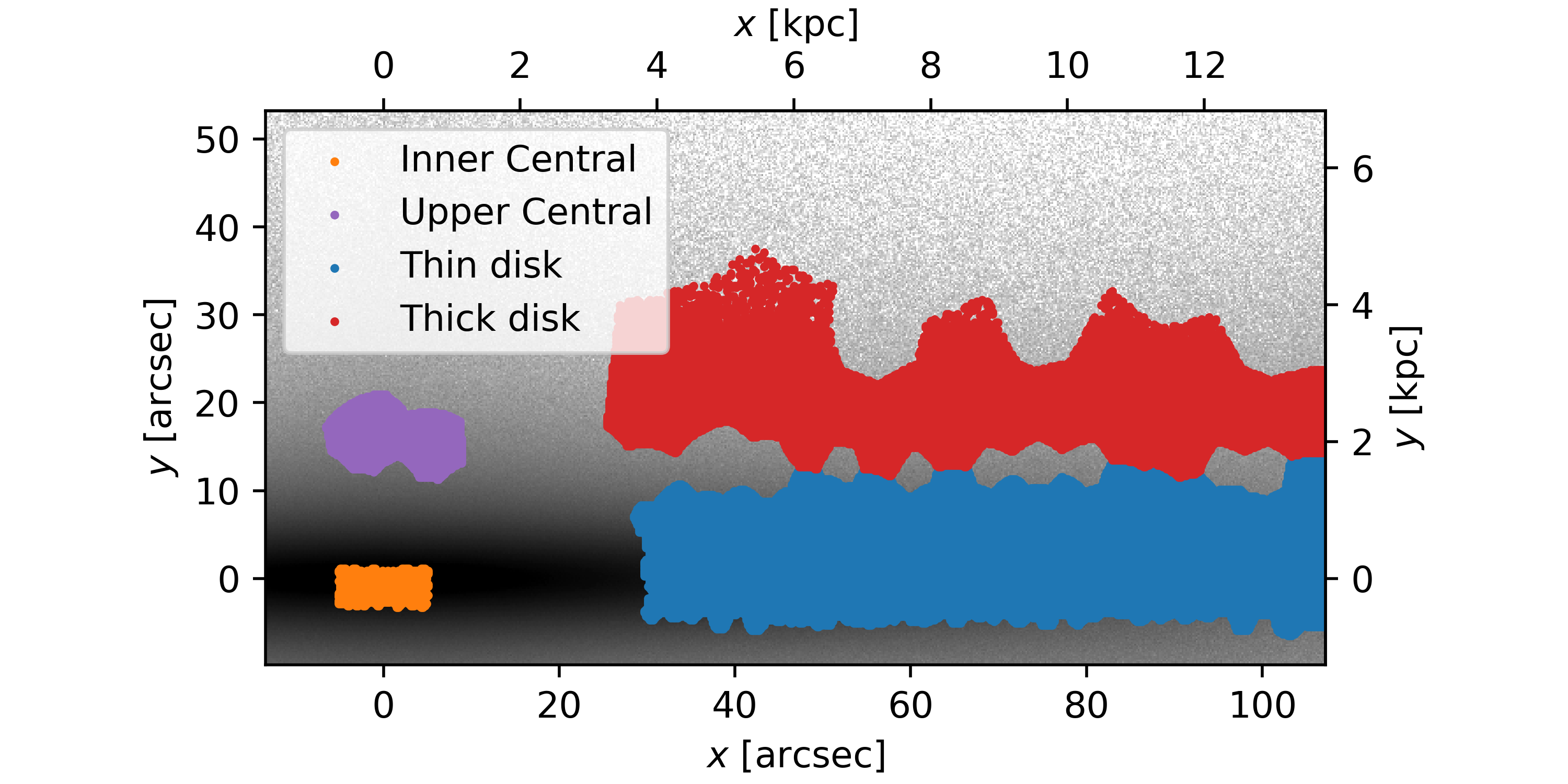}
\caption{Demonstration of the projected thin disk (blue), thick disk (red), upper central (purple), and inner central (orange) regions on top of a grey-scale image of the mock data cube generated in Section~\ref{sec:galaxymapmockcube}. 
}
\label{fig:galaxycomponents}
\end{figure}

In addition to calculating the light- and mass-weighted properties, we can also study the light/mass fraction distribution of stellar populations along the age and \mh{} dimension. This is done by using weights of templates from \ppxf{}. 
Because the flux of each template is normalized to either 1~\solarlum{} or 1~\solarmass{}, the weights array from \ppxf{} outputs in our tests are equivalent to stellar population light or mass fractions.
Therefore, we can study the weight distribution of any component of the mock MW. 
M21 employed multiple components morphological fitting to a \textit{Spitzer} 3.6~micron image of NGC~5746 to obtain regions dominated by the boxy/peanut bulge, nuclear disk, and thin and thick disks.
In Fig.~\ref{fig:galaxycomponents}, we artificially select similar regions based on the locations $(x, y)$ of different components of M21, and name them ``upper central'', ``inner central'', ``thin disk'' and ``thick disk'', as shown in different colors. 
We call them ``upper central'' and ``inner central'' because there is no boxy/peanut bulge and nuclear disk in the GCE of S21. 
Note these component definitions are purely following those in M21 to mock their data analysis.
In reality, radial scale lengths of the thin disk $(R^t)$ and thick disk $(R^T)$ in NGC~5746 and the MW are very different ($R^t_{\mathrm{MW}}=2.6\pm0.5$~kpc and $R^T_{\mathrm{MW}}=2.0\pm0.2$~kpc from \citealt{2016ARA&A..54..529B};
$R^t_{\mathrm{NGC~5746}}=6.1$~kpc and $R^T_{\mathrm{NGC~5746}}=8.2$~kpc from M21).
For each component, the light and mass weights of all the Voronoi bins are combined to represent its light and mass fraction distributions.

\begin{figure*}
\includegraphics[width=2\columnwidth]{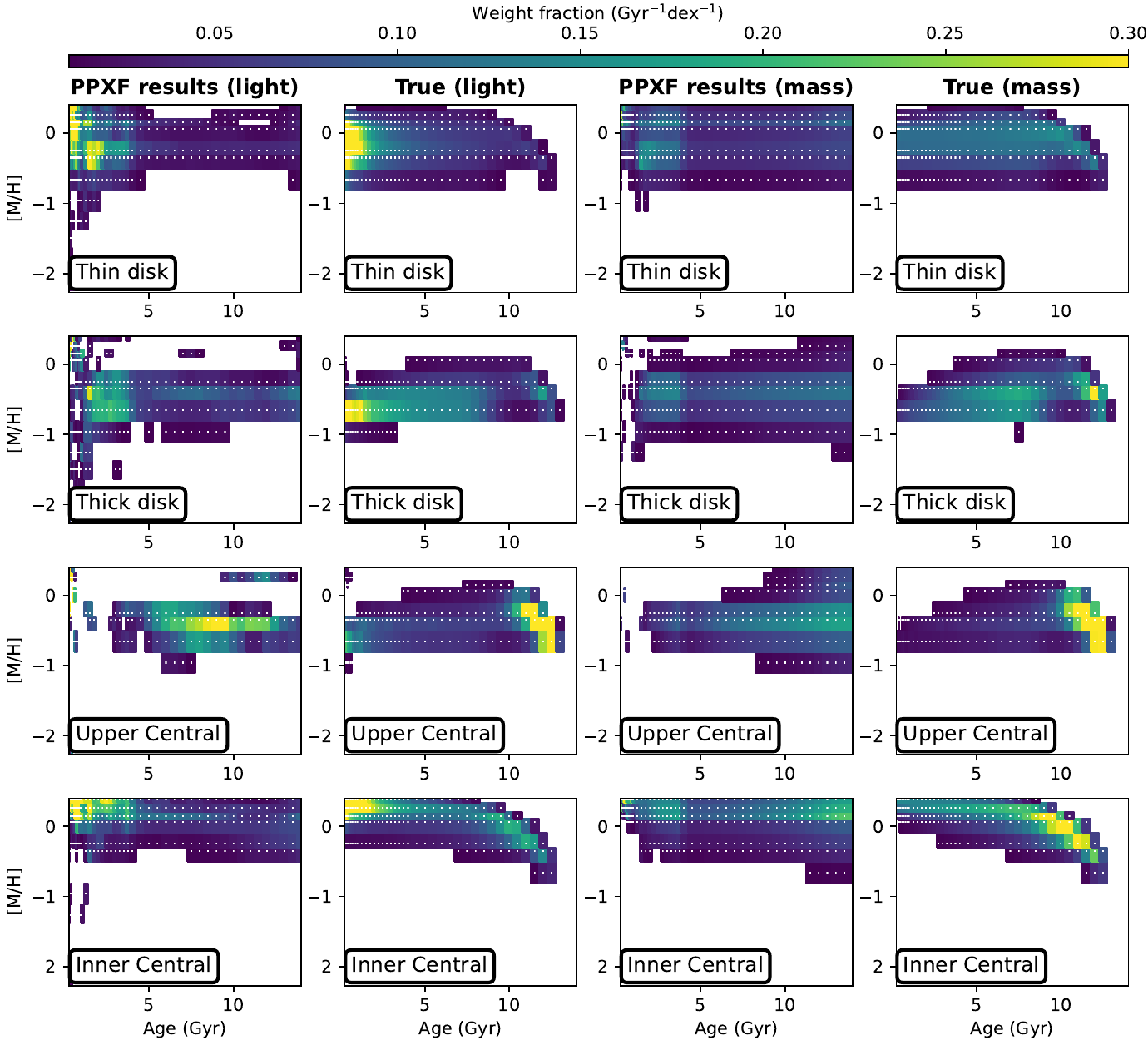}
\caption{
Light and mass fraction distributions of the thin disk, thick disk, inner central, and upper central of the mock data cube generated in Section~\ref{sec:galaxymapmockcube}. 
The total template weights are normalized to 1 for each panel. 
\textbf{First column}: Light fraction distributions from \ppxf{} with \texttt{regul}~$=5$. 
\textbf{Second column}: True light fraction values calculated using particles' properties in \eglx{} catalog. 
\textbf{Third column}: Mass fraction distributions from \ppxf{} with \texttt{regul}~$=5$. 
\textbf{Last column}: True mass fraction values calculated using particles' properties in \eglx{} catalog. 
The color bar is shown on the top. We only plot the weights with values above 0.001~\pergyrdex{}.
This figure indicates the broad trends of results from \ppxf{} are consistent with the true values for the thin/thick disk and inner central regions. 
However, both the light and mass fraction distribution from \ppxf{} are different from the true values for the upper central region.
In addition, there is an overdensity in $2-4$~Gyr in \ppxf{} mass distributions, and populations around 12~Gyr are smoothed towards older, more metal-rich regions, which are not seen in the true values. 
}
\label{fig:massfractioncomponents}
\end{figure*}

\begin{figure*}
\includegraphics[width=2.0\columnwidth]{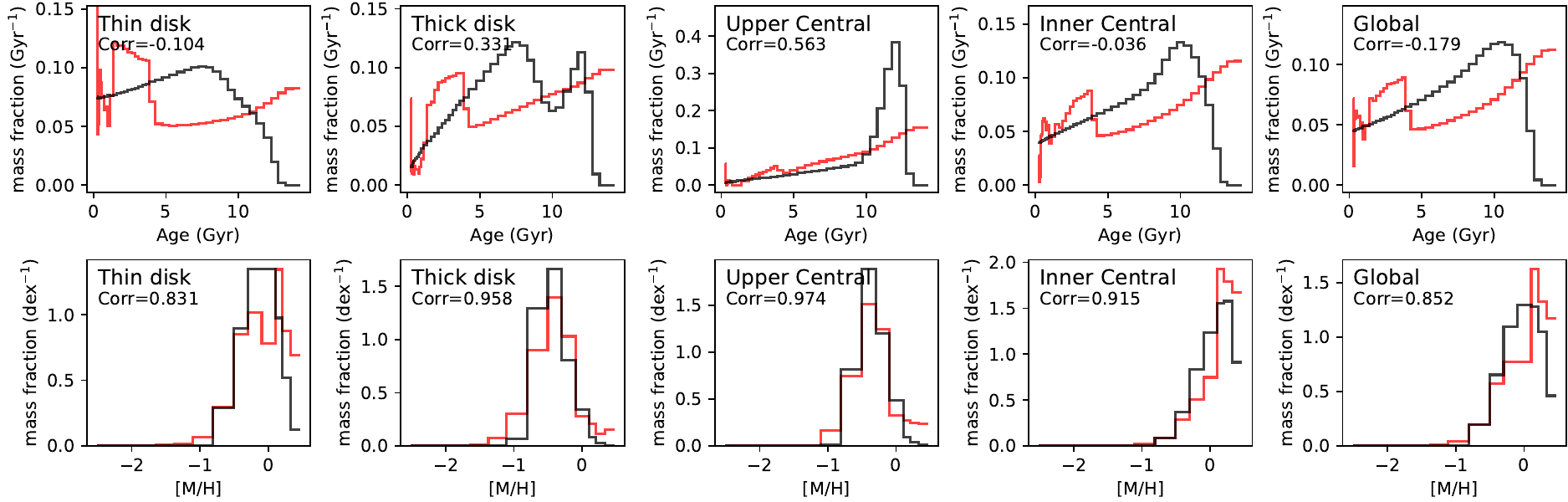}
\caption{
Age distribution (top row) and \mh{} distribution (bottom row) of the thin disk, thick disk, inner central, and upper central of the mock data cube generated in Section~\ref{sec:galaxymapmockcube}. 
These panels are obtained by integrating the mass fraction distributions in Fig.~\ref{fig:massfractioncomponents} along \mh{} and age axis, respectively.
Red lines are results from \ppxf{} with \texttt{regul}~$=5$. Black lines are the true values calculated using particles' properties in the mock \eglx{} catalog. 
The correlation coefficient of these two lines is written in the left top corner for each panel.
The overall trends from \ppxf{} are consistent with the true values for \mh{}. However, same as Fig.~\ref{fig:massfractioncomponents}, the mass fraction is overestimated in $2-4$~Gyr and $12-14$~Gyr and underestimated in $5-12$~Gyr for all the components, and the coefficients indicate inconsistency for age distributions.
We also find peaked features in regions less than 1~Gyr.
The mass weights of the most metal-rich bin are overestimated for all the components.
}
\label{fig:massfractioncomponents1d}
\end{figure*}

Fig.~\ref{fig:massfractioncomponents} shows the light and mass fraction distributions of these four components, respectively. 
The total weights are normalized to one for each panel. 
The left two columns are light fractions and the right two columns are mass fractions, respectively. 
The \ppxf{} results (first and third columns) are obtained with \texttt{regul}~$=5$.
The true values (second and last columns) are calculated using particles' properties in \eglx{} catalog weighted by luminosity or mass. 
We only plot the weights above 0.001~\pergyrdex{}.
For the thin disk, true light fractions are dominated by the youngest populations with ages less than 2~Gyr, and \ppxf{} can consistently recover its distribution. True mass fractions indicate a rapid metallicity enrichment history $\sim10$~Gyr ago, and it slowly increases later on. 
However, such a metallicity enrichment trend is indistinguishable in \ppxf{} mass fraction distributions. Moreover, \ppxf{} mass fractions are dominated at relatively young ($2-4$~Gyr) and old ($12-14$~Gyr) stellar populations. 
The same features are also seen in mass fraction distributions of other Galaxy components.
We will investigate these findings in detail in Section~\ref{sec:discussionmfd} to \ref{sec:discussionaddmfd}.
For the thick disk, both true light and mass fractions show the metallicity enrichment trend which is again not seen in \ppxf{} results.
In addition, the thick disk contains populations that are young and more metal-poor than the thin disk. 
One reason is the geometrical definition of the thick disk, which contains young and relatively metal-poor stars that are flared in the outer disk ($x\sim[60, 100]$~arcsec) due to radial gradients of age and metallicity.
Given that NGC~5746 in M21 is analysed to be four times more massive than the MW in total stellar mass, it has a larger scale length and the definition of its thick disk might not apply to the MW.
Another reason could be the projection effect, where young stars flared in the outer disk can have large Galactocentric (intrinsic) radius but small projected radius, so they could appear at the front and back of the line of sight in the region of $x\sim[30, 60]$~arcsec.
The upper central mass fractions show a similar metallicity enrichment trend with the thick disk and is more dominant in the old populations, but this domination is smoothed out in the \ppxf{} results. Moreover, light fractions from \ppxf{} are dominated at the age around 8~Gyr, which is inconsistent with the true light fractions.
For the inner central, true light and mass fractions are showing a clearer chemical enrichment trend and there is no new population born with $\mh{}<-0.2$~dex in the young region, while the \ppxf{} mass fractions show again two overdensities at $2-4$~Gyr and $12-14$~Gyr. 
Therefore, except for the overestimation in the young and old population regions in mass fraction distributions, and the inconsistency in light fractions of the upper central region, the light and mass fraction distributions of \ppxf{} are generally in agreement with true values.

In Fig.~\ref{fig:massfractioncomponents1d}, we integrate the mass fraction distributions in Fig.~\ref{fig:massfractioncomponents} along the two axes and derive age and \mh{} distributions for each component. The top panels are mass distributions as a function of age which is the definition of star formation history (SFH) or star formation rate (SFR), and the bottom panels are defined as metallicity distribution function (MDF). Results from \ppxf{} are in red lines and the true values are in black lines. 
For each panel, we calculate the correlation between these two lines to quantify their similarity. 
For the age distributions, we find the same as in Fig.~\ref{fig:massfractioncomponents}. Compared to the true values, the \ppxf{} results of all the components demonstrate an overestimation of weights in the ranges of $2-4$~Gyr and $12-14$~Gyr and underestimation in the range of $4-11$~Gyr. The underestimated regions seem to compensate for the overestimated regions. 
And the thin disk, inner central, and global panels show a peaked feature with age $<1$~Gyr.
For the \mh{} distributions, \ppxf{} results are consistent with true values for most regions, as indicated by the correlation coefficients.
However, mass fractions in the metal-rich region are overestimated.
Other than that, the overall trend of results from \ppxf{} is consistent with true values.

In conclusion, we find that spectral fitting methods can recover the broad trends of 2-D light and mass fraction distributions for different components, but with mass fractions overestimated in $2-4$~Gyr, $12-14$~Gyr and most metal-rich regions.
When integrating into 1-D age and metallicity distributions, these inconsistencies are significant.
According to correlation coefficients, \mh{} distributions are more consistent with the true values than age distributions.
We will investigate the reasons for such differences in detail in Section~\ref{sec:discussionmfd} to \ref{sec:discussionaddmfd}.

\subsection{Distributions of \texorpdfstring{\alphafe{}-\mh{}}~ along different galaxy locations $(R, z)$}
\label{sec:haydenplot}

\begin{figure*}
\includegraphics[width=2\columnwidth]{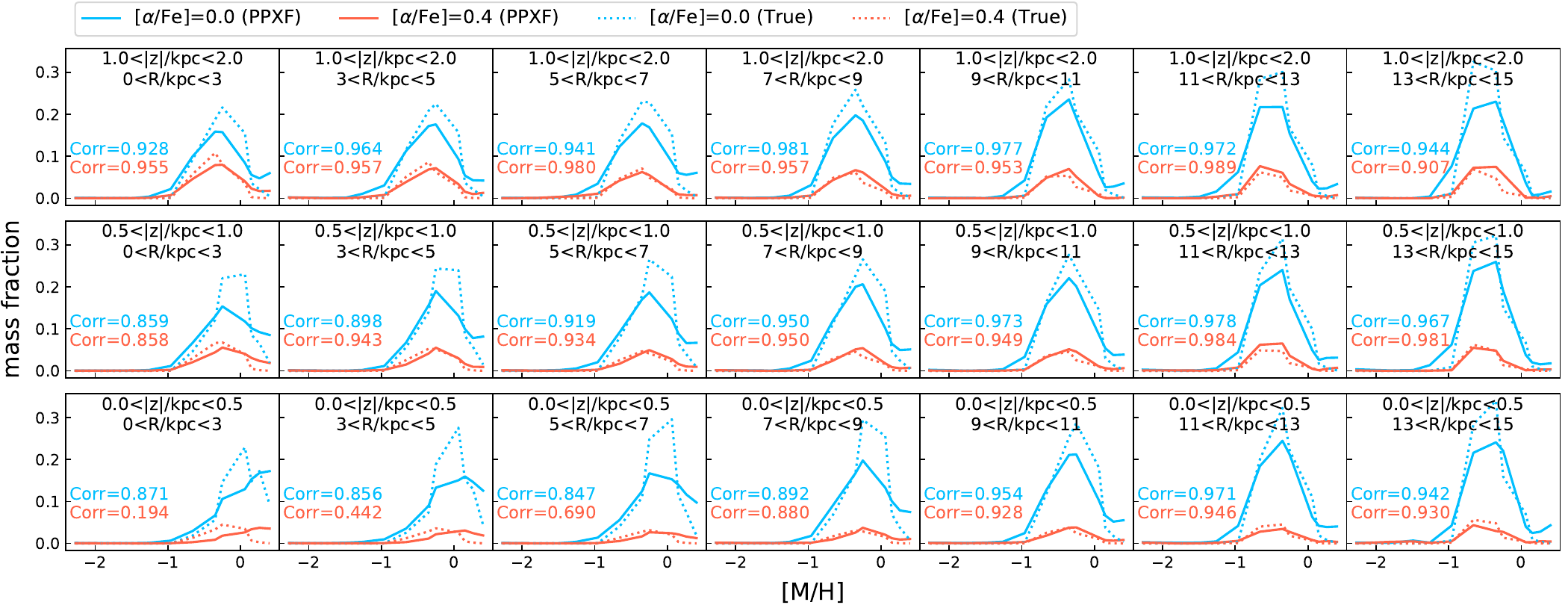}
\caption{
\mh{} distributions for different \alphafe{} components distributed along projected $R$ and $|z|$ of the mock cube  generated in Section~\ref{sec:galaxymapmockcube}. 
For each panel, the total mass fractions of true (dotted lines) and \ppxf{} (solid lines) are normalized to 1, respectively.
Blue lines represent mass fraction distributions for \alphafe{}~$=0.0$~dex while red lines represent the distributions with \alphafe{}~$=0.4$~dex. 
Solid lines represent the results from \ppxf{} and dotted lines are the true values. This figure is similar to the \alphafe{}-\feh{} distributions in \citealt{2015ApJ...808..132H} and \citealt{2021ApJ...913L..11S}. 
The correlation coefficient of the dotted and solid lines are written in the same color in the left bottom corner.
For most of the panels, \ppxf{} results are consistent with the true values. 
The difference appears at the inner central and thin disk, where both lines show a mass fraction underestimation at \mh{}~$\sim0.0$~dex and overestimation at the most metal-rich regions.}
\label{fig:haydenplot}
\end{figure*}

According to S21, different radial migration and kinematic heating efficiency will cause different fractions of stars in the distinct \alphafe{}-rich and \alphafe{}-poor sequences. 
A recent study by \cite{2021ApJ...913L..11S} derived this distribution from an external galaxy UGC~10738 using MILES $\alpha$-variable templates \citep{2015MNRAS.449.1177V} and concluded it has similar bimodality distributions to the MW.
Given we know the true values from the mock stellar catalog particles, here we explore the recovery ability of spectral fitting methods on measuring the changes of this bimodality at different projected $R$ and $|z|$. 
Different from \cite{2021ApJ...913L..11S} where they compared integrated values of UGC~10738 with individual stellar values of the MW, our integrated-to-integrated value comparison on stellar population level is more direct and there is no systematic bias due to different methodologies.
This direct comparison also provides an example for future studies on comparing an integrated version of the MW \alphafe{} bimodality with MW-like edge-on galaxies from MUSE observations (e.g., GECKOS survey \citealt{2024IAUS..377...27V}).

The results are shown in Fig.~\ref{fig:haydenplot} where we separate different locations in the same way as \cite{2015ApJ...808..132H}. 
For each panel, the total mass fraction given by true (dotted lines) and \ppxf{} (solid lines) is normalized to 1, respectively. 
Blue lines are mass fraction distributions for $\alphafe{}=0.0$~dex while red lines represent $\alphafe{}=0.4$~dex. 
The overall trends for \alphafe{}-rich and \alphafe{}-poor sequences from \ppxf{} are consistent with the true values for most of the panels, which indicates both sequences are well recovered by \ppxf{}, as also shown by correlation coefficients.
However, there are some discrepancies such as in the inner regions ($R<5$~kpc) and thin disk ($|z|$<0.5~kpc), where \ppxf{} show smoother metallicity distributions in both the blue and red lines compared to true values. 
In addition, the mass fraction is underestimated at $-1<\mh<0.0$~dex and overestimated at the most metal-rich regions, which is the same as metallicity distributions in Fig.~\ref{fig:massfractioncomponents1d}. 

When comparing different \alphafe{} sequences, the peaks of distributions are at the same metallicity position in both \ppxf{} and true results.
This differs from our understanding of \alphafe{}-\mh{} relation of the Milky Way.
It is more likely due to the limited number of \mh{} and \alphafe{} bins in MILES templates that cause the differences of metallicity distribution in different \alphafe{} bins to be indistinguishable. 
Similar findings are also mentioned by \cite{2021ApJ...913L..11S} when they analyzed the \alphafe{}-bimodality of UGC~10738.
Therefore, we emphasize that more \mh{} and \alphafe{} bins in the spectral templates are necessary to obtain more detailed distributions of this bimodality. 
It can help to make it more feasible to identify the effect of different radial migration and kinematic heating efficiency from the relative fractions of these two sequences when compared with other IFS observations.

\section{Discussion}
\label{sec:discussion}

\subsection{Systematic offsets in the kinematics recovery}
\label{sec:discussionkin}

\begin{figure*}
\includegraphics[width=2\columnwidth]{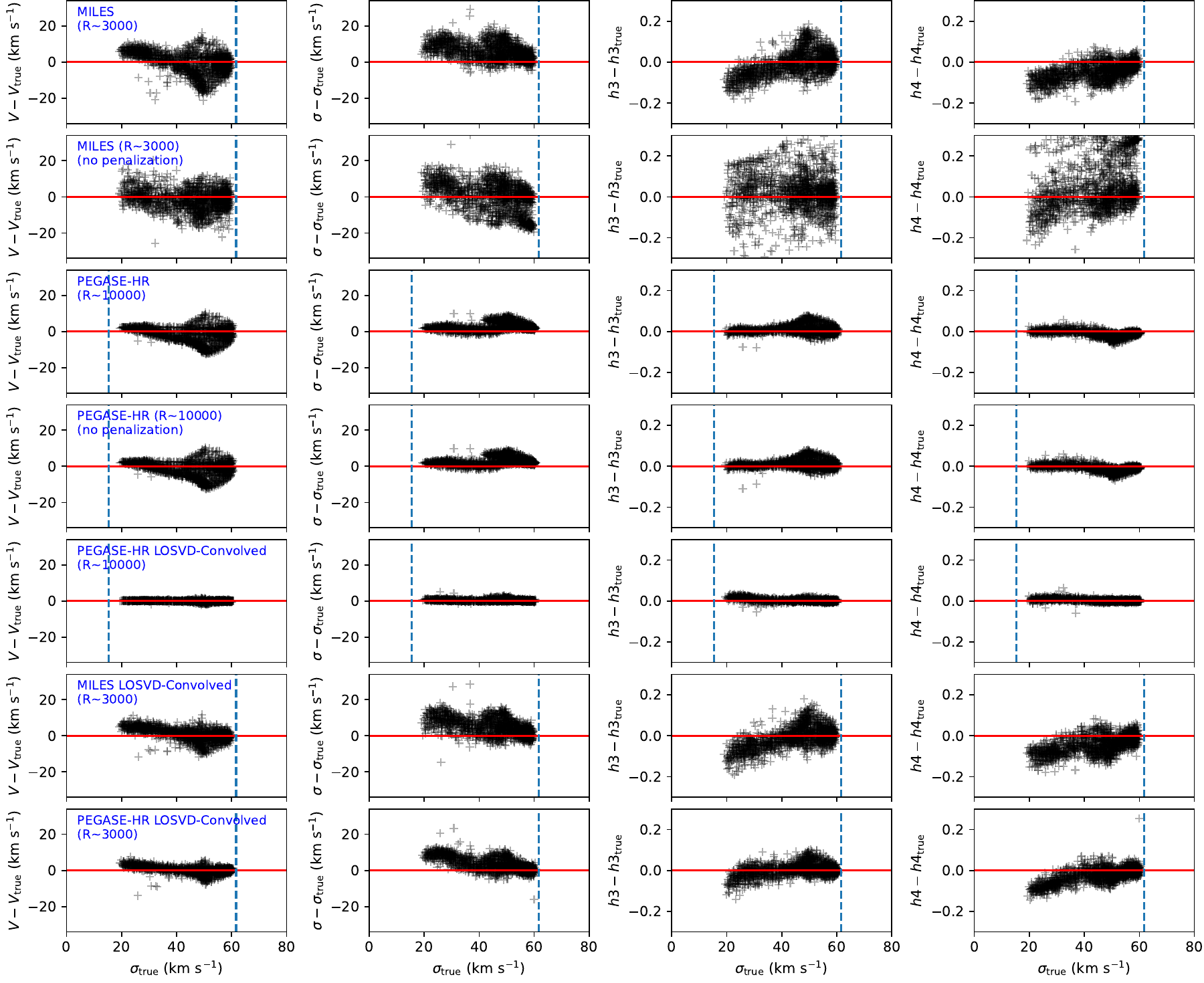}
\caption{
Residuals of four Gauss-Hermite moments $\Delta$(\vlos{}, \dispersion{}, $h_3$, $h_4$) as a function of true velocity dispersion.
Each data point represents a Voronoi bin. The red line at zero is to guide the eyes. The vertical dashed line indicates the instrumental dispersion \instdispersion{}.
\textbf{First row}: Results from \ppxf{} using MILES templates in MUSE spectral resolution (\instdispersion{}~$\sim62$~\kms{}), same as the residual panels in Fig.~\ref{fig:galaxymapkin}.
\textbf{Second row}: Similar to the first row but results are fitted by \ppxf{} with no penalization on $h_3$ and $h_4$.
\textbf{Third row}: Results from \ppxf{} on a mock cube similar to Section~\ref{sec:galaxymap} but generated and fitted using PEGASE-HR \citep{2004A&A...425..881L} templates (\instdispersion{}~$\sim15$~\kms{}).
\textbf{Fourth row}: Similar to the third row but penalization was turned off during \ppxf{} fitting.
\textbf{Fifth row}: Similar to the third row but for LOSVD-Convolved mock cubes.
\textbf{Sixth row}: Similar to the first row but for LOSVD-Convolved mock cubes.
\textbf{Seventh row}: Similar to the fourth row but all the analysis is in MUSE spectral resolution ($\mathrm{FWHM}\sim2.65\Angstrom$). 
This figure indicates that both the spectral resolution and the relation between LOSVD and age, \mh{} create the kinematics systematics.
}
\label{fig:kinvsdispersionallcase}
\end{figure*}

In Fig.~\ref{fig:galaxymapkin}, we show the deviations between the kinematics map from \ppxf{} and the true values. 
To better illustrate the offsets of each moment, we plot their residuals as a function of true velocity dispersion $\sigma_{\rm{true}}$ in Fig.~\ref{fig:kinvsdispersionallcase}, in which the y-axis is calculated by \ppxf{} fitted value subtracting the true value. 
Each data point represents one Voronoi bin. 
The blue dotted lines represent the instrumental dispersion (\instdispersion{}) based on spectral resolution.
We also plot the zero-line in red to guide the eyes.
The first row demonstrates the results in Fig.~\ref{fig:galaxymapkin}, and it clearly shows the systematic offsets for each moment: $\Delta$\vlos{} increases with $\sigma_{\rm{true}}$, and \dispersion{} is overestimated for most of the Voronoi bins; as for $h_3$ and $h_4$, the residuals have a slight positive gradient. 

From Fig.~\ref{fig:galaxymapkin}, we find that several Voronoi bins have large absolute $h_3$ and $h_4$ true values (e.g., central region and the thick disk), which will not be well constrained when SNR is relatively low \citep{2004PASP..116..138C, 2021ApJ...921....8P}.
Therefore, \ppxf{} with keyword \texttt{bias}
will penalize them towards lower absolute values during the fitting to make the LOSVD towards a Gaussian shape. 
However, using different penalizations can result in different kinematic measurements (see full analysis using SAMI galaxies by \citealt{2017ApJ...835..104V}). 
In our case, when the LOSVD is strongly non-Gaussian, the Gauss-Hermite approximation no longer works well and large $h3$ and $h4$ could be obtained.
Therefore, we turn off the penalization of these higher-order moments and re-measure the kinematics, and show the results in the second row of Fig.~\ref{fig:kinvsdispersionallcase}. 
A spatially distributed map can also be seen in Fig.~\ref{appfig:galaxymapkin_nopenal} of the Appendix~\ref{appendix:morefigures}.
After turning off penalization, both $h_3$ and $h_4$ from \ppxf{} are visually closer to the true values in Fig.~\ref{appfig:galaxymapkin_nopenal}. 
However, the second row of Fig.~\ref{fig:kinvsdispersionallcase} still shows similar trends to the first row, and the four moments are more scattered, which indicates penalization is not the cause of the systematic offsets.

Another possible reason causing the kinematic inconsistency could be the low spectral resolution of the MUSE instrument. 
As shown in the first row of Fig.~\ref{fig:kinvsdispersionallcase}, \truedispersion{} values for most of the Voronoi bins are smaller than the MUSE instrumental dispersion \instdispersion{}$\sim62$~\kms{}. 
In this case, the recovery of kinematics will be uncertain. This has been pointed out in Fig. 3 of \cite{2017MNRAS.466..798C}, and the only way to avoid it is to increase the instrumental spectral resolution.
The explanation is when \dispersion{}~$<$~\instdispersion{}, during \ppxf{} fitting, the broadening of one sharp spectral feature is less than the distance to its nearby wavelength pixels.
In this case, the nearby wavelength pixels have minor changes and this brings difficulties for \ppxf{} to measure \dispersion{} correctly, and $h_3$ and $h_4$ will go towards zero because there are also not enough pixels to identify the skewness and kurtosis. 

To remove the effect of low spectral resolution, we tested the kinematics recovery again by using PEGASE-HR templates \citep{2004A&A...425..881L} generated by Kroupa IMF and PADOVA 1994 isochrones, which has a higher spectral resolution ($\rm{FWHM}=0.55\Angstrom$) than MUSE ($\mathrm{FWHM}\sim2.65\Angstrom$). 
The instrumental velocity scale of PEGASE-HR is \instdispersion{}$\sim15$~\kms{}, which is smaller than the minimum $\sigma_{\rm{true}}$ of our Voronoi bins. 
We repeat all the procedures in Section~\ref{sec:galaxymapdatareduction} to generate new cubes using PEGASE-HR, and then apply them to the GIST pipeline to measure the kinematics. We keep the original PEGASE-HR spectral resolution throughout the whole process.
The third row of Fig.~\ref{fig:kinvsdispersionallcase} shows the kinematics recovery using PEGASE-HR (also see the kinematics map in Fig.~\ref{appfig:galaxymapkin_pegasehr}). 
Compared to the first row, the residuals of each panel are slightly better or clearer when using higher spectral resolution templates. However, all these four moments from \ppxf{} still have the same systematic offsets to the true values. 
Most improvements are for $h_3$ and $h_4$, with less bias to zero values because the 
higher spectral resolution helps identify skewness and kurtosis. 
We also show the results by turning off penalization in the fourth row (also see the kinematics map in Fig.~\ref{appfig:galaxymapkin_pegasehr_nopenal}), and find the $h_3$ and $h_4$ residuals are identical to the third row.
Therefore, turning off penalization helps on low-spectral-resolution kinematics measurements, but not for higher-resolution spectra.
The higher spectral resolution improves the kinematics recovery but does not fix the systematics offsets.

\begin{figure*}
\centering
\includegraphics[width=1.98\columnwidth]{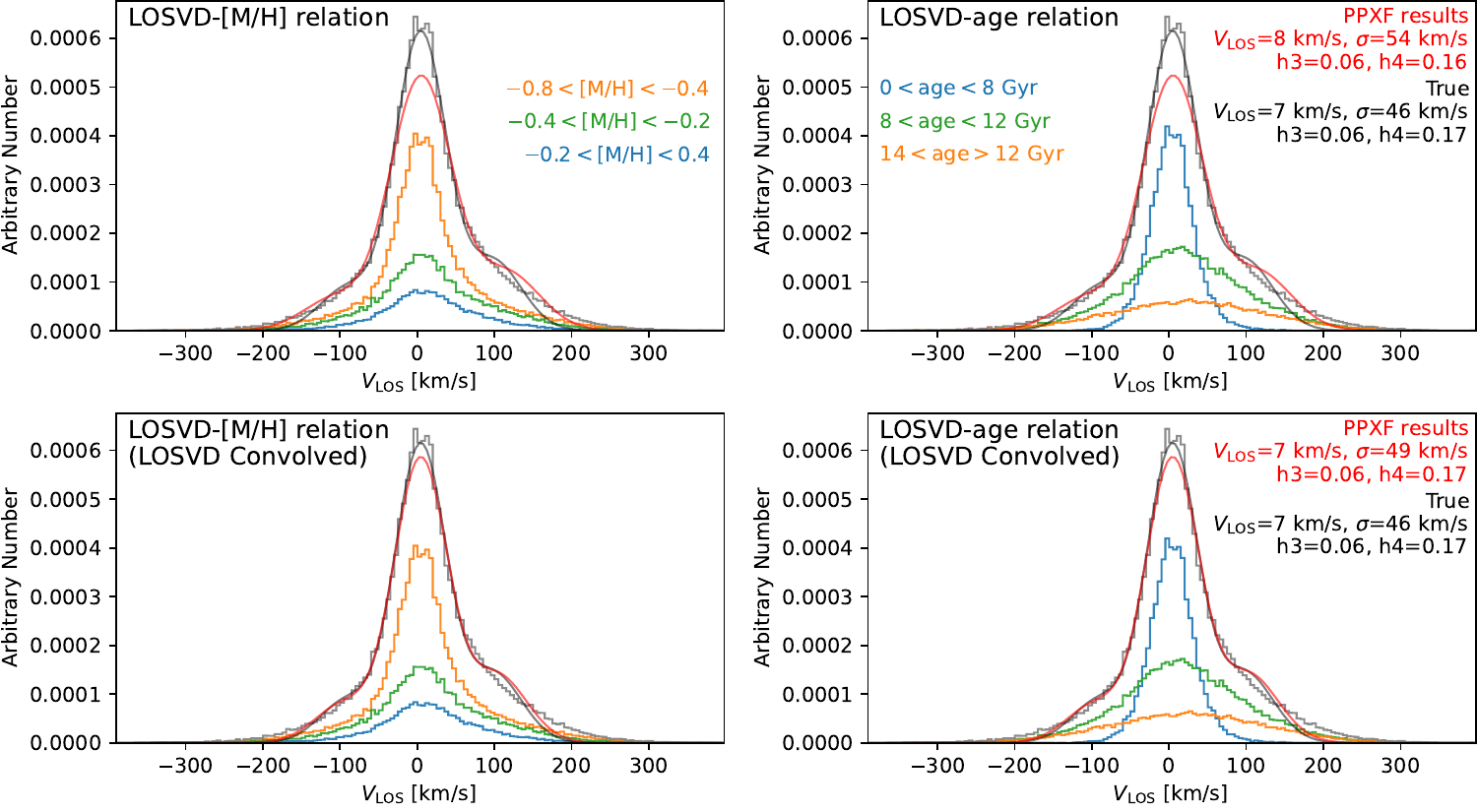}
\caption{
LOSVD of particles with different ages and \mh{} in one Voronoi bin. 
Particles are divided into three \mh{} and age groups, then their \vlos{} histograms are plotted on the left and right panels in different colors, respectively.
The grey histograms are the total \vlos{} distributions of the three groups.
The top rows are results using the original mock cubes.
while the bottom rows use LOSVD-Convolved mock cubes.
The red lines are \ppxf{} results for mock cubes generated and fitted using PEGASE-HR templates in PEGASE-HR spectral resolution ($\rm{FWHM}=0.55\Angstrom$). The black lines are the true values calculated by fitting the Gauss-Hermite function with particles' light-weighted LOSVD in the black histogram.
The kinematic parameters for these two lines are written in the top right corner.
This figure indicates LOSVD changes with age and \mh{}. After using LOSVD-Convolved cubes, this relationship can be artificially eliminated, and \ppxf{} can measure kinematics consistently.
}
\label{fig:losvdmetalage}
\end{figure*}

Fundamentally, during the spectral fitting process, SSP templates are loaded and convolved with a Gauss-Hermite function by FFT to match the observed spectrum \citep{2017MNRAS.466..798C}, and the code returns one series of LOSVD moments which are best fitted to the observed spectrum. 
This means all the templates are assumed to have the same kinematics (\vlos{}, \dispersion{}, $h_3$, $h_4$) in a line-of-sight, i.e., stellar populations with different \mh{} and ages are assumed to have the same LOSVD. 
But in real galaxies, due to the joint process of chemical enrichment and dynamical movements of stars, populations with different \mh{} and ages are different in LOSVDs.
This effect is non-negligible for edge-on projected galaxies because metallicity and age gradients along the disk are the strongest.
Therefore, \vlos{} should change with \mh{} and age, which disagrees with the analysis used in most of the previous studies when employing \ppxf{}, even though \ppxf{} allows such variations. 
Because of the relation between \alphafe{} and \mh{}, stars with different \alphafe{} should also have different \vlos{}.

To investigate it in our mock stellar catalog, we select one Voronoi bin and split the included particles into three groups using \mh{} and age, respectively. 
We reserve this investigation on \alphafe{} for the future due to the limited number of \alphafe{} templates.
Then, we plot the LOSVDs of these groups of particles, weighted by each particle's total flux, and show them in the first row of Fig.~\ref{fig:losvdmetalage}. 
On the top left panel, particles with $-0.8<\mh{}<-0.4$~dex have the sharpest distribution, while particles with $-0.2<\mh{}<0.4$~dex show a nearly normal distribution. 
In the top-right panel, particles with $12<\rm{age}<14$~Gyr have the broadest distribution, while particles with $0<\rm{age}<8$~Gyr have the narrowest and most peaked distribution. 
The red line and black line represent \ppxf{} fitted and the true LOSVD for this Voronoi bin, with kinematic parameters written in the top right corner, respectively. 
Compared to the true LOSVD, the width of \ppxf{} fitted curve in red is close to the youngest populations ($0<\rm{age}<8$~Gyr), which is reasonable because these populations dominate the light. 
However, the top right panel shows that \ppxf{} tends to also care for the oldest populations by having a more skewed tail of their LOSVD at \vlos{} around $150-300$~\kms{}.
Even though the youngest populations dominate the spectral light, the old populations have the most features along the whole fitted wavelength region.
Therefore, the differences of the skewness in the region of \vlos{} at $150-300$~\kms{} indicate the effect caused by different spectral features having different LOSVDs. 
This figure strongly suggests there are limitations on current techniques to obtain unbiased kinematics due to the dependence of \vlos{} on age and \mh{}.

We made a new experiment to verify this effect on spectral fitting results. 
Firstly, we re-generate mock MUSE cubes following the same procedures as above, but not using each particle's \vlos{} to shift the spectrum.
Then we obtain non-kinematic mock data cubes.
Next, for each Voronoi bin, we select all the particles included and obtain its \vlos{} histogram, which is the true LOSVD as shown in grey in Fig.~\ref{fig:losvdmetalage}.
Then, we directly use this histogram as the LOSVD kernel function and convolve it with all the spectra in this Voronoi bin to add kinematic effect.
We do it for all the Voronoi bins and obtain new mock cubes, where spectral templates with different \mh{} and age have the same LOSVDs.
Therefore, in this way, we eliminated the relation of \vlos{} with \mh{} and age as mentioned above.
Finally, we employ \ppxf{} to the new cubes (hereafter LOSVD-Convolved cubes) and measure the kinematics. 
The fifth row of Fig.~\ref{fig:kinvsdispersionallcase} shows the kinematics recovery results. 
It is clear that all the Gauss-Hermite moments are well recovered with all the data points aligned to the zero line. 
The red and black lines in the bottom row of Fig.~\ref{fig:losvdmetalage} also indicate the correction of \ppxf{} measurements.
We also plot the kinematics maps in the same way as Fig.~\ref{fig:galaxymapkin} in Appendix (Fig.~\ref{appfig:galaxymapkin_pegasehrshufflevrRE}) to better demonstrate the consistency.

We also tested the kinematics recovery on LOSVD-Convolved cubes generated using MILES templates, as shown in the sixth row of Fig.~\ref{fig:kinvsdispersionallcase} (see also Fig.~\ref{appfig:galaxymapkin_shufflevrRE}). 
In this low spectral resolution, even though we make different populations have the same LOSVD for the spectra, there are still systematic offsets appearing in all the panels.
We also demonstrate in the last row the results on LOSVD-Convolved cubes generated using PEGASE-HR templates, but in MUSE spectral resolution to remove the effects of different SSP templates (see kinematics maps in Fig.~\ref{appfig:galaxymapkin_pegasehrloresshufflevrRE}).
The appearance of offsets indicates the effect purely due to insufficient spectral resolution.

In conclusion, all these results strongly suggest that both the low spectral resolution and variance of LOSVDs in different age and \mh{} contribute to the systematic offsets we see in residual panels of Fig.~\ref{fig:galaxymapkin}. 
In the future, to remove the systematic offsets of stellar kinematics, one needs to use an instrument with higher spectral resolution than MUSE. 
This means more exposure time is needed to achieve the same SNR as MUSE observations. 
As we know \vlos{} changes with \mh{} and age in the line-of-sight of real galaxy observations, it might be necessary to allow different templates to have different LOSVDs during the spectrum fitting analysis to obtain more accurate results. 
However, kinematic and stellar population measurements from spectral fitting methods have already been a highly degenerate problem. 
Allowing variable LOSVDs will increase the degeneracy by a factor of the number of templates.
In this case, it is easy for the algorithm to obtain incorrect results.
One possible way would be simply assuming a quantitative relation between LOSVD and metallicity and age at different locations of the galaxy and that this relation can be expressed by an analytical equation. 
This is equivalent to adding a prior to the fitting process to improve the accuracy of kinematics measurement without adding degeneracy.
Even though this way has many limitations, it will be still useful for the analysis of Milky-Way-like galaxies, and \galcraft{} can help provide this prior for galaxies in different projections.

Apart from the above analysis on inconsistencies caused by spectral fitting methods, Fig.~\ref{fig:losvdmetalage} also shows the deviations between \vlos{} histogram (grey line) and the parametrically best-fitted true LOSVD (black line).
In the future, we will explore using non-parametric techniques (e.g., BAYES-LOSVD \citealt{2021A&A...646A..31F}) to recover the true kinematic parameters rather than Gauss-Hermite equation, which might result in better fits when the LOSVD is strongly non-Gaussian.

\subsection{Inconsistency of mass fraction distributions recovery}
\label{sec:discussionmfd}

\begin{figure*}
\includegraphics[width=2\columnwidth]{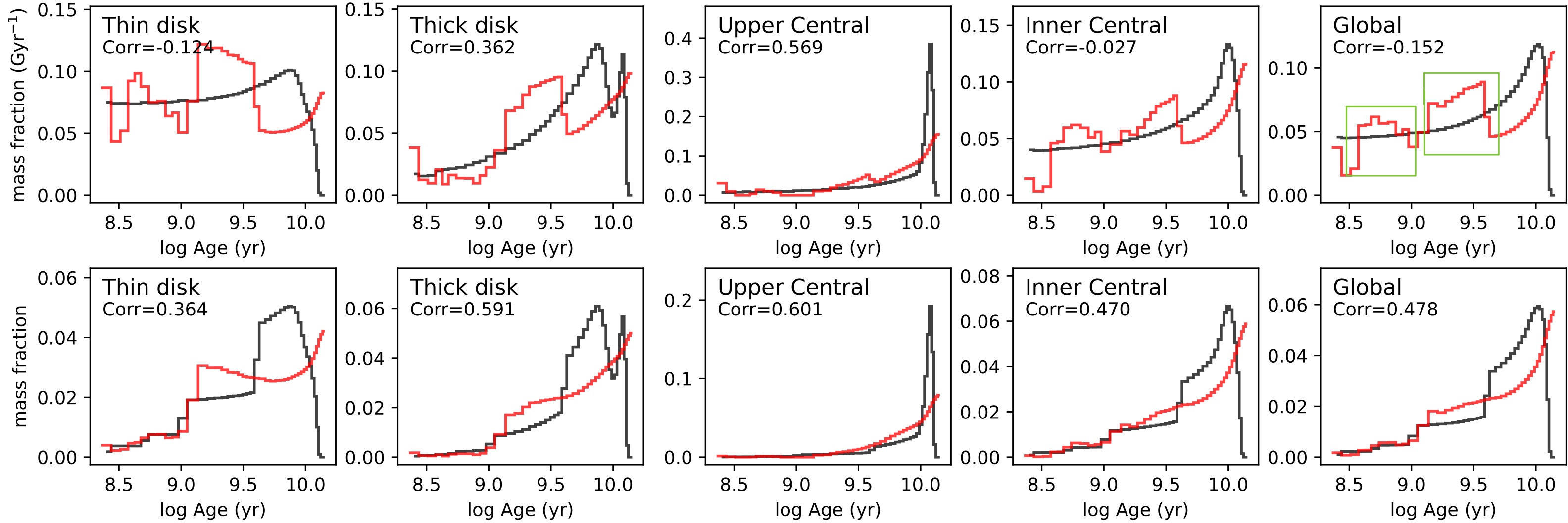}
\caption{
Age distribution recovery of different components of the galaxy from \ppxf{} on the mock MUSE cube generated in Section~\ref{sec:galaxymapmockcube}. This figure is similar to Fig.~\ref{fig:massfractioncomponents1d} but the x-axis is in logarithmic scale. 
Black lines are the true values and red lines are the results from \ppxf{}.
The first row shows mass fraction (\pergyr{}), which is the representation of star formation history, and the two green boxes in the ``Global'' panel are the two overdensity regions. The second row shows just the mass fraction, which is the direct output from \ppxf{}. 
The correlation coefficients of the red and black lines are written on the top right corner for each panel. 
}
\label{fig:massfractioncomponents1dlogage}
\end{figure*}

\begin{figure*}
\centering
\includegraphics[width=1.8\columnwidth]{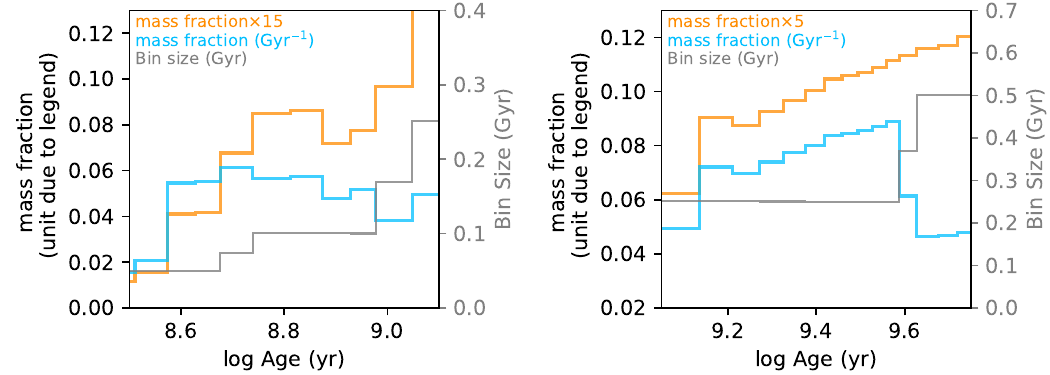}
\caption{
Zoom-in of the two overdensity regions pointed out in Fig.~\ref{fig:massfractioncomponents1dlogage}. The blue and orange lines are mass fraction (\pergyr{}) and mass fraction distributions, equivalent to the first and second row of Fig.~\ref{fig:massfractioncomponents1dlogage}, respectively. The mass fraction for each panel is multiplied by a factor for better visualization. We also plot MILES templates' age bin size as the grey line and the scale in the right y-axis. 
}
\label{fig:zoomin_massfractioncomponents1dlogage}
\end{figure*}

\begin{figure}
\centering
\includegraphics[width=0.99\columnwidth]{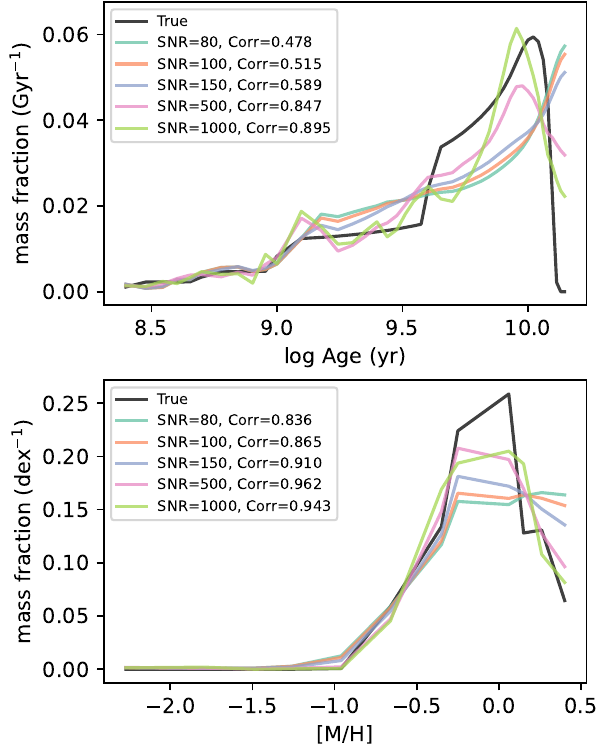}
\caption{
Global age (top panel) and \mh{} (bottom panel) distributions at different SNRs, which are obtained by running \ppxf{} on the mock MUSE cube generated in Section~\ref{sec:galaxymapmockcube} with different target SNRs of Voronoi bins, and then the mass fraction of all the bins are added together to represent the global mass distributions of the galaxy. Black lines are the true values.
``Corr'' is the correlation coefficient value of the \ppxf{} and true values at each SNR.
This figure indicates that increasing SNR can help obtain better mass distributions compared to the true values, especially the issue of metal-rich, old stellar population overestimation is alleviated.
}
\label{fig:sfhmetalsnr}
\end{figure}

\begin{figure*}
\includegraphics[width=2.0\columnwidth]{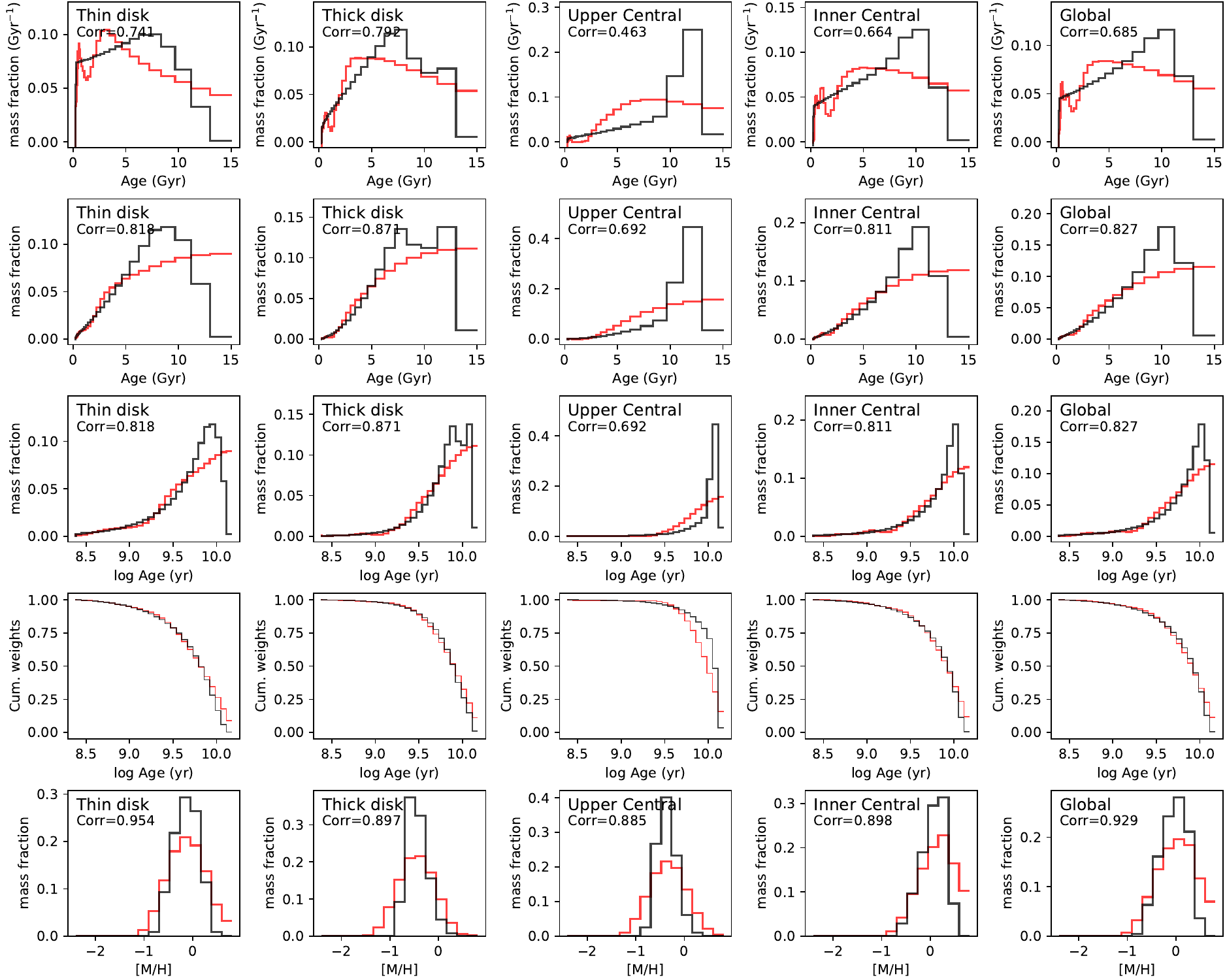}
\caption{
Age and \mh{} distributions of each component for the mock cubes generated using PEGASE-HR templates and degraded to MUSE resolution ($\rm{FWHM}=2.65\Angstrom$). 
The first row demonstrates mass fraction (\pergyr{}) changes with age (SFH).
The second row shows the mass fraction changes with age (MDF). 
The third row shows mass fraction changes with \logage{}.
The fourth row shows cumulative mass fraction distributions and the fifth row shows metallicity distributions. 
}
\label{fig:massfractioncomponents1d_pegasehr_lores}
\end{figure*}

In general spectral fitting, measuring mass/light fraction distributions of stellar populations is a highly degenerate problem, especially when one wants to obtain the star formation history at different components.
In this study, we follow the data analysis procedures in M21 to divide the template's age and metallicity bin size due to the definition of star formation history.
However, most of the previous studies did not consider bin size of template grids. Nevertheless, the mass fractions increasing from young to old and metal-poor to metal-rich populations found in Fig.~\ref{fig:massfractioncomponents1d} is still seen in some components of the galaxies (e.g., \citealt{2019A&A...623A..19P, 2019A&A...625A..95P}). 
In the mock and real data tests of \cite{2018MNRAS.480.1973K}, the mass fraction distributions are also smoothed towards old and metal-rich regions (see their Fig. 5).
The over-dense feature of young populations at around $2-4$ and $12-14$~Gyr in our Fig.~\ref{fig:massfractioncomponents1d} is also seen in some of the above-mentioned studies.
In this section, we re-investigate all these tests more comprehensively and discuss the differences in the mass fraction distribution from \ppxf{}, including whether or not to divide the bin size, effects caused by SNR and different SSP models, and try to address potential reasons causing the inconsistency to the true values.
From now on, we refer ``mass fraction'' as \ppxf{} obtained mass weights of SSP templates normalized by the total weights (for each panel in the figures). And we refer ``mass fraction (\pergyr{})'' and ``mass fraction (\perdex{})'' as mass fraction divided by the bin size of age and metallicity grids, respectively, where mass fraction (\pergyr{}) is also equivalent to star formation history (SFH).

We first convert the age distribution x-axis of Fig.~\ref{fig:massfractioncomponents1d} from linear to log-scale, because MILES templates are closer to being evenly scaled in \logage{} than linear age, then the width of each step is close to being equal. 
Next, we re-plot the age distribution recovery in Fig.~\ref{fig:massfractioncomponents1dlogage} and calculate the correlation coefficients using the original age grids. 
The first row shows age distribution in mass fraction (\pergyr{}), which is the SFH of each component. The thin disk, inner central, and global panels clearly show two overdensity regions, which are pointed out by two green boxes. The thick disk and upper central panels also have peaked features in these boxes, but not significant. 
The second row shows just the mass fraction without dividing the bin size, and all the panels show relatively smoother distributions than the first row.
Therefore, given the differences between the first and second rows are whether or not dividing bin size, we speculate the two overdensity regions are artifacts due to the bin size arrangements of the MILES template grids.

To verify our speculation, we plot in Fig.~\ref{fig:zoomin_massfractioncomponents1dlogage} a zoom-in of the two overdensity regions pointed out on the top right panel of Fig.~\ref{fig:massfractioncomponents1dlogage}. 
The blue and orange lines are mass fraction (\pergyr{}) and mass fraction distributions, respectively, which are equivalent to the first and second row of Fig.~\ref{fig:massfractioncomponents1dlogage}, respectively. The mass fraction for each panel is multiplied by a factor for better visualization. We also plot the age bin size of MILES templates as the grey line and the scale in the right y-axis.
From these two panels, we find the peaked features of mass fraction (\pergyr{}) (blue line) appear when the age bin size experiences a decrease from old to young ages, and the mass fraction (orange line) is still relatively smooth.
Therefore, we can confirm the peaked features of mass fraction (\pergyr{}) in our results are due to the non-regularized spacing of MILES age grid.
This also explains the better recovery of \mh{} distributions (bottom row of Fig.~\ref{fig:massfractioncomponents1d}) which is due to the almost linearly spaced \mh{} grids.

Previous studies applied regularization \citep{2015MNRAS.448.3484M} in \ppxf{} to obtain smooth weight fraction distributions. 
The regularization works as an extra term in the equation to be minimized during the optimization to damp high-frequency variations in the weight distributions along spectral templates grid (see details in Section 3.5 of \citealt{2017MNRAS.466..798C}) and leads to smooth weights distributions.
However, this smoothness has an effect on weights without dividing the bin size. 
This explains why mass fraction distributions are smooth in our results, and whether mass fraction (\pergyr{}) distributions are smooth depends on how the template grids are spaced. 
For the rest of this section and  Section~\ref{sec:discussionregul}, we only use the mass fraction without dividing the bin size to justify the recovery consistency (Fig.~\ref{fig:sfhmetalsnr} and Fig.~\ref{fig:1dmfreg}), because this is expected to be achieved by spectral fitting methods with regularization.
And we leave investigations on mass fraction (\pergyr{}) to Section~\ref{sec:discussinterpmfd} and~\ref{sec:discussionaddmfd}.

Although the current \ppxf{} regularization algorithm does not consider the bin size of the template age grid and different regularization strategies (e.g., \texttt{regul} values, order of regularization) can affect the mass fraction recovery (will be discussed in Section~\ref{sec:discussionregul}), we still see the inconsistency between spectral fitting results and true mass fraction distributions of the oldest populations in the bottom row of Fig.~\ref{fig:massfractioncomponents1dlogage}. 
The main reason is that spectral templates in the oldest and metal-rich regions are very similar and indistinguishable at the current SNR level. 
Therefore, it is difficult for \ppxf{} to correctly recover the mass distributions in these age-\mh{} regions when performing fitting using templates normalized to 1~\solarmass{}, even though the algorithm has already reached the global minimum. The mass fraction uncertainties in these regions can be much larger (relative uncertainties from 1$\sigma$ values dividing the mean of MC realizations are $\sim20$\% for regions less than $8$~Gyr and $50-65$\% in $10-14$~Gyr).
We also confirm in Appendix Fig.~\ref{appfig:massfractioncomponents1d_shufflevrRE} that using LOSVD-Convolved cubes does not improve the results significantly, so the effect of systematic offsets in kinematics measurement in Fig.~\ref{fig:galaxymapkin} is not the cause of mass fraction distribution inconsistency.

One way to increase the quality of the mass fraction recovery in old and metal-rich population regions could be to increase the SNR of the spectrum. 
This can be achieved by requiring a higher target SNR during the process of Voronoi binning but with the cost of reducing the number of Voronoi bins, i.e., details in the spatial distributions. 
We show in Fig.~\ref{fig:sfhmetalsnr} the age and \mh{} distributions at different SNRs (shown in different colors), compared to the true values shown in black lines.
This figure is obtained by running the GIST pipeline on the mock MUSE cube generated in Section~\ref{sec:galaxymapmockcube} with different target SNRs during Voronoi binning, and then the mass fraction distributions of all the Voronoi bins are added together to represent the global mass distributions of the mock galaxy, and finally integrated through age and \mh{} axes, respectively. To make it a fair comparison, we make sure that the spatial pixels used to generate Voronoi bins are the same. 
According to the correlation coefficients, this figure indicates that with the increase of Voronoi bin SNR, the global mass fraction distributions in both panels are more consistent with the true values. 
Specifically, the weights of metal-rich and old stellar populations are going lower with higher SNR.
Therefore, Fig.~\ref{fig:sfhmetalsnr} indicates that increasing SNR can improve SFH recovery.
However, we note that in reality, an integrated spectrum with SNR larger than 1000 is impractical. In addition, even with SNR at 200 \perpixel{} is pointless because in this case numerical issues and spectral library uncertainty can become the dominant factor.

To study the effect of different choice of SSP templates, we re-plot age and \mh{} distributions of mock cubes generated by PEGASE-HR templates in Fig~\ref{fig:massfractioncomponents1d_pegasehr_lores}.
PEGASE-HR templates are evenly spaced in \logage{}. 
During the \ppxf{} fitting, we use the same templates and fitting strategies (fitting wavelength region, regularization, etc.) as those in Section~\ref{sec:galaxymapdatareduction} and also degrade to MUSE spectral resolution to remove the resolution differences (even though we find increasing spectral resolution does not improve the recovery significantly after comparing to Fig.~\ref{appfig:massfractioncomponents1d_pegasehr} of a PEGASE-HR spectral resolution version). 
The Voronoi bin allocation is the same as mock cubes generated by MILES templates.
According to the correlation coefficients, mass fraction using PEGASE-HR in Fig.~\ref{fig:massfractioncomponents1d_pegasehr_lores} are more consistent than those using MILES in Fig.~\ref{fig:massfractioncomponents1d} and Fig.~\ref{fig:massfractioncomponents1dlogage} for both age and \mh{}.
Moreover, the mass fraction - \logage{} panels (second row) have almost no overdensities than the results using MILES (second row of Fig.~\ref{fig:massfractioncomponents1dlogage}), and the oldest populations have a more modest increase.
Therefore, in our test, PEGASE-HR performs better during the mass fraction recovery and could provide more consistent mass fraction distributions than MILES.
We think this is likely due to the templates' perfectly even spacing in \logage{}, which leads to smoother input and is preferred by \ppxf{} during regularization.
However, the PEGASE-HR templates used in our tests are generated using PADOVA 1994 \citep{1994A&AS..106..275B} isochrones but MILES are generated using BASTI isochrones \citep{2004ApJ...612..168P, 2006ApJ...642..797P}.
Whether isochrones differences can affect mass fraction distribution recovery ability requires further investigations.
Moreover, the current limitation of PEGASE-HR templates is they only have one bin in \alphafe{}, we emphasize here again the need for multiple-\alphafe{} templates.

In the future, non-parametrically, it is worthwhile to test the mass fraction recovery ability using templates with linearly spaced grids (see details in Section~\ref{sec:discussinterpmfd}) or modifying regularization to apply smoothness effect in terms of mass weights (\pergyr{} \perdex{}). 
Another approach is to take the mean of Monte Carlo realization (e.g., \citealt{2023A&A...673A.147P}), which is more consistent with the true values than regularization based on simple tests (see Fig.~1 of \citealt{2023MNRAS.526.3273C}). We can verify whether this is true for a mock MW cube in future studies.
Parametrically, one can try to recover both chemical enrichment history and star formation history by taking into account chemical evolution theories, which will help remove counter-intuitive features in the mass fraction distributions. One approach is to use Bayesian spectral fitting codes, such as Prospector \citep{2017ApJ...837..170L} or BAGPIPES \citep{2018MNRAS.480.4379C}.
Another example is the semi-analytic spectral fitting method from \cite{2022MNRAS.513.5446Z}. 
This method applies spectrum fitting with the predicted best-fit spectrum from chemical evolution models, which is similar to adding a prior on \ppxf{} during the fitting.
Fig.~6 of \cite{2022MNRAS.513.5446Z} showed that this method has better consistency on mass fraction recovery than \ppxf{}, and the over-estimations in metal-rich, $2-4$ and $12-14$~Gyr regions are fixed successfully.
Therefore, it provides a way to measure accurate chemical enrichment and SFH for future studies.
One important caveat is that their chemical evolution model is a close-box model, which does not apply to galaxies dominated by frequent passive merger events (e.g. NGC 7793 in \citealt{2018MNRAS.480.1973K}). In addition, this method does not take into account the relation of chemical abundances with velocity, which are important indicators for radial migration and kinematic heating.

\subsection{Effect of regularization}
\label{sec:discussionregul}

\begin{figure*}
\includegraphics[width=2.0\columnwidth]{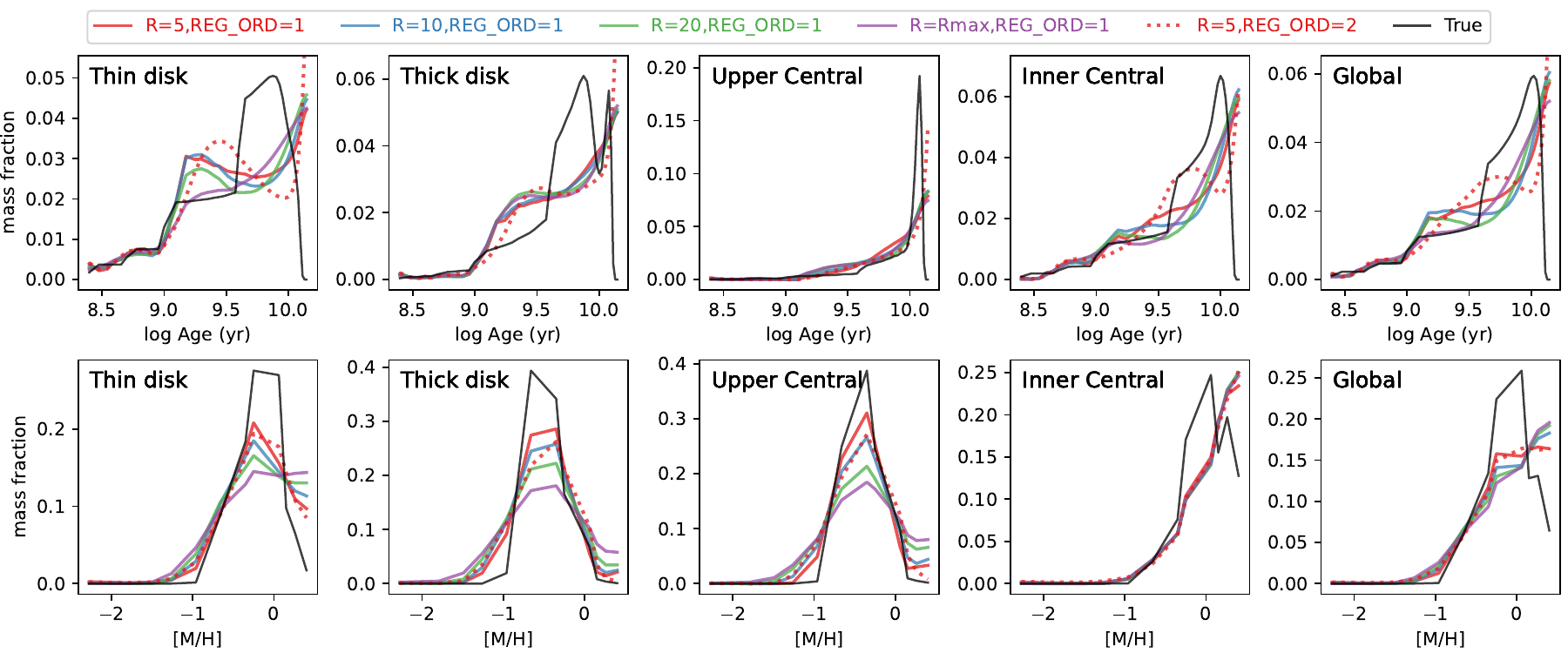}
\caption{
Age (top panel) and \mh{} (bottom panel) distributions for different components by running \ppxf{} on the mock MUSE cube generated in Section~\ref{sec:galaxymapmockcube} with different regularization strategies (shown in different colors). Black lines are the true values. The last column shows global age and \mh{} distributions. 
This figure indicates that results using the first order of regularization are better than results using the second order of regularization; increasing \texttt{regul} can help obtain smoother results but the mass fractions will go towards metal-rich regions for all the components.}
\label{fig:1dmfreg}
\end{figure*}

In this section, we focus on the effects of different regularization strategies on mass fraction distributions. There are two parameters controlling the regularization process, which are \texttt{regul} and \texttt{reg\_ord}, where \texttt{regul} applies linear regularization to the weights during the \ppxf{} fit and \texttt{reg\_ord} controls the order of regularization.
\cite{2018MNRAS.480.1973K} investigated mass fraction differences of mock data with different order of regularization (see their Fig.~5 and Fig.~C2) and concluded that the choice of order of regularization does not affect the results over a significant level.
\cite{2020MNRAS.491..823B} tested the mass fraction distributions recovery of the second- and third-order of regularization and compared them to the true values using stellar particles from an EAGLE simulated galaxy. They found results from third-order regularization are more consistent with the true values.
\cite{2023MNRAS.526.3273C} compared the mass fraction recovery consistency of non-regularized \ppxf{} fit, an average of 100 \ppxf{} fits to MC realizations, and single regularized \ppxf{} fit (see their Fig.~1), and concluded that regularized results are comparable to that of the average of multiple realizations. 

The selection of \texttt{regul} value was briefly discussed in Section~\ref{sec:galaxymapdatareduction}.  Here we re-investigate this question in detail by applying different \texttt{regul} and \texttt{reg\_ord} strategies on our mock cubes generated in Section~\ref{sec:galaxymapmockcube}. 
The results are shown in Fig.~\ref{fig:1dmfreg}, where we plot the age distributions on the top panels and \mh{} distributions on the bottom panels for different components of the galaxy. We also plot the global mass distributions in the last column.
We apply five different strategies during the fitting: the first three cases use first-order regularization with \texttt{regul}~$=5$, \texttt{regul}~$=10$, \texttt{regul}~$=20$, and \texttt{regul}~$=$~\texttt{regul$_{max}$}~$(30-100)$, respectively; the last uses \texttt{regul}~$=5$ and second-order regularization. Black lines are the true values.
When comparing results with the same order of regularization but different \texttt{regul}, we find results with larger \texttt{regul} are smoother, and their weights at old populations are relatively smaller than results with smaller \texttt{regul}. However, in the \mh{} distribution panels, weights are going more toward metal-rich populations and the distributions become less consistent with the true values for all the components.
For results with the same \texttt{regul} but different orders of regularization, we see results with \texttt{reg\_ord}~$=1$ show better consistency with the true values than those with \texttt{reg\_ord}~$=2$. Specifically, results with \texttt{reg\_ord}~$=2$ have more mass fractions in the oldest populations.

In conclusion, Fig.~\ref{fig:1dmfreg} indicates that for this mock cube, results using the first order of regularization are better than results using the second order of regularization; increasing \texttt{regul} can help obtain smoother results but the weights will go towards metal-rich regions.
Given we do not have the codes to perform third-order regularization, it is hard to say which order is the best. Intuitively, it seems higher-order regularization tends to have more mass fraction in the old populations. Therefore, we think the third-order regularization is preferred for EAGLE simulated galaxy test \citep{2020MNRAS.491..823B} because the particles' mass weights from the EAGLE simulations are naturally dominated by old and metal-rich populations, then the third-order regularization tends to fit better in this region. 
Our \eglx{} mocked catalog has no particle in the metal-rich and oldest regions, so the first-order regularization is better.
However, all the above discussions need further verification and we will test the third-order regularization results in the future.
For the real galaxy fitting results of \cite{2019A&A...623A..19P, 2019A&A...625A..95P} and M21, they all used second-order regularization and showed an overdensity of mass weights in the oldest, metal-rich populations for different components of the galaxies, which makes it difficult to discuss differences in the formation epoch between these components, and it is not possible to tell if the overdensity is true or instead from full-spectrum fitting artifacts.

\subsection{Mass fraction distribution recovery using interpolated MILES templates}
\label{sec:discussinterpmfd}

\begin{figure*}
\includegraphics[width=2\columnwidth]{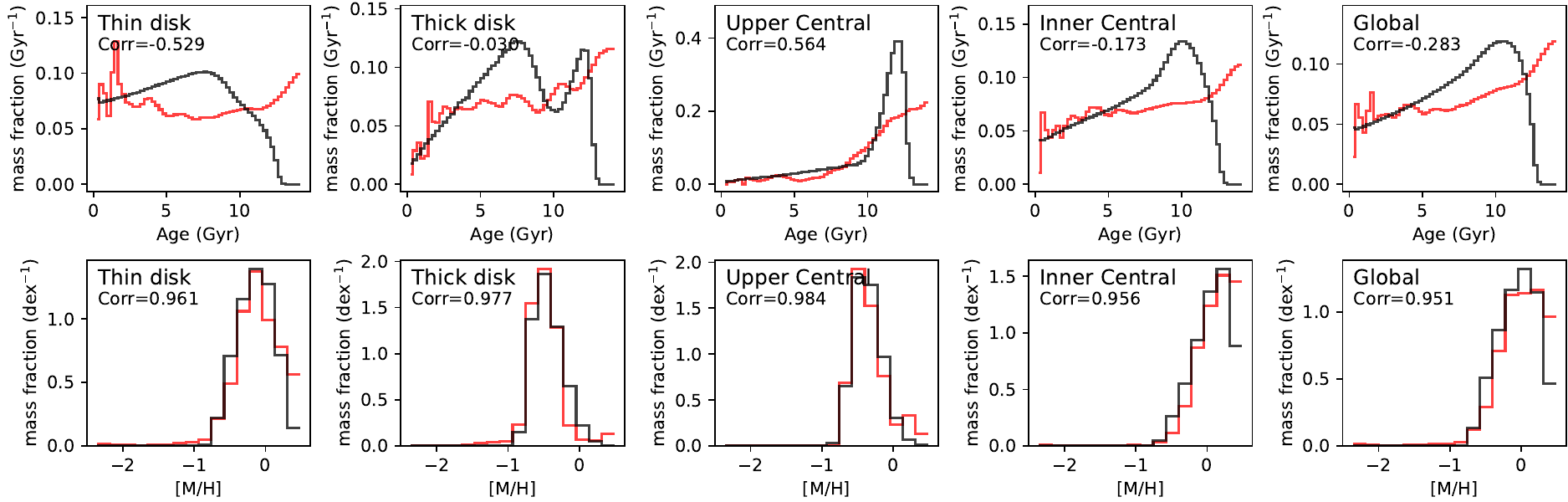}
\caption{
Same as Fig.~\ref{fig:massfractioncomponents1d} but for the mock cube generated using MILES templates interpolated linearly in age and \mh{} in MUSE spectral resolution.}
\label{fig:massfractioncomponents1dinterp}
\end{figure*}

According to discussions in Section~\ref{sec:discussionmfd} and \ref{sec:discussionregul}, the overdensity features of age distributions in mass fraction (\pergyr{}) shown in Fig.~\ref{fig:massfractioncomponents1dlogage} are likely due to not considering templates bin size during regularization, and using different regularization strategies can barely alleviate the inconsistencies with the true values.
Therefore, we investigate the improvements of age and \mh{} distributions using interpolated MILES templates in this section.
Fig.~\ref{fig:massfractioncomponents1dinterp} shows the age and \mh{} distributions of mock cubes generated by \galcraft{} and fitted by \ppxf{} using MILES templates that are interpolated linearly in age and \mh{} grid.
The fitting strategies are the same as those in Section~\ref{sec:galaxymapdatareduction}.
We can clearly see that the overdensities in age distributions (in \pergyr{}) at $2-4$~Gyr are alleviated to a large extent in the ``thick disk'', ``inner central'', and ``global'' panels compared to the results using original MILES templates in Fig.~\ref{fig:massfractioncomponents1d}.
In addition, \mh{} distributions (in \perdex{}) from \ppxf{} are also more consistent to the true values in terms of correlation coefficients.
The increase of old populations still exists, which will be discussed in detail in Section~\ref{sec:discussionaddmfd}.
We also investigate the results using MILES templates interpolated linearly in $\log$~age and \mh{} (shown in 
Fig.~\ref{appfig:massfractioncomponents1dloginterp}) and find the overdensities still exist with the shape being more similar to the results using PEGASE-HR (Fig.~\ref{fig:massfractioncomponents1d_pegasehr_lores}).
Given PEGASE-HR template grid is also linearly spaced in $\log$~age, this similarity is expected.
In conclusion, we confirm that performing spectral fitting using linearly interpolated templates in age and \mh{} can alleviate the overdensities in $2-4$~Gyr. 
In addition, the new version of \ppxf{} considering bin size during regularization is under development (private communication with Michele Cappellari).
Given this is equivalent to interpolating templates as we did in Fig.~\ref{fig:massfractioncomponents1dinterp}, one can choose this approach in the future to avoid additional template uncertainties due to interpolation. 

\subsection{Potential effects of effective age resolution on star formation history recovery}
\label{sec:discussionaddmfd}

In Section~\ref{sec:discussionmfd} to \ref{sec:discussinterpmfd}, we investigated the effects of template grid spacing and regularization on the recovery of mass fraction distribution.
The comparison between Fig.~\ref{fig:massfractioncomponents1d} and Fig.~\ref{fig:massfractioncomponents1dinterp} confirms that the over-estimation of age distribution (SFH) in the $2-4$ Gyr range is significantly alleviated by using templates with a linearly spaced age grid.
However, in Fig.~\ref{fig:massfractioncomponents1dinterp}, which uses the linearly spaced age grid, narrow peaks at around $1$ and $2$ Gyr can still be seen prominently in the ``thin disk'', and moderately in the ``thick disk'', and ``inner central''. 
These peaks differ from the broad $2-4$ Gyr overdensities seen in Fig.~\ref{fig:massfractioncomponents1d}, but are more similar to peaks seen for the case of the  \logage{} grid (Fig.~\ref{fig:massfractioncomponents1d_pegasehr_lores} and Fig.~\ref{appfig:massfractioncomponents1dloginterp}), although they suffer from the additional effect of inconsistent regularization.  
When using the template grid with linear age spacing, there is an over-estimation in $12-14$~Gyr range that is present across all components (Fig.~\ref{fig:massfractioncomponents1dinterp}). 
These results suggest that even after fixing the issue of inconsistent regularization with the linearly spaced age grid and trying out different SSP models, the recovery of SFH is far from perfect and a possible reason for this could be the fact that weights of some SSP templates are degenerate with respect to the observed spectra.


To examine the effect of template weight degeneracies on SFH recovery, \cite{2024MNRAS.528.2790Z} estimated and explored the age resolution of SSP templates that can be achieved through non-parametric full-spectrum fitting, which we refer to as effective age resolution. 
The effective age resolution is defined as the smallest width of independent age bins in a reconstructed SFH, and it is limited by the amount of information that is possible to extract from a spectrum. 
We refer to their Section~2 for a detailed analytical definition.
\cite{2024MNRAS.528.2790Z} found that the effective age resolution is generally no better than 40\% for most age grids older than 100~Myr (see their Fig.~4).
This fraction is roughly consistent with the mass weights uncertainties in our results.
More importantly, \cite{2024MNRAS.528.2790Z} found that the age resolution does not vary smoothly but shows some well defined maxima, e.g. at 10 Myr and 1 Gyr. The origin of this 
could be related to the fact that this is the stage where some unique spectral features develop due to significant changes in stellar evolution.
They speculated (their Sec 5.5) that these maxima in age resolution could be responsible for the 
peaks seen in SFH of galaxies extracted using non-parametric full-spectrum fitting 
on IFS data (e.g., CALIFA \citealt{2013A&A...557A..86C, 2017A&A...607A.128G} and MaNGA \citealt{2019MNRAS.482.1557S}). These peaks in SFH appear across a wide variety of galaxies and span of cosmic time, which lends further support to the idea.
Our work where we try to recover the SFH of simulated galaxies further supports this idea-- even though the SFH of the simulated galaxy is smooth, the recovered SFH has peaks/overdensities (e.g., see the age distributions in mass fraction (\pergyr{}) shown in \ref{fig:massfractioncomponents1dlogage}, \ref{fig:massfractioncomponents1d_pegasehr_lores}, \ref{fig:massfractioncomponents1dinterp},
\ref{appfig:massfractioncomponents1d_pegasehr}, \ref{appfig:massfractioncomponents1dloginterp}).
The over-estimation of SFH at $12-14$~Gyr in our results could be due to a combination  of a number of factors. 
Our input SFH as specified in \eglx{} drops sharply to zero at 12.8~Gyr which makes it challenging to recover the SFH. 
Unfortunately, the spectra of templates at old ages are quite similar to each other (low effective age resolution), this would lead the fitting machinery to assign non-zero weights for old templates. 
The regularisation which favours smoothness will also favour non-zero weights for old populations, even with the input truncating at 12.8~Gyr. Additionally, due to the low luminosity of these old SSPs, to account for a given flux in spectra, a significant mass fraction can be assigned to these old ages as compared to young ages. 
The linear age grid has more templates at old ages as compared to a \logage{} grid.
This suggests that bias in SFH (which is mass fraction per unit age) due to the above effects is expected to be higher for a grid with linear age as compared to log age.
This is evident in our results and can be seen by comparing the SFH at old ages in Fig.~\ref{fig:massfractioncomponents1dinterp} and Fig.~\ref{appfig:massfractioncomponents1dloginterp}.

\subsection{Caveats when using \galcraft{}}
\label{sec:caveats}

In this section, we give some caveats when using the \galcraft{} code.
The \eglx{} mock stellar catalog is generated based on the GCE model of S21. 
The success of this model is that it can reproduce \alphafe{}-\feh{} bimodality which is quantitatively consistent with the APOGEE observations of the MW \citep{2015ApJ...808..132H} by employing radial migration and kinematic heating.
However, more complex kinematic and dynamical processes are likely to exist in real galaxies, such as non-asymmetric perturbations due to spiral arms, bars, interlopers, etc.
The current model does not have these processes and features, which should be added in the future.
In addition, the interpretation of \alphafe{} bimodality in S21 due to radial migration and kinematic heating is opposite to findings from simulations which suggest the need of mergers (e.g., \citealt{2019MNRAS.484.3476C, 2020MNRAS.491.5435B}), and this model lacks stars in the Galactic halo.
Therefore, this code is not applicable for comparing the merger features or the halo distributions;
This model also lacks descriptions of parameter distributions in the bulge, especially the nuclear disk in the MW center (e.g., \citealt{2021A&A...650A.191S}) and chemodynamics of the B/P bulge (e.g., \citealt{2021A&A...653A.143W}), which could be improved in the future.

\section{Future Studies}
\label{sec:future}

In this study, we did not apply extinction on the mock data cube because this work is purely on testing the kinematics and stellar population recovery via full-spectrum fitting.
In real edge-on galaxies, extinction can have an essential effect because it can obscure nearly 50\% of the total light from stars in the thin disk, making SNR in this region lower. 
With the \galcraft{} code, we can test the kinematics and stellar population recovery of mock cubes with extinction.
The extinction model in \eglx{} is assumed as a double exponential distribution mainly in the thin disk, which can be integrated through the line-of-sight, and each particle will have a \ebv{} value.
Then, we can add the extinction effect to the spectrum by using a specific reddening curve (e.g., \citealt{1989ApJ...345..245C, 2000ApJ...533..682C}), which is also the strategy that \ppxf{} applies to estimate extinction.
\cite{2018MNRAS.478.2633G} has already tested the effect of extinction on full-spectrum fitting results using one stellar population template, we can re-investigate this question with comprehensive mock IFS Milky-Way data from the Galactic chemical evolution model S21 and also study how different extinction models can affect the projected galaxy properties.
In addition, several non-asymmetric features should be added to the chemical evolution model like the spiral arms, bars, mergers, and halo particles to mimic a more realistic MW.
It is also necessary to add gas emissions spectral features.

There are several improvements can be made to the choice of SSP models.
The current version of \galcraft{} has embedded MILES and PEGASE-HR templates. In the future, we will add other empirical models such as BC03 \citep{2003MNRAS.344.1000B}, X-Shooter \citep{2022A&A...661A..50V}, and theoretical models like STARBURST99 \citep{1999ApJS..123....3L} and BPASS \citep{2017PASA...34...58E}, which can satisfy different science goals since different models have different advantages.
Some interesting tests can also be executed, e.g., using one SSP model to generate mock cubes and fit them with another model, which can help study the effect of SSP template uncertainties on full-spectrum fitting results.

In addition, even though we showed in Section~\ref{sec:galaxymapchemppxf} that the \alphafe-rich and \alphafe-poor populations can be well identified using two \alphafe{} bins, it is still essential to add more bins to obtain detailed \alphafe{} distributions.
The MW results in \cite{2015ApJ...808..132H} have shown that stars in the thin and thick disk have \alphafe{} values about $0.0$ and $0.2$~dex, respectively, and there are some stars with negative \alphafe{}, which is also shown in the \eglx{} catalog. 
However, it is challenging to use current MILES $\alpha$-variable templates to identify these two \alphafe{} sequences because most of the weights are in the bin of \alphafe{}~$=0$, and it is also difficult to assign a spectrum to particles with negative \alphafe{}. 
Therefore, templates with more \alphafe{} bins such as sMILES \citep{2023MNRAS.523.3450K} and FSPS \citep{2009ApJ...699..486C, 2010ApJ...712..833C} will help on the study of \alphafe{}-\mh{} distributions for external galaxies at different components.

According to Section~\ref{sec:discussionkin}, given that spectral resolution is important in stellar kinematics measurements, it is also necessary to increase both the instruments' and stellar population models' spectral resolution. Current IFS observations such as MUSE perform well on kinematics measurements of galaxies in the central region. However, to go deeper and obtain distributions of the outer regions, especially the outer thin disk, which normally has lower dispersion, one might need to use instruments and SSP templates with a higher spectral resolution to obtain more accurate kinematics maps.

As discussed in Section~\ref{sec:discussionkin}, other than increasing spectral resolution, a prior can be added to the spectral fitting process which assumes the relation between LOSVD and metallicity and age at different locations of the galaxy. Then it can allow templates to have different LOSVDs and the degeneracy is not increased. 
The other way is to add the evolution of stellar kinematics to the semi-analytic model like the one in \cite{2022MNRAS.513.5446Z} and make it derive the chemical evolution histories along with kinematic processes directly by fitting with the integrated spectra, where the fitting process is constrained by several model parameters to be more physical.
We will test the feasibility of these methods in the future.

As for weight fraction distributions, Section~\ref{sec:discussionmfd} demonstrated how the bin size can artificially create peak features in the star formation history, and in Section~\ref{sec:discussinterpmfd} we verified that fitting the spectra using templates that are interpolated with grids linearly spaced in age and \mh{} can alleviate this effect.
Given interpolating semi-empirical spectral libraries is not recommended when fitting with real observational spectra, our tests indicate the need for considering bin size during regularization.
In the future, we will also test modifying the regularization algorithm by considering the bin size in age and metallicity. Parametrical methods like \cite{2022MNRAS.513.5446Z} can also be used to test recovery abilities.
Moreover, in this work, we input a smooth SFH for the tests. Whether the mass fraction distribution inconsistencies we found also apply to merger-dominated galaxies with multiple starburst phases can be investigated using \galcraft{} generated mock cubes of N-body/hydrodynamical simulations.

The aim of \galcraft{} code is to apply the knowledge we learned from the MW to other MW-like galaxies by mocking the IFS observations as a bridge. Especially, we want to interpret physical processes such as radial migration, kinematic heating, and \alphafe{}-\mh{} distributions of other galaxies to answer the question of whether all the MWAs have similar processes to the MW, and if not, how much difference can be.
However, we have not yet performed any comparison of the mock MW IFS data cube with existing MWA observations due to the lack of extinction and some Galactic components in the current version of \eglx{}, and the artifacts created by full-spectrum fitting methods we found in this work.
Therefore, when all these aspects get improved in the future, we will compare our mock MW data cube from \galcraft{} with real IFS observations from the GECKOS survey \citep{2024IAUS..377...27V}, which will observe 35 edge-on MW-like galaxies, and see whether the MW is similar to the so-called MWAs according to their kinematics and stellar population properties.

\section{Summary and Conclusions}
\label{sec:summary}

In this work, we present the \galcraft{} code which uses simple-stellar population models and the mock MW catalog from \eglx{} to generate a mock MW IFS data cube in the same format as extragalactic observations. 
We aim to eliminate the differences in analysis techniques between the Galactic and extragalactic studies such as the use of individual stars vs. integrated stellar populations, the number density distributions of stellar parameters vs. integrated mass- or light-weighted distributions, the results with or without projection effect, etc.
The mock data cube can be put into the GIST pipeline to perform data analysis in the same way as extragalactic observations and the results can be compared directly to external galaxies to study the similarities and differences between the MW and its analogs.
Therefore, this code is a bridge to link the Galactic and extragalactic studies to understand the formation and evolution of disk galaxies.

The \galcraft{} code is flexible, allowing users to choose their preferred SSP templates, galaxy distance and inclination, spatial/spectral resolution, and field-of-view of the instrument. 
\galcraft{} is also designed to mock current existing instrumental observations such as MUSE, SAMI, or MaNGA, as well as test the performance of future instrument designs.
Moreover, it can also be applied to N-body/hydrodynamical simulations that contain MW-like galaxies and generate mock observations for comparisons.

For the rest of the paper, we applied \ppxf{} (included in the GIST pipeline) as a representation of full-spectrum fitting methods on \galcraft{} generated mock cubes to test the ability of \ppxf{} to recover kinematics and stellar population properties.
The mock MUSE data cubes have a distance of $26.5$~Mpc and inclination $86^{\circ}$, which is the same as MUSE observations on NGC~5746 from M21.

After comparing the true values calculated from \eglx{} particles' properties with \ppxf{} results, we found that there are systematic offsets between kinematic moments (\vlos{}, \dispersion{}, $h_3$, $h_4$) and the true values.
We confirm two reasons are causing it:
\begin{itemize}
    \item The velocity dispersion \dispersion{} of most Voronoi bins are smaller than the instrument velocity dispersion \instdispersion{} in MUSE spectral resolution ($\rm{FWHM}=2.65\Angstrom$).
    \item \vlos{} changes with age and \mh{} for particles through the line-of-sight, but most previous studies assumed all stellar population templates have the same LOSVD during the spectral fitting.
\end{itemize}
By using the higher spectral resolution templates PEGASE-HR ($\rm{FWHM}=0.55\Angstrom$) and applying the LOSVD-Convolved cubes which eliminate degeneracies of \vlos{} with age and \mh{}, we can obtain consistent kinematic results with the true values. 
Therefore, we indicate the need to allow different templates to have different LOSVDs or assume a quantitative relation between LOSVD and metallicity and age at different locations of the galaxy, where the equation coefficients can be measured during spectral fitting. 
The latter method is equivalent to adding a prior to the fitting process to measure kinematics without adding degeneracy. 
We will perform tests in the future to verify if kinematics recovery can be improved.
In addition, our tests also indicate the need to use non-parametric methods such as BAYES-LOSVD \citep{2021A&A...646A..31F} rather than Gauss-Hermite equation for more accurate measurements of \vlos{} distributions.

In terms of stellar population properties, we verified that spectral fitting methods can recover light- and mass-weighted age, \mh{}, and \alphafe{} with good consistency, except the mass-weighed age and \mh{} are overestimated in the outer and inner regions of the galaxy, respectively. 
Both the \alphafe{}-rich and \alphafe{}-poor disk structures can be identified with reasonable regularization values during the fitting.
Line-strength indices method can also identify these structures.
We found the weight fraction distributions of stellar populations from spectral fitting using MILES templates on MILES-generated mock cubes are mostly consistent with the true values. 
However, some deviations are shown in mass fraction distributions, where mass fractions normalized by the bin size are overestimated in regions of $2-4$~Gyr, $12-14$~Gyr, the most metal-rich regions, and underestimated in regions of $5-10$~Gyr. 
This is be due to many reasons including having low SNR that the flux error is larger than templates' similarities in the region of old and metal-rich populations, the uneven spacing of age grids in MILES templates, not considering bin size in the current regularization algorithm.
We verified that having a regularly spaced age grid (linear or log) resolved the sharp discontinuities originally observed in the SFH with the default MILES library.
The \logage{} grid is still incompatible with the way the regularization is implemented in \ppxf{}, so one should either alter the regularization scheme or use linear age spacing.
Even when using the linear age grid, we see narrow peaks at around $1-2$~Gyr in the SFH.
This is most likely due to the fact that the precision of estimating the age of an SSP varies significantly with age as pointed out by \cite{2024MNRAS.528.2790Z}.
In addition, we found that the first-order regularization can obtain better mass fraction distributions than the second order regularization. 
Our tests and conclusions are helpful in identifying limitations of extragalactic data analysis using current methods and provide a reference for potential improvements in the future.

Even though \galcraft{} provides a bridge to link the Galactic and extragalactic studies by transferring MW to mock IFS observations, there remains a need for future improvements to facilitate more accurate measurements of external galaxy properties such as \alphafe{}-bimodality and to enable detailed comparisons of the MW with MW-like galaxies.
These improvements include more accurate SSP models with higher spectral resolution and more \alphafe{} grids, more advanced spectral fitting algorithms, and instruments with deeper observations.
Future tests using \galcraft{} are required including modifying spectral fitting codes to improve the recovery ability of kinematics and stellar population properties, using templates with more \alphafe{} grids, and adding extinction, to achieve measurements of known parameters such as radial migration and kinematic heating efficiencies of the MW in external galaxies.

\section*{Acknowledgements}

We thank James Binney, Alina Boecker, Martin Bureau, Andy Casey, Michele Cappellari, Scott Croom, Eric Emsellem, Jesus Falcon-Barroso, Jianhui Lian, Richard McDermid, Martin Roth, Anil Seth, Juntai Shen, Yuan-Sen Ting, Glenn Van de Ven, Gail Zasowski, Ling Zhu and Zhuyun Zhuang for useful discussions.
We thank all the attendees who joined the discussions of this work during the conference Linking the Galactic and Extragalactic (Wollongong, Australia) and GECKOS-Oxford conference (Oxford, UK). We also thank the referee for providing useful comments.

ZW acknowledges the HPC service at The University of Sydney for providing HPC resources that have contributed to the research results reported in this paper.
Works in this paper were also done using \texttt{Yoga}\footnote{\url{https://yoga-server.github.io/}}, a privately built Linux server for astronomical computing.

ZW is supported by 
Australian Research Council Centre of Excellence for All Sky Astrophysics in Three Dimensions (ASTRO-3D) through project number CE170100013. 
MRH acknowledges support from ARC DP grant DP160103747 and ASTRO-3D.
SS is funded by ASTRO-3D Research Fellowship and JBH’s Laureate Fellowship from the Australian Research Council.
JBH is supported by an ARC Australian Laureate Fellowship (FL140100278) and ASTRO-3D.
JvdS acknowledges the support of an Australian Research Council Discovery Early Career Research Award (project number DE200100461) funded by the Australian Government. 
FP acknowledges support from the Spanish Ministry of Science and Innovation (MICINN) through the project PID2021-128131NB-I00 / 10.13039/501100011033. 

This research has also made use of Astropy\footnote{\url{http://www.astropy.org}}, a community-developed core Python package for Astronomy \citep{2013A&A...558A..33A, 2018AJ....156..123A}, numpy \citep{2020Natur.585..357H}, scipy \citep{2020NatMe..17..261V}, matplotlib \citep{2007CSE.....9...90H} and SpectRes \citep{2017arXiv170505165C}.

\section*{Data Availability}

\galcraft{} source code is publicly available via \url{https://github.com/purmortal/galcraft}. 
The test data and figures can be shared with reasonable requests.



\bibliographystyle{mnras}
\bibliography{mnras_paper} 



\appendix

\section{Modifications of the GIST pipeline}
\label{appendix:gistrevision}

We provide a list of modifications on the GIST pipeline \citep{2019A&A...628A.117B} to improve the flexibility of this software:
We add more templates such as the original and interpolated PEGASE-HR \citep{2004A&A...425..881L} to the software and allow the option to oversample the spectra when degrading to lower spectral resolution, which is for the same reason as we do for the \galcraft{} code in Section~\ref{sec:mwcubesteps};
We also added the option to select SSP templates with a certain age and \mh{} range in some special cases.
When measuring stellar kinematics and stellar population properties, we add the options to choose \texttt{velscale\_ratio}, \texttt{reg\_ord}, and use real spectra noise during the fitting and normalizing the integrated spectrum and noise by the median of the spectrum, which will be important to perform the iteration for estimating \texttt{regul$_{max}$};
For regularization, we add an option to estimate \texttt{regul$_{max}$} following the procedures in \cite{2015MNRAS.448.3484M} and save the stellar population results with \texttt{regul}~$=$~\texttt{regul$_{max}$}. 
We also allow further degrading to lower spectral resolution during the fitting when measuring SFH, which is mainly for the reduction procedures in M21 where they wanted to compare the mass-weighted stellar population parameters from \ppxf{} with the SPP-equivalent parameters from line-strength indices in the same spectral resolution;
We add the option to change the penalization value \texttt{bias} during \ppxf{} fitting and measure templates weights uncertainties using Monte Carlo realizations similar to procedures in \cite{2018MNRAS.480.1973K}, M21 and \cite{2023MNRAS.526.3273C}.

\section{Complementary Figures}
\label{appendix:morefigures}

\begin{figure*}
\centering
\includegraphics[width=2\columnwidth]{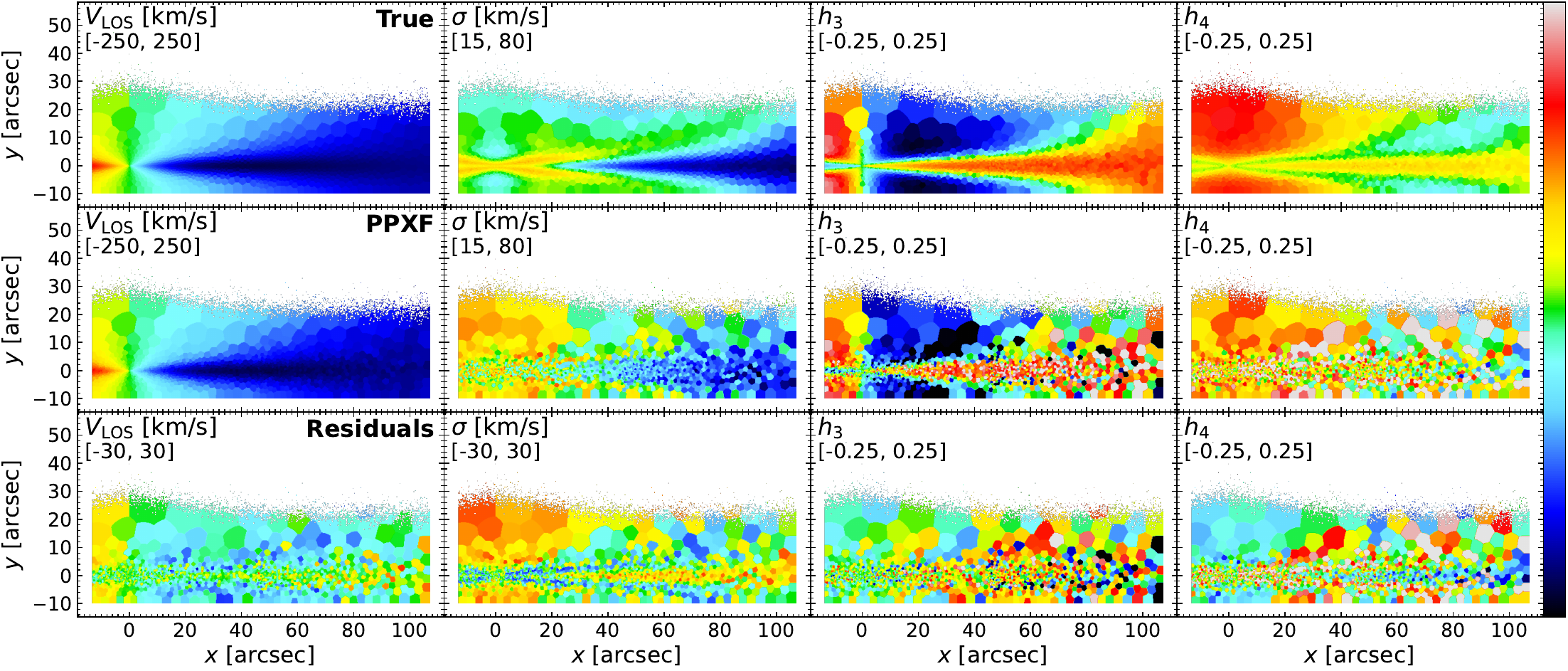}
\caption{
Same as Fig.~\ref{fig:galaxymapkin} but turning off the kinematics penalization during \ppxf{} fitting. 
}
\label{appfig:galaxymapkin_nopenal}
\end{figure*}

\begin{figure*}
\centering
\includegraphics[width=2\columnwidth]{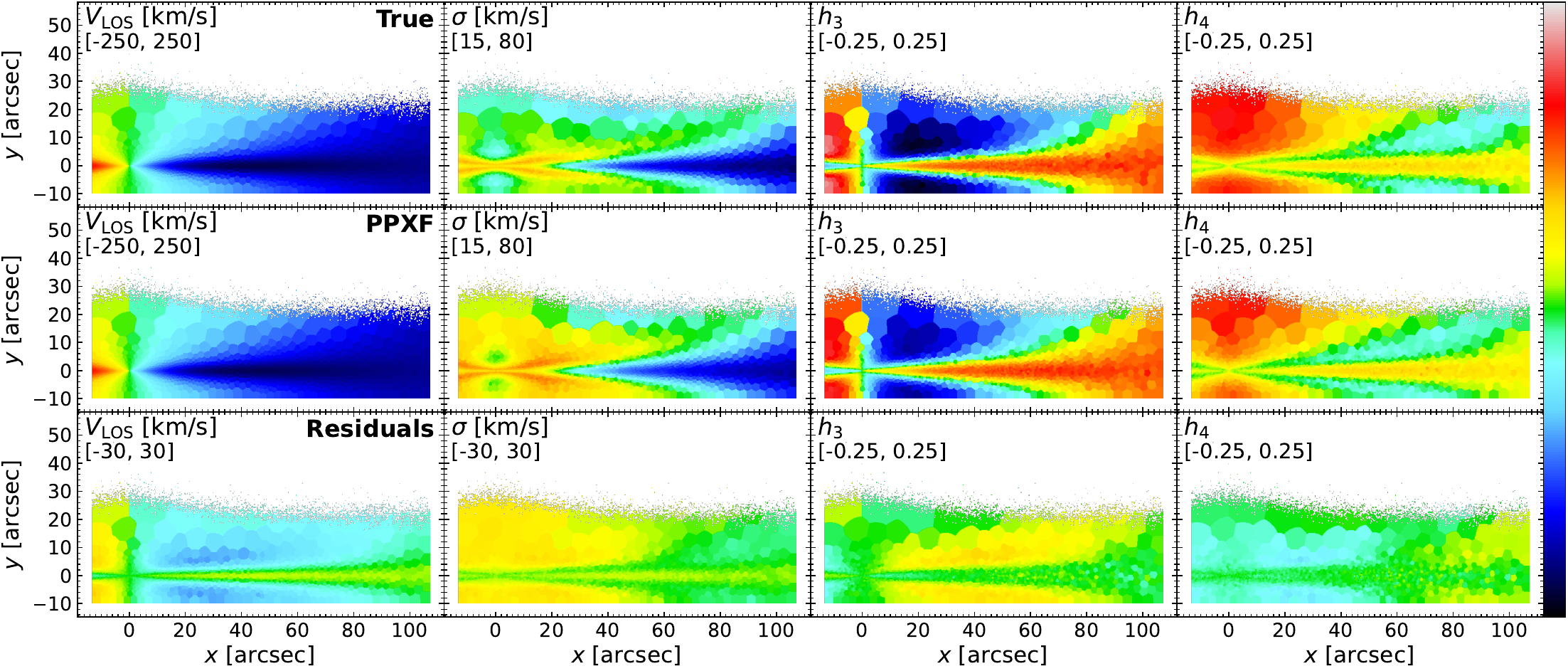}
\caption{
Same as Fig.~\ref{fig:galaxymapkin} but for the mock cube generated and fitted by PEGASE-HR templates in PEGASE-HR spectral resolution ($\rm{FWHM}=0.55\Angstrom$). 
}
\label{appfig:galaxymapkin_pegasehr}
\end{figure*}

\begin{figure*}
\centering
\includegraphics[width=2\columnwidth]{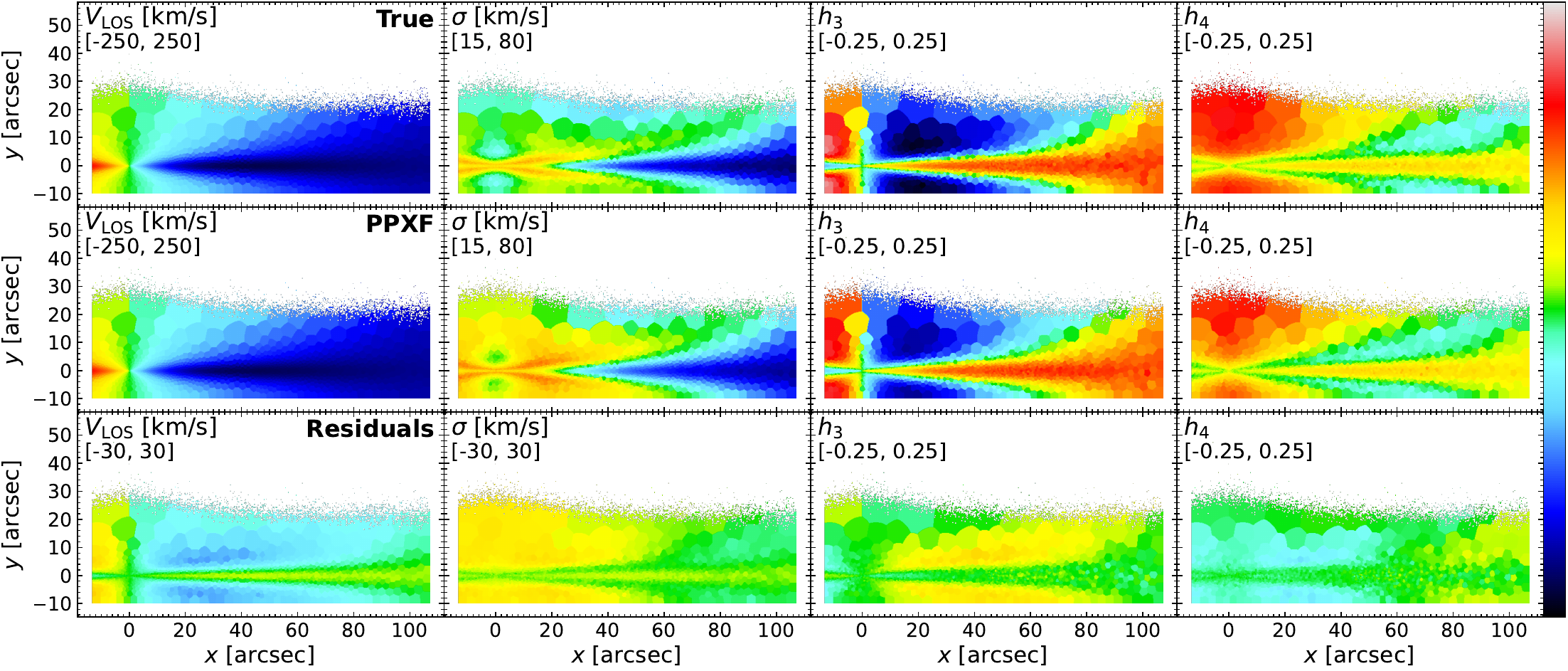}
\caption{
Same as Fig.~\ref{appfig:galaxymapkin_pegasehr} but turning off the kinematics penalization during \ppxf{} fitting. 
}
\label{appfig:galaxymapkin_pegasehr_nopenal}
\end{figure*}

\begin{figure*}
\centering
\includegraphics[width=2\columnwidth]{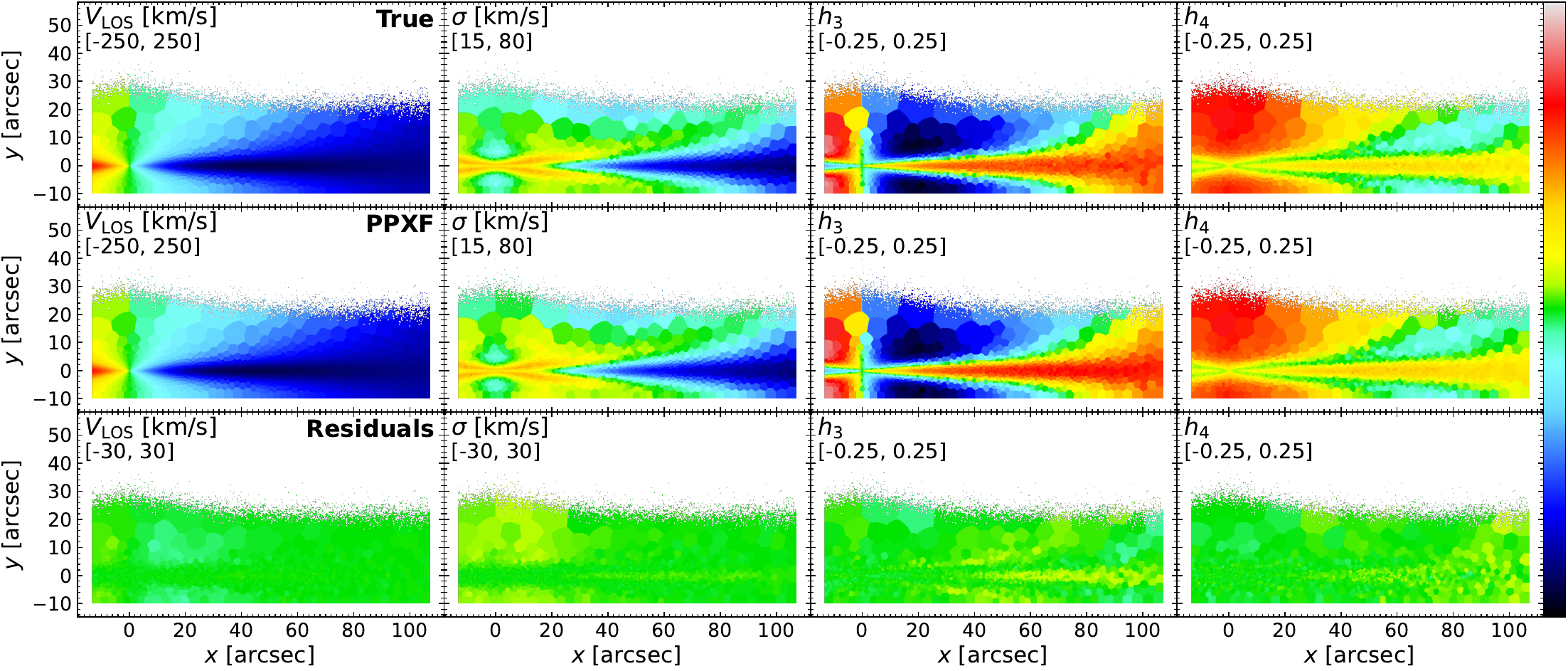}
\caption{
Same as Fig.~\ref{fig:galaxymapkin} but for LOSVD-Convolved mock cubes generated and fitted by PEGASE-HR templates in PEGASE-HR spectral resolution ($\rm{FWHM}=0.55\Angstrom$).
}
\label{appfig:galaxymapkin_pegasehrshufflevrRE}
\end{figure*}

\begin{figure*}
\centering
\includegraphics[width=2\columnwidth]{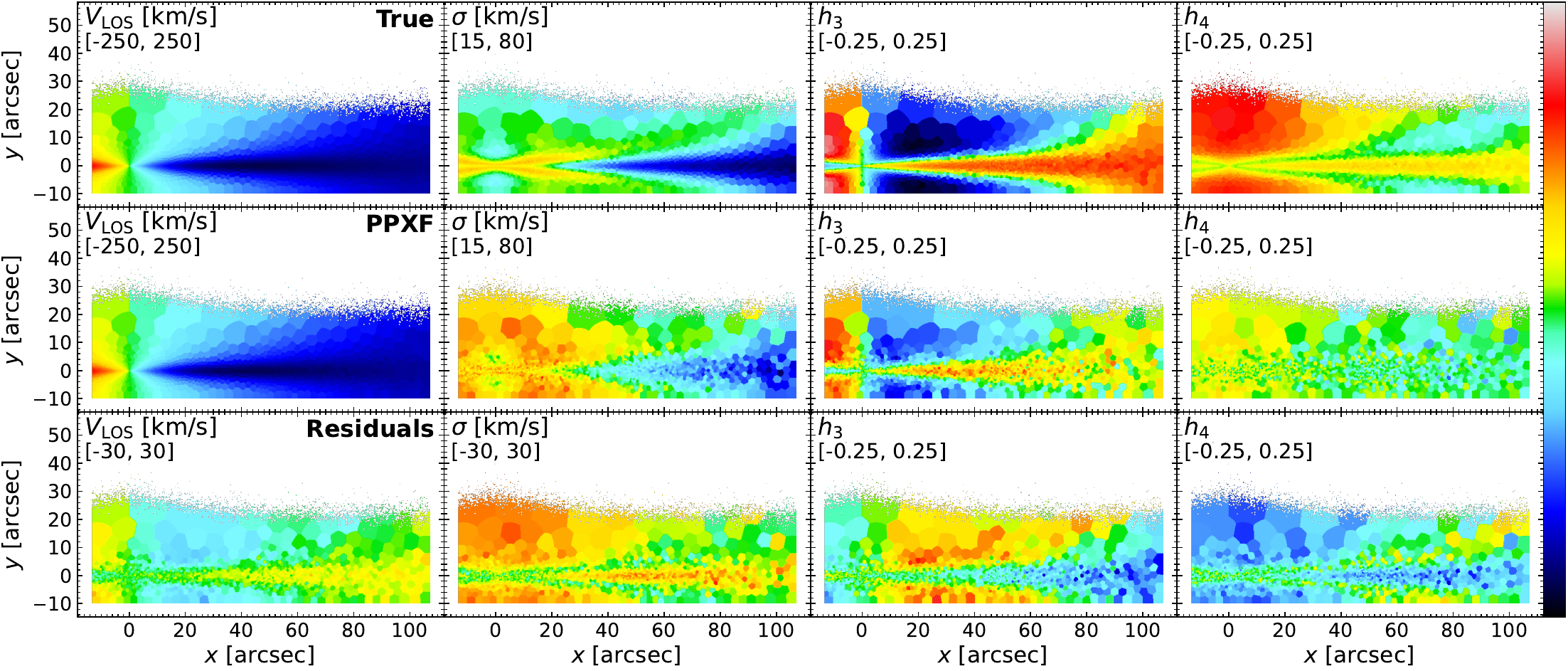}
\caption{
Same as Fig.~\ref{fig:galaxymapkin} but for LOSVD-Convolved mock cubes generated and fitted using MILES templates in MUSE spectral resolution ($\rm{FWHM}=2.65\Angstrom$).
}
\label{appfig:galaxymapkin_shufflevrRE}
\end{figure*}

\begin{figure*}
\centering
\includegraphics[width=2\columnwidth]{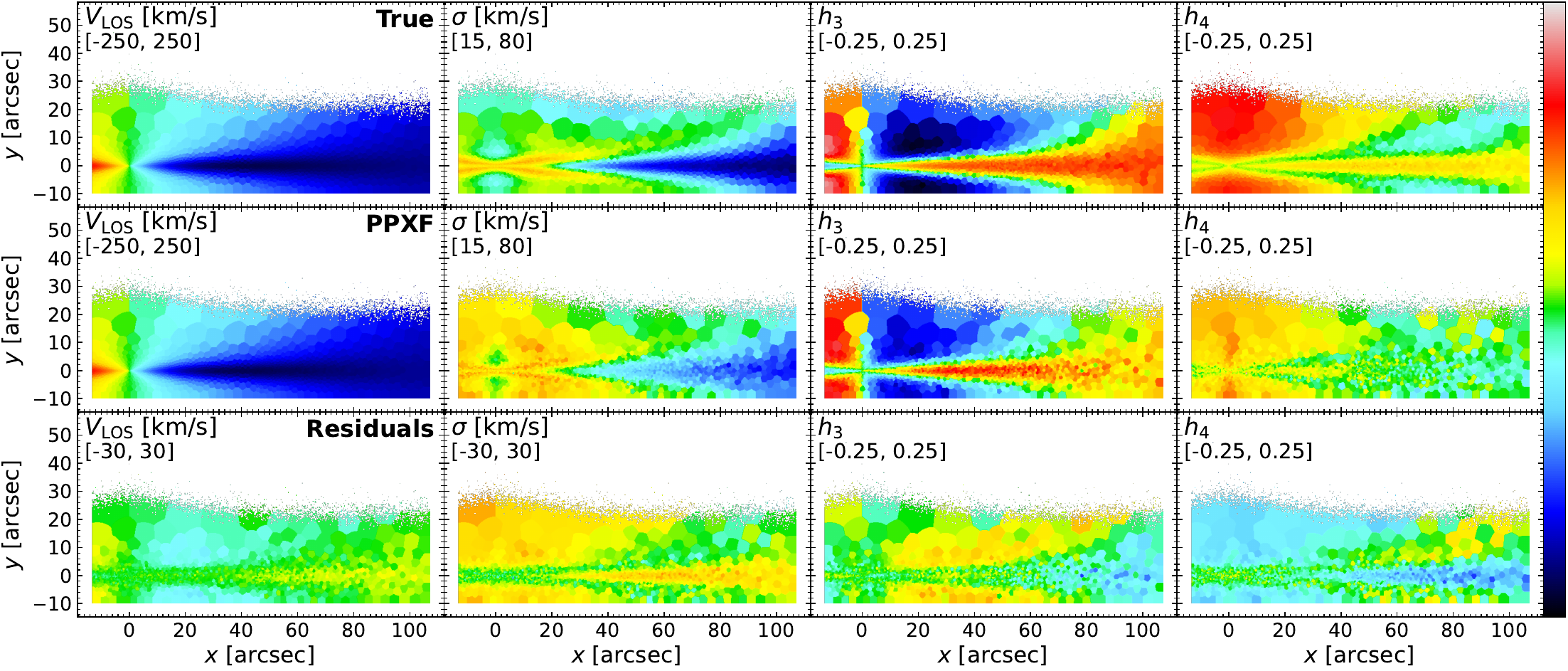}
\caption{
Same as Fig.~\ref{fig:galaxymapkin} but for LOSVD-Convolved mock cubes generated and fitted by PEGASE-HR templates in MUSE spectral resolution ($\rm{FWHM}=2.65\Angstrom$).
}
\label{appfig:galaxymapkin_pegasehrloresshufflevrRE}
\end{figure*}

\begin{figure*}
\includegraphics[width=2.0\columnwidth]{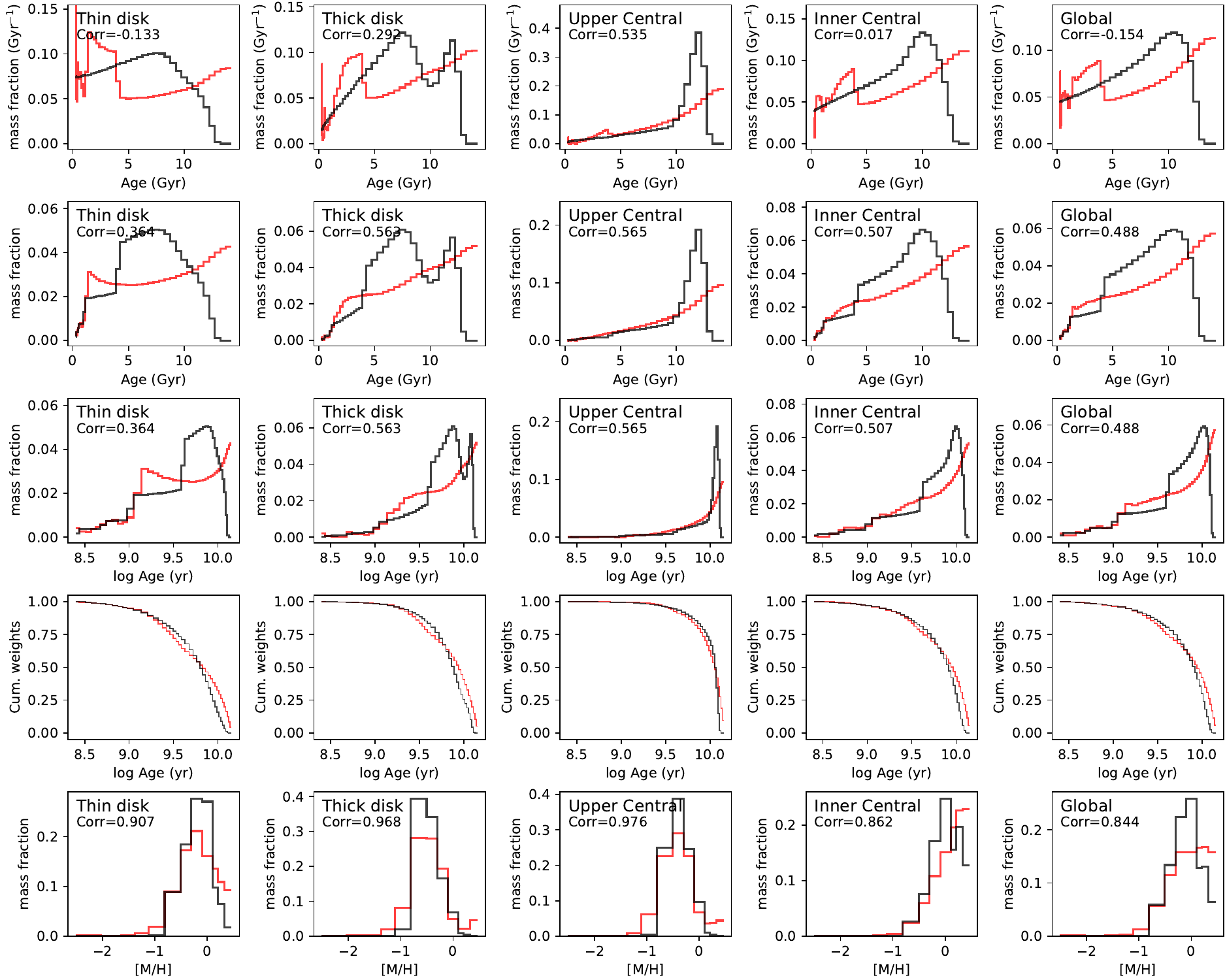}
\caption{
Same as Fig.~\ref{fig:massfractioncomponents1d} and Fig.~\ref{fig:massfractioncomponents1dlogage} but for LOSVD-Convolved mock cubes generated using MILES templates in MUSE spectral resolution ($\rm{FWHM}=2.65\Angstrom$).
}
\label{appfig:massfractioncomponents1d_shufflevrRE}
\end{figure*}

\begin{figure*}
\includegraphics[width=2.0\columnwidth]{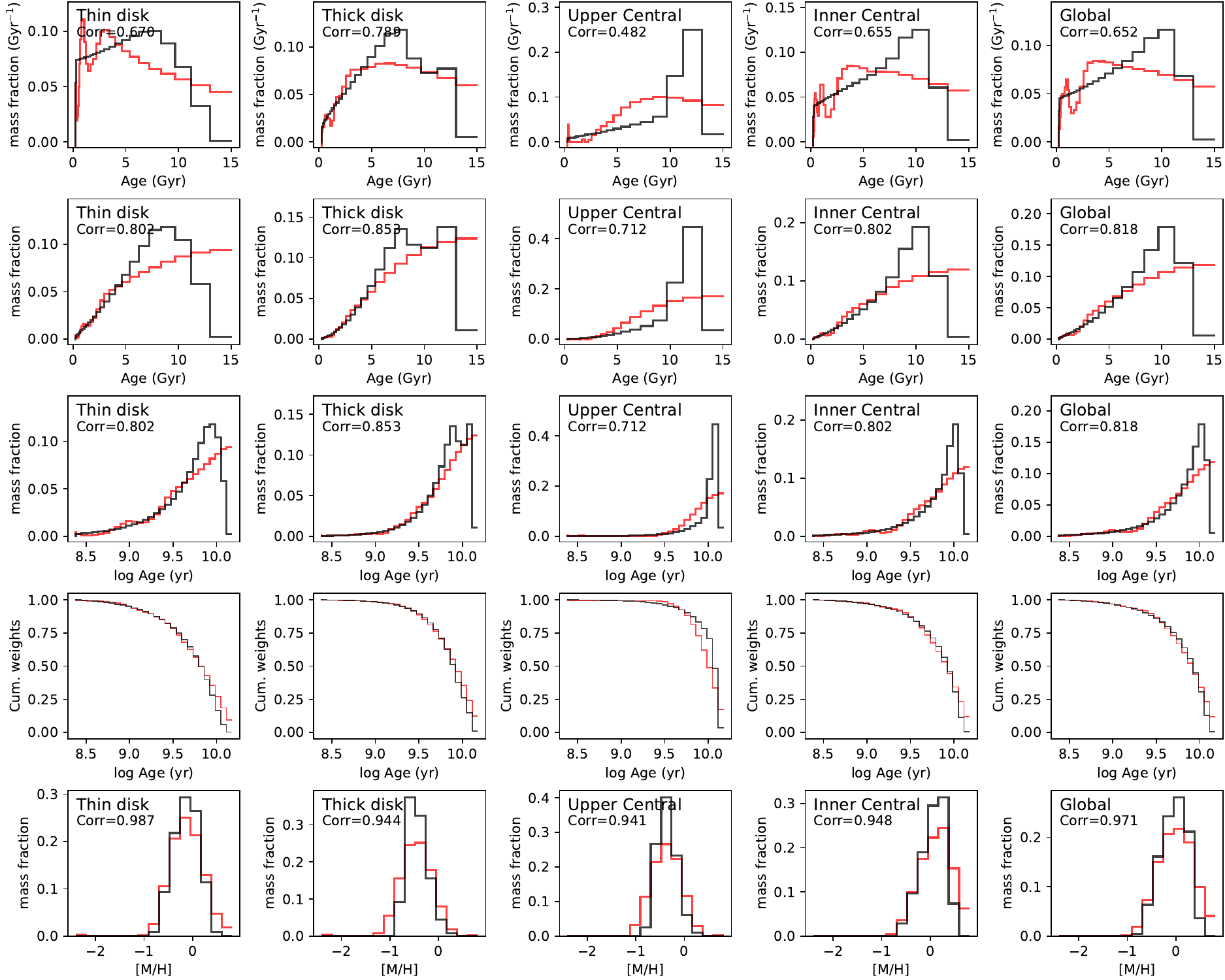}
\caption{
Same as Fig.~\ref{fig:massfractioncomponents1d_pegasehr_lores} but for mock cubes generated using PEGASE-HR templates in PEGASE-HR spectral resolution ($\rm{FWHM}=0.55\Angstrom$).
}
\label{appfig:massfractioncomponents1d_pegasehr}
\end{figure*}

\begin{figure*}
\includegraphics[width=2.0\columnwidth]{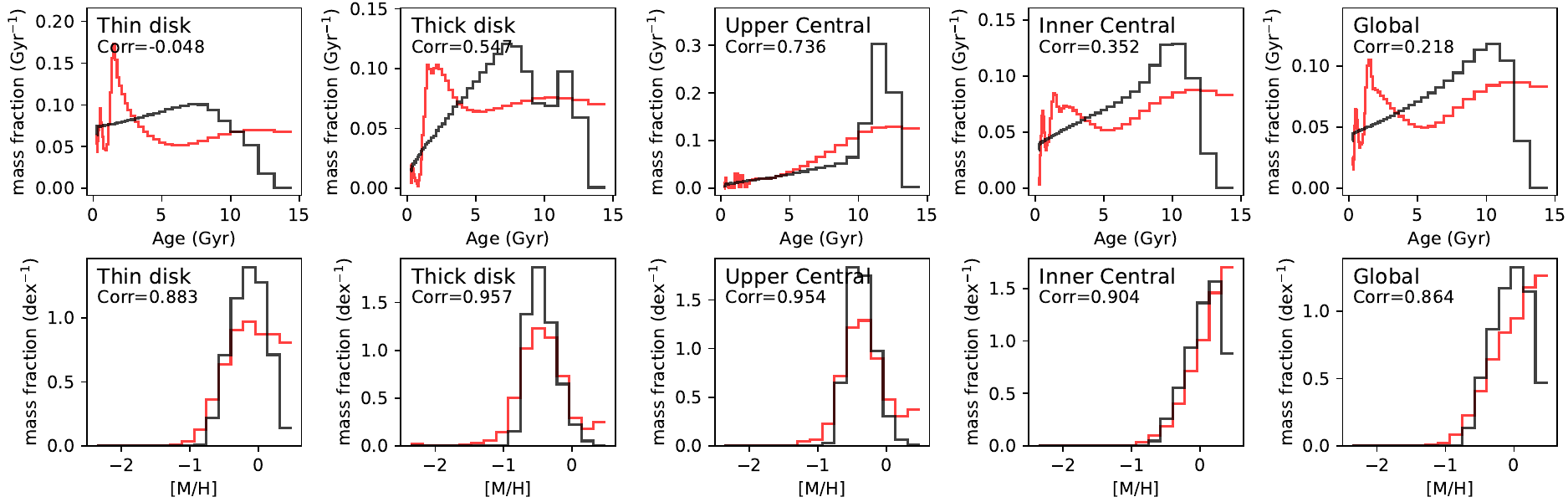}
\caption{
Same as Fig.~\ref{fig:massfractioncomponents1d} but for mock cubes generated using MILES templates interpolated linearly in $\log$~age and \mh{} in MUSE spectral resolution.
}
\label{appfig:massfractioncomponents1dloginterp}
\end{figure*}


\bsp	
\label{lastpage}
\end{document}